\documentclass{jpsj3}
\usepackage{graphicx}
\usepackage{bm}

\title{Symmetry and Topology in Superconductors \\
- Odd-frequency pairing and edge states - }
\author{$^1$\name{Yukio \surname{Tanaka}}
\thanks{E-mail address ytanaka@nuap.nagoya-u.ac.jp},
$^2$\name{Masatoshi \surname{Sato}}
\thanks{E-mail address msato@issp.u-tokyo.ac.jp},
and $^{3,4}$\name{Naoto \surname{Nagaosa}}
\thanks{E-mail address nagaosa@ap.t.u-tokyo.ac.jp}
}
\inst{\address{$^1$Department of Applied Physics, Nagoya University, Nagoya,
464-8603, Japan\\
$^2$The Institute for Solid State Physics, 
The University of Tokyo, 
Kashiwanoha 5-1-5, Kashiwa, Chiba, 277-8581, Japan\\
$^3$Department of Applied Physics, The University of Tokyo, 
7-3-1 Hongo,
Bunkyo-ku, Tokyo, 113-8656, Japan\\
$^4$Cross-Correlated Materials Research Group (CMRG) 
and Correlated Electron Research Group (CERG), 
RIKEN-ASI, Wako, Saitama 351-0198, Japan
}}
\abst{Superconductivity is a phenomenon where the
macroscopic quantum coherence appears due to the pairing of
electrons. This offers a fascinating arena to study the physics of
symmetry breaking, {\it i.e.}, broken gauge symmetry. However, the important
symmetries in superconductors are not only the gauge
invariance. Especially, the symmetry properties of the pairing, {\it i.e.},
the parity and spin-singlet / spin-triplet, determine the physical
properties of the superconducting state. Recently it has been recognized
that there is the important third symmetry of the pair amplitude,
{\it i.e.}, even or odd parity with respect to the frequency. The conventional
uniform superconducting states correspond to the even-frequency pairing,
but the recent finding is that the odd-frequency pair amplitude
arises in the spatially non-uniform situation quite
ubiquitously. Especially, this is the case in the Andreev bound state
(ABS) appearing at the surface/interface of the sample. The other
important recent development is on the nontrivial topological aspects of
superconductors. As the band insulators are classified by topological
indices into (i) conventional insulator, (ii) quantum Hall insulator,
and (iii) topological insulator, also are the gapped
superconductors. The influence of the nontrivial topology of the bulk
states appears as the edge or surface of the sample, {\it i.e.}, bulk-edge
correspondence. In the superconductors, this leads to the formation of
zero energy ABS (ZEABS). Therefore, the ABSs at the surface/interface of
the superconductors are the place where the symmetry and topology meet
each other which offer the stage of rich physics. In this review, we
discuss the physics of ABS from the viewpoint of the odd-frequency
pairing, the topological bulk-edge correspondence, and the interplay of
these two issues. It is described how the symmetry of the pairing and
topological indices determine the absence/presence of the ZEABS, its
energy dispersion, and properties as the Majorana fermions. 
Various
related issues such as the Helium 3, transport of Majorana fermions, and
tunneling spectroscopies are also briefly discussed.
}

\kword{Odd-frequency pairing, Andreev bound state, Bulk-edge correspondence, Majorana fermion, Topological index}

\begin{document}
\maketitle

\section{Introduction}
Two of the most important principles in physics are 
symmetry and topology, both of which are related to the 
quantum numbers and energy degeneracy of states \cite{Thouless}. 
For the compact
symmetry group such as the $2 \pi$ rotation, the discrete quantum numbers
result. However, even a slight breaking of  
the symmetry destroys these features, {\it i.e.}, the conservation law is 
violated and 
the degeneracy is lifted. In the many-body systems, it often happens that
the symmetry is {\it spontaneously} broken as a collective phenomenon.
Ferromagnetism and superconductivity are two representative 
examples of this spontaneous symmetry breaking characterized by
the order parameters. The magnetization, which breaks the rotational
symmetry in the spin space, and the pair amplitude, which breaks the
gauge symmetry, are the corresponding order parameters, respectively. 
In sharp contrast to this, the topological quantum numbers  are more robust 
against the perturbations. For example, the quantization of the magnetic flux 
captured by one vortex in a superconductor is quantized to be 
$\phi_0 = hc/(2e)$ 
even in the disordered case without the cylindrical symmetry. 
This is guaranteed by the single-valued property of the superconducting 
order parameter and the stability of its winding number.  It is evident 
that the winding number around a singular point
( the center of the vortex in the above example) 
does not change for the continuous deformation  
as long as the path does not cross the singular point. 
Therefore, the topological properties often lead to the
robustness of the system against the weak and moderate 
perturbations protected by the topological quantum number. 

It is not always clear how to separate the two origins of 
the quantum numbers, {\it i.e.}, symmetry and topology. 
Even more interesting is the interplay between these two principles.
It often happens that
they are entangled to give a rich variety of physical phenomena. 
In this review article, we describe how these two principles are 
merged into the novel properties of superconductors. \par

\subsection{Symmetry classifications of bulk superconductors 
and relevant quantum phenomena}
Symmetry of Cooper pair is a central topic in the physics of 
superconductivity.
Here we discuss uniform bulk systems.
It is established that Cooper pair is formed between two electrons. 
In accordance with the Fermi-Dirac statistics, 
pair amplitude, which is a wave function of Cooper pair, must have 
a sign change with the exchange of two electrons. 
Symmetry of 
Cooper pair is customary classified into 
spin-singlet even-parity and spin-triplet odd-parity 
where even (odd) refers to the orbital part of 
the pair amplitude. For example, $s$-wave and $d$-wave pairings 
belong to the former case while $p$-wave pairing 
belongs to the latter \cite{Sigrist}. 
With broken inversion symmetry, 
spatial parity is no more a good quantum number. 
In that case, spin-singlet even-parity 
and spin-triplet odd-parity pairings can mix with each other. 
The mixed parity pairing has become an important issue 
in non-centrosymmetric superconductors \cite{Bauer,Sigrist07}. 
It is noted that 
both in spin-singlet even-parity and spin-triplet odd-parity pairings, 
the pair amplitude does not have a sign change 
with the exchange of two time variables for 
the two electrons forming the Cooper pair.
However, there is a possibility that 
pair amplitude has a sign change by this operation 
\cite{exchange}.  
The latter type of pairing is so called odd-frequency pairing 
originally proposed by Berezinskii \cite{Berezinskii}. 
There are two possibilities of odd-frequency pairing, 
$i.e.$, spin-singlet odd-parity and spin-triplet even-parity 
pairings. 
Thus, in the presence of both the inversion and spin-rotational 
symmetries, Cooper pairs are classified into 
(i)even-frequency spin-singlet even-parity (ESE), 
(ii)even-frequency spin-triplet odd-parity (ETO),  
(iii)odd-frequency spin-triplet even-parity (OTE), 
and 
(iv)odd-frequency spin-singlet odd-parity (OSO). 
In strongly correlated systems, due to 
the retardation effect of the electron interaction, 
there is a possibility 
for odd-frequency pairing in the bulk. 
After the prediction by Berezinskii\cite{Berezinskii}, 
there have been several theoretical proposals 
about odd-frequency pairing in the bulk 
\cite{Balatsky,Balatsky2,Coleman,Coleman2,Vojta,Fuseya,Hotta,Shigeta,Shigeta2,
Kusunose,Kusunose2}. 
However, up to now, 
odd-frequency bulk superconductor has not been established 
experimentally yet.  
Although it is not easy to realize odd-frequency 
gap function in uniform bulk systems, 
it is more promising to induce odd-frequency pair amplitude 
in the non-uniform system with lower  
symmetry. 

\begin{center}
\begin{table}[h]
\begin{tabular}{|p{1.3cm}|p{1.3cm}|p{1.3cm}|p{1.3cm}|p{1.3cm}|}
\hline
Class & Time & Spin & Orbital & Total \\ \hline
ESE & + & $-$ & + & $-$ \\ \hline
ETO & + & + & $-$ & $-$ \\ \hline
OTE & $-$ & + & + & $-$ \\ \hline
OSO & $-$ & $-$ & $-$ & $-$ \\ \hline
\end{tabular}%
\caption{The symmetry of pair amplitude 
with respect to the exchange of spins, spatial coordinates, 
and time variables 
for possible four classes.   
ESE, ETO, OTE and OSO denote 
even-frequency spin-singlet even-parity, 
even-frequency spin-triplet odd-parity, 
odd-frequency spin-triplet even-parity, 
and 
odd-frequency spin-singlet odd-parity.  
 }
\label{table:1}
\end{table}
\end{center}
%
Bergeret, Volkov and Efetov have clarified that 
in ferromagnet/superconductor heterostructures with 
inhomogeneous magnetization, 
odd-frequency pairing is generated in ferromagnet \cite{Efetov1}. 
They have predicted the long range 
proximity effect by the odd-frequency pairing. 
Furthermore, one of the authors (Y.T.) has revealed that 
the odd-frequency pairing  is possible
in inhomogeneous superconducting systems  
even without magnetic ordering \cite{odd1,odd3,odd3b}. 
We will show ubiquitous presence of odd-frequency pairing in \S \ref{sec:2} 
\cite{odd3,Eschrig2007}. \par
Symmetry of the Cooper pair influences tunneling phenomena 
in superconducting heterostructures. 
It is known that Andreev bound state (ABS) 
is generally generated at the surface of  anisotropic superconductor 
where pair potential (energy gap function) has a 
sign change on the Fermi surface 
\cite{ABS,ABSb,ABSR1,ABSR2,Hu,TK95,MS95,TK96a,ATK04}. 
By properly taking into account of the 
Andreev reflection in anisotropic superconductors
\cite{Andreev64,BTK82,Bruder90}, 
it has been proven that  the ABS manifests itself  as a zero bias 
conductance peak (ZBCP) of quasiparticle 
tunneling spectroscopy \cite{TK95,ABSR1,KT96}.  
The presence of ZBCP has been 
observed in tunneling spectroscopy of 
high $T_{c}$ cuprates 
\cite{Experiment1,Experiment2,Experiment3,Experiment4,Experiment5,
Iguchi00,Experiment6,Experiment7}. 
It has been established that a dispersionless zero energy ABS (ZEABS)
\cite{ABSR1} is generated 
at the surface for spin-singlet $d_{xy}$-wave superconductors.  
The existence of the ABS influences seriously on 
charge transport \cite{Josephson,Josephson2,Josephson3}, 
spin transport \cite{Spin} and magnetic responses
\cite{Meissner,Meissner2,Meissner3}.  \par
ABS is also expected for spin-triplet $p$-wave superconductors. 
It induces ZBCP in the 
tunneling spectroscopy \cite{YTK97,YTK98}. 
However, proximity effect via ABS 
is completely different between 
spin-singlet superconductors and spin-triplet 
superconductors. 
The ABS cannot penetrate into a diffusive normal 
metal (DN) attached to the superconductor in the former case, 
while it is possible 
in the latter one. 
The underlying physics can be expressed by the symmetry of 
the induced odd-frequency pairing \cite{odd1}. 
We will discuss these exotic phenomena in \S \ref{sec:2}. 

\subsection{Topological orders and topological quantum numbers}
\label{sec:TOandTQN}

Up to now, we have discussed the classifications of the 
superconductors by symmetry as an example of quantum states. 
Then, is there any other way to characterize the quantum 
phases of electronic systems ? 
The answer is yes, and 
the topological orders 
and topological quantum numbers serve this purpose \cite{Wen}. 
Historically, the first topological ordered state 
in condensed matter physics is the quantum Hall 
system, $i.e.$, the two-dimensional
electronic systems under strong magnetic field at
low temperature, where the Hall conductance $\sigma_H$ is quantized to be
the integer times the unit $e^2/h$ (integer quantum Hall effect (IQHE))
 or its fractions (fractional quantum Hall effect (FQHE)) 
 ($e$: unit charge, $h$: Planck constant). 
Here, the topological 
quantum number $C_1$ (integer, see eq.(\ref{eq:C1})) characterizing the 
topological order is related to the physical observable, {\it i.e.}, the
Hall conductance $\sigma_H$ as
\begin{equation}
\sigma_H = - { {e^2} \over h} C_1,
\label{eq:Hall1}
\end{equation}
which explains why $\sigma_H$ is quantized as the integer 
multiple of $e^2/h$~\cite{QHE}. 

Another important issue related to the IQHE is the 
one-dimensional channels along the edge of the sample~\cite{HalperinEdge}.
A classical picture for these edge channels is that the cyclotron motion 
of the electrons in the bulk is reflected at the boundary of the sample,
forming the one-dimensional motion in one-direction. 
This property of "one-direction" is called "chiral". 
In the quantum mechanical picture, the two-dimensional  
electronic states under magnetic field form the Landau levels,
whose energies are pushed up near the edges of the sample by 
the confining potential. As a consequence, some of the edge channels
cross the Fermi energy, which carry current. 
A remarkable fact is that the direction of the propagating 
waves along these edge channels is one-way, {\it i.e.}, chiral, and
those with the opposite direction are on the other side of the 
sample, {\it i.e.}, far away separated by the bulk. 
(Consider the cylindrical sample where the two edges are 
separated by the bulk.) 
Therefore, the backward scattering
is strictly forbidden even if the impurity potential is there. 
This leads to the perfect one-dimensional conductance of each channel,
and consequently the Hall conductance is given by  
\begin{equation}
\sigma_H =  { {e^2} \over h} n
\label{eq:Hall2}
\end{equation}
where $n$ is the number of the edge channels including the sign 
representing the chirality. 
Comparing eqs.(\ref{eq:Hall1}) and (\ref{eq:Hall2}),
one concludes that there is a relation between the bulk
topological quantum number ($C_1$) 
and the edge channels ($n$) \cite{Hatsugai93, Hatsugai97}. This is called 
"bulk-edge correspondence" and will be explored more in details later.
One can say that the one-dimensional electronic states are 
"split into halves" at the two edges of the sample according to
the chiralities, and the fractionalized electrons become robust.
This fractionalization of the electrons at the edge is 
realized/supported by the topological order in the bulk state. 
  
FQHE is a more complex phenomenon where 
the electron-electron interaction plays the essential role. 
Therefore, one needs 
to study the many-body wave functions, and the topological quantum numbers
characterizing the quantum states are much more intriguing.
The readers are referred to the textbooks for the details \cite{Thouless,Wen}. 
However, conceptually the story is quite analogous and similar to the
IQHE. The topological order in the bulk reflects itself as the edge channels
at the boundary of the sample, {\it i.e.}, the bulk-edge correspondence occurs.
 
 A recent breakthrough is the discovery of the topological insulators which 
preserve the time-reversal symmetry $T$, {\it i.e.}, without the external magnetic 
field or the spontaneous magnetization \cite{HK10,QZ10}. 
With $T$-symmetry, the
Hall response and hence the Chern number is zero. Therefore, it is required
to define new topological quantum numbers to characterize 
the topological orders. 

  The concept of topologically nontrivial states of the single-particle
Hamiltonian can be generalized to the superconducting state as long as
one considers the BCS Hamiltonian in the Nambu representation; 
\begin{equation}
H_{BCS} = \sum_{\bm k} \psi^\dagger_{\bm k} h({\bm k}) \psi_{\bm k}
\end{equation}
where 
$
\psi_{\bm k} = ( c_{n {\bm k} \uparrow}, c_{n {\bm k} \downarrow}, 
c^\dagger_{n {\bm k} \uparrow},  c^\dagger_{n {\bm k} \downarrow} )^T.
$
A naive way is to consider the matrix elements of $h({\bm k})$
as that of the Bloch states discussed above. 
However, the situation is not so trivial since  
the degrees of freedom are redundant in 
$\psi_{\bm k}$ and $\psi^\dagger_{\bm k}$.  
Consequently, the edge channels and surface states 
originating from the bulk-edge correspondence 
has the same redundancy,  
which sometimes leads to Majorana fermion.  
The present edge channels and surface states 
appear as the ABSs characterized by the 
symmetry of Cooper pair as discussed in 
the former subsection. 
Therefore, the physics is
much richer in this topological superconductor
than the non-interacting electron systems.

\begin{figure}[tb]
\begin{center}
\scalebox{0.8}{
\includegraphics[width=7cm]{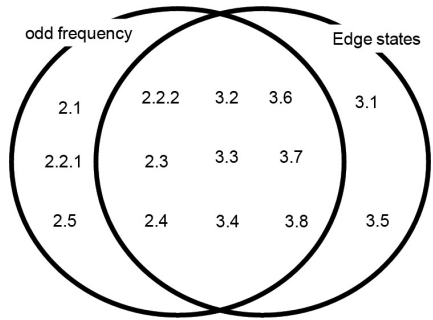}
}
\end{center}
\caption{The relations between each subsection and two key concepts, 
{\it i.e.}, odd-frequency pairing and edge states, in the present review. 
}
\label{fig:plan}
\end{figure}

In this review, we discuss the recent developments on the 
topological superconductors as the merging point of the
two principles, {\it i.e.}, symmetry and topology. The in-depth study of 
this issue will clarify the interplay between these two principles 
in physics. The plan of this review follows. In \S \ref{sec:2},  
the odd-frequency superconducting pairing 
induced by the extrinsic symmetry breaking 
such as spatial non-uniformity and/or the time-reversal symmetry breaking
is studied 
to demonstrate how the superconductivity is influenced by the 
symmetries.  In \S \ref{sec:3}, we discuss the topology of 
the bulk states in superconductors and its relation to the
surface/edge states (bulk-edge correspondence).
Various related issues are described in \S \ref{sec:4}. \S
\ref{sec:4} is also 
devoted to the summary and future perspectives.
Hereafter, we choose the unit $\hbar = k_{B}=1$ 
unless explicitly written. 
Two key concepts in this review are odd-frequency pairing and 
edge states. 
In Fig.\ref{fig:plan}, we show the relations between 
each subsection
and above two key concepts explicitly. 
The readers who want to know the essence of the review urgently 
would be appreciated to look at Table IX in \S \ref{sec:4}.

\section{Extrinsic symmetry breaking and odd-frequency pairing 
in superconductors}
\label{sec:2}

\subsection{What is extrinsic symmetry breaking ?}
 In this section, we discuss what is an extrinsic symmetry breaking. 
Typical examples are 
breaking of the translational invariance or 
that of spin-rotational symmetry. 
In general, Cooper pair is formed between position 
${\bm r}_{1}$ and ${\bm r}_{2}$. 
Pair amplitude is a function of 
${\bm r}=({\bm r}_{1} + {\bm r}_{2})/2$ and 
$\bar{{\bm r}}=({\bm r}_{1} - {\bm r}_{2})$.  
When we focus on the symmetry with respect to the exchange of 
two coordinates ${\bm r}_{1}$ and ${\bm r}_{2}$, 
{\it i.e.}, $\bar{{\bm r}} \rightarrow -\bar{{\bm r}}$,  
we can define the parity of the Cooper pair. 
We employ the  Fourier transformation from 
$\bar{{\bm r}}$ to ${\bm k}$. 
If we consider inhomogeneous superconducting 
systems, like vortex and junctions, translational symmetry 
is broken. 
In that system, parity of the Cooper pair 
is no more a good quantum number. Then, 
near the vortex core or the interface / surface of the 
superconductor, mixed parity state can be realized. 
To be more specific, let us consider normal metal / 
superconductor (N/S) junction where the symmetry of the 
superconductor belongs to ESE. 
Near the interface or in the normal metal, 
due to the breaking of translational symmetry, 
odd-parity state can be mixed. 
Since the spin rotational symmetry is not broken, 
pairing symmetry of spin should be singlet. 
To be consistent with Fermi Dirac statistics, 
the resulting Cooper pair should be odd in frequency. 
Then, we can expect OSO paring is generated \cite{odd1} near the interface and also inside normal metal. 
On the other hand, if symmetry in the bulk is ETO,  then the 
induced pairing symmetry by breaking of the translational invariance or the 
spatial inversion symmetry is OTE. 

\par
Next, we discuss the breaking of the spin rotational symmetry only. 
By this symmetry breaking, 
spin-singlet and spin-triplet parings are mixed with each other. 
Let us consider superconductor with ESE symmetry.  
In the presence of Zeeman magnetic field, 
spin-triplet pairing state is generated. 
Since spatial parity is not broken, the induced 
pair amplitude  should be OTE. 
On the other hand, if the pairing in the bulk is ETO,  then the 
induced pair amplitude  by  breaking of 
spin rotational symmetry is OSO. \par
From this viewpoint, 
paring symmetry in ferromagnet / superconductor (F/S) junction is very 
interesting. 
Here, we specify that the symmetry of the superconductor is 
spin-singlet $s$-wave one which belongs to 
ESE. 
In the light of above discussion, OSO pairing is induced in F 
by translational symmetry breaking. Also by the spin rotational symmetry 
breaking, OTE (ETO) pairing is generated from 
ESE (OSO) pairing. Thus, in F, 
all four kinds of pairings are basically possible \cite{Eschrig2007}. 
In the standard case accessible in experiments, 
F is in the diffusive regime. 
Then, basically only $s$-wave pairing is possible 
since it is robust against impurity scattering \cite{Anderson}.  
The resulting paring symmetries in F are only ESE $s$-wave and OTE $s$-wave. 
It has been shown that OTE pairing seriously influences on the 
density of states of quasiparticles. 
When OTE $s$-wave pairing dominates, 
quasiparticle density of state does not have a gap-like structure around 
zero energy\cite{Yokoyama2007}. 
In the extreme case, it has a zero energy peak (ZEP) structure 
\cite{Yokoyama2007,Yokoyama2006}. 

In \S \ref{sec:2b} and \S \ref{sec:2c}, we restrict 
our discussion to ballistic transport regime. 
We will discuss relevant problems in diffusive regime in 
\S \ref{sec:2d} and \S \ref{sec:2e}. 

\subsection{N/S junction with ballistic system} 
\label{sec:2b}

In this subsection, we consider superconducting proximity effect and 
resulting ubiquitous presence of odd-frequency pairing 
in N / S junctions \cite{TGKU07}. 
In general, physical quantities have spatial dependence near the 
interface. 
There are two characteristic length scales: 
i) The inverse of the Fermi wave length and 
ii) coherence length $\xi$ given by 
 $\xi=v_{F}/\Delta$, where $v_{F}$ is the 
Fermi velocity and $\Delta$ is the energy gap (pair potential) in the 
superconductor. 
%
%
Here, we are interested in physical quantities
varying slowly over the coherence length.
In the present situation, it is appropriate  to use 
quasiclassical Green's function method where high energy 
degrees of freedom are integrated out\cite{kopnin,Quasi,Quasi1,Serene}. 
The quasiclassical Green's function is obtained from Gor'kov Green's function  
by integrating the energy and is 
concentrating on the low energy scale measured from the Fermi energy level. 
In this approximation, 
quasiparticles feel the pair potential 
depending on their direction of their motions on the Fermi surface. \par
\begin{figure}[tb]
\begin{center}
\scalebox{0.8}{
\includegraphics[width=7cm,clip]{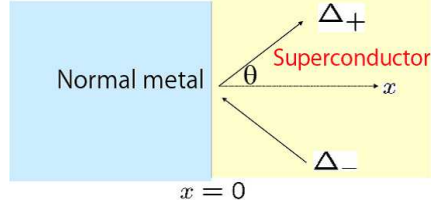}
}
\end{center}
\caption{(Color online)Schematic illustration of 
normal metal (semi-infinite) 
/ superconductor (semi-infinite）junction. 
$\Delta_{+}$,$\Delta_{-}$ denote pair potentials 
felt by right going (toward positive $x$) and 
left going (toward negative $x$) quasiparticles, respectively.  
}
\label{fig:1}
\end{figure}

In the following, 
we consider two-dimensional N ($x<0$) / 
S ($x>0$) junctions in the 
ballistic limit (Fig. \ref{fig:1}). 
For the bulk superconductor, 
we choose even-frequency ESE or ETO pairings. 
As regards the 
spin-triplet ETO pairing, we choose $S_{z}=0$ for simplicity. 
There is no essential difference even if we consider spin-triplet pairing 
with $S_{z}=1$. 
We assume a thin insulating barrier located at the N/S interface ($x=0$) 
modeled by a delta functional potential $H\delta (x)$. 
The reflection coefficient and transmission coefficient 
of the junction for the quasiparticle with the
injection angle $\theta $ is given by 
$R=Z^{2} /(Z^{2}+4\cos ^{2}\theta)$ and 
$T_{m}=1-R$, respectively, 
with $Z=2H/v_{F}$, where $\theta $ $(-\pi /2<\theta <\pi /2)$ is
measured from the normal to the interface.

Here, we introduce quasiclassical retarded (advanced) 
Green's functions 
$g^{{\rm R(A)}}_{\pm}(\varepsilon,\theta,{\bm r})$, 
and $f^{{\rm R(A)}}_{\pm}(\varepsilon,\theta,{\bm r})$ where $\theta$ is given
by ${\bm k}=k_{F}(\cos\theta,\sin\theta)$. 
$g^{{\rm R(A)}}_{+}(\varepsilon,\theta,{\bm r})$
($g^{\rm{R(A)}}_{-}(\varepsilon,\theta,{\bm r})$)
is the Green's function of quasiparticle 
at position ${\bm r}$, energy $\varepsilon$ 
with injection angle $\theta$ ($\pi-\theta$).
$f^{\rm{R(A)}}_{\pm}(\varepsilon,\theta,{\bm r})$ denotes anomalous 
Green's function  which becomes zero in the normal state.
$f^{\rm{R(A)}}_{\pm}(\varepsilon,\theta,{\bm r})$ can be decomposed into 
$f^{\rm{R(A)}}_{1\pm}(\varepsilon,\theta,{\bm r})$ and $f^{\rm{R(A)}}_{2\pm}(\varepsilon,\theta,{\bm r})$, 
\[
f^{\rm{R(A)}}_{\pm}(\varepsilon,\theta,{\bm r})=
f^{\rm{R(A)}}_{1\pm}(\varepsilon,\theta,{\bm r}) 
-i f^{\rm{R(A)}}_{2\pm}(\varepsilon,\theta,{\bm r}) 
\]
with 
\[
f^{\rm{R}}_{1\pm}(\varepsilon,\theta,{\bm r}) 
= -f^{\rm{A}}_{1\pm}(-\varepsilon,\theta,{\bm r}), 
\]  
\begin{equation}
f^{\rm{R}}_{2\pm}(\varepsilon,\theta,{\bm r}) 
= f^{\rm{A}}_{2\pm}(-\varepsilon,\theta,{\bm r}). 
\end{equation}
From this definition, it is clear that 
$f^{\rm{R(A)}}_{1\pm}(\varepsilon,\theta,{\bm r})$ 
($f^{\rm{R(A)}}_{2\pm}(\varepsilon,\theta,{\bm r})$) corresponds to 
the odd (even)-frequency pairing. 
Due to the translational invariance in the direction parallel 
to the 
interface $x=0$, 
we can simply replace ${\bm r}$ by $x$. 
$g^{\rm{R(A)}}_{\pm}(\varepsilon,\theta,x)=g^{\rm{R(A)}}_{\pm}$,
$f^{\rm{R(A)}}_{1\pm}(\varepsilon,\theta,x)=f^{\rm{R(A)}}_{1\pm}$,
and $f^{\rm{R(A)}}_{2\pm}(\varepsilon,\theta,x)=f^{\rm{R(A)}}_{2\pm}$
satisfy following so called the Eilenberger equation \cite{Eilenberger}

\begin{equation}
\mp i \mid v_{Fx} \mid \frac{\partial }{\partial x} \hat{g}^{\rm{R(A)}}_{\pm}
=[\hat{H},\hat{g}^{\rm{R(A)}}_{\pm}]
\end{equation}
with 
\[
\hat{H}=\varepsilon \hat{\tau}_{3} + i \hat{\tau}_{2}
\bar{\Delta}_{\pm}(x)
\]
and 
\[
\hat{g}^{\rm{R(A)}}_{\pm}=g^{\rm{R(A)}}_{\pm}\hat{\tau}_{3}
+ f^{\rm{R(A)}}_{1\pm}\hat{\tau}_{1}
+ f^{\rm{R(A)}}_{2\pm}\hat{\tau}_{2}. 
\]
Here, $\tau _{i}$ ($i=1-3$) is the Pauli matrix in the electron-hole space. 
If we write it explicitly decomposing each $\tau_{i}$ component, 
following three equations are derived 
\begin{equation}
\mp i\mid v_{Fx} \mid \partial_{x} f_{1\pm}=-2i \varepsilon f_{2\pm} 
-2\bar{\Delta}_{\pm}(x)g_{\pm},
\label{Eilen1}
\end{equation}
\begin{equation}
\mp i \mid v_{Fx} \mid\partial_{x} f_{2\pm}=2i\varepsilon f_{1\pm},
\label{Eilen2}
\end{equation}
\begin{equation}
\mp i\mid v_{Fx}\mid \partial_{x} g_{\pm}=2\bar{\Delta}_{\pm}(x)f_{1\pm}. 
\label{Eilen3}
\end{equation}
Here, we omit the  superscripts R and A. 
All the Green's functions in eqs. (\ref{Eilen1})-(\ref{Eilen3}) 
have the same analytical properties, $i.e.$, R and A are not mixed. 
In the above, $v_{Fx}=v_{F}\cos \theta$ 
is the $x$-component of the Fermi velocity and 
$\bar{\Delta}_{\pm}(x)$ is the 
pair potential felt by $\pm$ direction in Fig. \ref{fig:1}. 
In general, quasiparticle 
is scattered from electron(hole)-like quasiparticle 
to hole(electron)-like one by the pair potential. 
As seen from the Eilenberger equations, the scattering of quasiparticle by 
the pair potential is a driving force to produce 
pair amplitudes $f_{1\pm}$ and $f_{2\pm}$. 
It should be remarked that $\bar{\Delta}_{\pm}$ and 
$f_{2(1)\pm}$ are different physical quantities. 
Even if $\bar{\Delta}_{\pm}$ is absent in a certain place $x$, 
$f_{2(1)\pm}$ may take a nonzero value by the proximity effect. 
$\bar{\Delta}_{\pm }(x)$ can be expressed by 
$\bar{\Delta}_{\pm}(x)=\Delta (x)\Phi _{\pm }(\theta )\Theta(x)$ 
\cite{Quasi1}. 
For $s$, $d_{x^{2}-y^{2}}$,
$d_{xy}$, $p_{x}$ and $p_{y}$-wave superconductors, 
$\Phi_{+(-) }(\theta )$ is given by  
1 (1), $\cos 2\theta $ ($\cos 2\theta $), 
$\sin 2\theta $ ($-\sin 2\theta $), 
$\cos \theta$ $(-\cos\theta)$, and $\sin \theta $  
($\sin \theta $), respectively. 
For sufficiently large magnitude of $|x|$ with $x>0$, 
$\Delta(x)$ converges to $\Delta_{0}$, which is the bulk value of the 
pair potential. 
$g_{\pm}$, $f_{1\pm}$ and $f_{2\pm}$ satisfy the normalization condition
\begin{equation}
g_{\pm}^{2} + f_{1\pm}^{2} + f_{2\pm}^{2}=1. 
\label{Eilen4}
\end{equation}
We first note the general features of  
the pair amplitudes $f_{1\pm}$ and $f_{2\pm}$.
In normal state without $\bar{\Delta}_{\pm}$, 
$f_{1\pm}=0$, $f_{2\pm}=0$, and 
$g_{\pm}=1$. 
This describes the 
quasiparticle without pair amplitudes. 
In bulk superconductors with 
$\bar{\Delta}_{\pm}=\Delta_{0}$ for $\Delta_{0}>0$, 
we obtain $f_{1\pm}=0$ since there is no spatial dependence of 
$\bar{\Delta}_{\pm}$.
For spin-singlet $s$-wave superconductor, 
$g_{\pm} =\varepsilon/\sqrt{\varepsilon^{2} -\Delta_{0}^{2}}$, 
$f_{2\pm} =\Delta_{0}/\sqrt{\Delta_{0}^{2} -\varepsilon^{2}}$. 
At $\varepsilon=0$, $g_{\pm}=0$ and $f_{2\pm}=1$. 
This means the absence of quasiparticle and only the 
presence of Cooper pair at $\varepsilon=0$. \par

In order to discuss the parity with respect to
the frequency much more clearly, 
it is useful to use the Matsubara frequency representation:
The real energy $\varepsilon$ is replaced by $i\omega_n$ where  
the Matsubara frequency $\omega_n$ is given by $ \omega_{n}=2\pi T(n + 1/2)$ 
with temperature $T$ and integer $n$. 
Here, we pay attention to the $i \omega_{n}$ and $\theta$ dependences of 
$f_{1\pm }=f_{1\pm }(i\omega_{n},\theta,x)$, 
$f_{2\pm }=f_{2\pm }(i\omega_{n},\theta,x)$.  
For $x=\infty$, 
eqs. (\ref{Eilen1})-(\ref{Eilen3}) lead to
$f_{2\pm}(i\omega_{n},\theta,\infty)
=\Delta_{0}\Phi_{\pm}
/\sqrt{\Delta_{0}^{2}\Phi_{\pm}^{2} +\omega_{n}^{2}}$, 
$f_{2\pm }(i\omega_{n},\theta,x )=f_{2\pm }(-i\omega_{n},\theta,x )$ and 
$f_{1\pm}(i\omega_{n},\theta,x)=0$. 
This means that only even-frequency pairing 
is possible in the bulk. 
Although $f_{1\pm}(i \omega_{n},\theta,x )$ is absent in the 
bulk, it has a nonzero value 
due to the spatial change of the 
pair potential $\bar{\Delta}_{\pm}$.   
As seen from eqs. (\ref{Eilen1})-(\ref{Eilen3}), 
$f_{1\pm}(i\omega_{n},\theta,x )
=-f_{1\pm }(-i\omega_{n},\theta,x )$ and 
 $f_{2\pm }(i\omega_{n},\theta,x )=f_{2\pm }(-i\omega_{n},\theta,x )$ 
are always satisfied.
This means that the symmetry of  
$f_{1\pm }(x,i\omega_{n},\theta )$ belongs to 
odd-frequency pairing \cite{odd3}. 
Furthermore, 
as seen from eqs. (\ref{Eilen1})-(\ref{Eilen3}), 
\begin{equation}
v_{fx}^{2} \partial_{x}^{2} f_{1\pm}
-4(\omega_{n}^{2} + \bar{\Delta}_{\pm}^{2}(x))f_{1\pm} 
\pm 2i\mid v_{Fx} \mid [\partial_{x} \bar{\Delta}_{\pm}(x)] g_{\pm}
=0
\end{equation}
is satisfied 
in the Matsubara representation. 
It is clear that 
the spatial dependence of the pair potential $\bar{\Delta}_{\pm}(x)$ 
is a source of the generation of odd frequency pair amplitude 
$f_{1\pm}$. 
\par


We discuss the parity of 
$f_{1\pm}(i\omega_{n},\theta,x )$ and 
$f_{2\pm }(i\omega_{n},\theta,x )$. 
$\Phi _{\pm }(-\theta )=\Phi _{\mp }(\theta)$ 
is satisfied for an even-parity superconductor, 
while $\Phi _{\pm }(-\theta )=-\Phi _{\mp }(\theta)$ 
is satisfied for an odd-parity one. It follows 
from eqs. (\ref{Eilen1})-(\ref{Eilen3}) that 
$f_{1\pm }(i\omega _{n},\theta,x)=-f_{1\mp }(i\omega _{n},-\theta,x )$ 
and $f_{2\pm }(i\omega _{n},\theta)=f_{2\mp }
(i\omega _{n},-\theta,x )$ for an even-parity superconductor and 
$f_{1\pm }(i\omega _{n},\theta,x )=f_{1\mp }(i\omega _{n},-\theta,x )$ and 
$f_{2\pm}(i\omega _{n},\theta,x )=-f_{2\mp }(i\omega _{n},-\theta,x )$ for 
an odd-parity 
superconductor. 
Note that the parity of the odd-frequency pair amplitude 
$f_{1\pm}(i\omega _{n},\theta,x )$ is different from that of 
the bulk superconductor.  \par

Next, we discuss the 
values of the pair amplitudes at the interface $x=0$. 
For this purpose, it is convenient 
to express the above anomalous Green's function as \cite{Eschrig00}
\begin{equation}
f_{1\pm }=\frac{\pm i(F_{\pm }+D_{\pm })}{1-D_{\pm }F_{\pm }}, 
\quad
f_{2\pm }=%
\frac{D_{\pm }-F_{\pm }}{1-D_{\pm }F_{\pm }},  \label{eq2}
\end{equation}
where $D_{\pm }$ and $F_{\pm }$ satisfy the Eilenberger equations in the
Riccati parameterization \cite{Eschrig00}
\begin{equation}
v_{Fx}\partial _{x}D_{\pm }=-\bar{\Delta}_{\pm }(x)(1-D_{\pm }^{2})+2\omega
_{n}D_{\pm }  \label{eq3}
\end{equation}%
\begin{equation}
v_{Fx}\partial _{x}F_{\pm }=-\bar{\Delta}_{\pm }(x)(1-F_{\pm }^{2})-2\omega
_{n}F_{\pm }.  \label{eq4}
\end{equation}%
Since the interface is flat, 
$F_{\pm }=-RD_{\mp }$ for $\omega_{n}>0$ and 
$F_{\pm }=-R^{-1}D_{\mp }$ for $\omega_{n}<0$ 
hold at $x=0$ 
\cite{Eschrig00} with reflection coefficient $R$. 
If we denote $\omega_{n}$ dependences of 
$D_{\pm}$ and $F_{\pm}$ explicitly, 
$D_{\pm}(\omega_{n})=1/D_{\pm}(-\omega_{n})$ 
and $F_{\pm}(\omega_{n})=1/F_{\pm}(-\omega_{n})$ are satisfied. \par

 We concentrate on two extreme cases with (A) $\Phi _{+}(\theta )=\Phi
_{-}(\theta )$ and (B) $\Phi _{+}(\theta )=-\Phi _{-}(\theta )$. 
In the first case, there is no
sign change of the pair potential felt by the quasiparticle reflected 
at the interface.
Then, ABS is absent. 
On the other hand, in the second case, due to the sign change of the pair potential \cite{TK95}, 
ABS is generated near the interface. 
By using Riccati parameters defined above, 
it is easy to express that 
\begin{eqnarray}
f_{1\pm }=\pm i(1-R)D_{+}/(1+RD_{+}^{2}), 
\quad
f_{2\pm }=(1+R)D_{+}/(1+RD_{+}^{2}), \quad \omega_{n}>0 
\nonumber
\\
f_{1\pm }=\pm i(R-1)D_{+}/(R+D_{+}^{2}), 
\quad
f_{2\pm }=(1+R)D_{+}/(R+D_{+}^{2}), \quad \omega_{n}<0 
\label{eq:Riccatia}
\end{eqnarray}
for the case (A) and 
\begin{eqnarray}
f_{1\pm}=i(1+R)D_{+}/(1-RD_{+}^{2}), 
\quad
f_{2\pm }=\pm (1-R)D_{+}/(1-RD_{+}^{2}),
\quad \omega_{n}>0 
\nonumber
\\
f_{1\pm}=i(1+R)D_{+}/(R-D_{+}^{2}), 
\quad
f_{2\pm }=\mp (1-R)D_{+}/(R-D_{+}^{2}),
\quad \omega_{n}<0 
\label{eq:Riccatib}
\end{eqnarray}
for the case (B), respectively.
In the low transparent limit with $T_{m}\rightarrow 0$ ($R\rightarrow 1$),
only $f_{2\pm }$ is nonzero for the case (A) and $f_{1\pm }$ is nonzero
for the case (B) where $T_{m}$ is a transmission coefficient. 
However, even in the case (A), the mixture of odd and even-frequency pairings 
occur at the interface for general $R$.  
The underlying physics behind 
these results can be qualitatively explained as follows. Due to the
breakdown of the translational invariance, even- and odd-parity pairings 
are coupled near the interface. 
To be consistent with the Fermi-Dirac statistics, the
interface-induced pair amplitude $f_{1\pm}$ 
should be odd in frequency with
odd(even)-parity, when the pair potential has an even(odd)-parity 
\cite{odd3,odd3b,Eschrig2007}.
Note that $f_{1\pm}$ is pure imaginary and $f_{2\pm}$ is a real number, and 
there is a $\pm \pi/2$ phase between them.
This twist of the phase is forced by the coexistence of the even- 
and odd-frequency components near the interface so as to be 
compatible with the time-reversal  symmetry in the bulk superconductor
\cite{odd3}.
We summarize above results in Table \ref{table:2}.  \par

\begin{center}
\begin{table}[h]
\begin{tabular}{|c|p{2.9cm}|p{1.6cm}|p{2cm}|}
\hline
& Symmetry in bulk & Sign change & Interface \\ \hline
(1-1)& ESE ($s$, $d_{x^{2}-y^{2}}$ wave ) &   No  &  ESE +(OSO) \\ \hline
(1-2)& ESE ($d_{xy}$ wave ) &  Yes &  OSO +(ESE) \\ \hline
(2-1)& ETO ($p_{y}$ wave ) &  No &  ETO +(OTE) \\ \hline
(2-2)& ETO ($p_{x}$ wave) &  Yes &  OTE +(ETO) \\ \hline
\end{tabular}
\caption{Paring symmetry in the bulk and 
that at the N/S interface. The pair amplitude in the 
bracket is suppressed in the low transparency limit. 
}
\label{table:2}
\end{table}
\end{center}
Finally, we comment that analytical solution of Eilenberger equation is 
possible in a special case for fixed $\theta$.  
If we choose spatial dependence of $\bar{\Delta}_{+}(x)$ as 
\begin{equation}
 \Delta_{+}(x)=-\Delta_{-}(x)=\Delta_{0} \tanh (x/\xi_{0})
\end{equation}
with a certain constant $\xi_{0}$, 
we obtain following results, 
\[
g_{+}=g_{-}=\frac{1}{\omega_{n}^{2}+\Delta_{0}^{2}}
[\omega_{n} + \frac{\Delta_{0}^{2}}{2\omega_{n}}
{\rm sech}^{2}(\frac{x}{\xi_{0}})],
\]

\[
f_{1+}=f_{1-}=\frac{i}{\omega_{n}^{2}+\Delta_{0}^{2}}
\frac{\Delta_{0}^{2}}{2\omega_{n}}
{\rm sech}^{2}(\frac{x}{\xi_{0}}), 
\]
\begin{equation}
f_{2+}=-f_{2-}=
\frac{\Delta_{0}}{\omega_{n}^{2}+\Delta_{0}^{2}}
\tanh(\frac{x}{\xi_{0}}). 
\end{equation}
It is evident that odd-frequency pair amplitude $f_{1\pm}$ 
is localized near $x=0$ similar to zero energy quasiparticle state
as seen from $g_{\pm}$. 
Similar spatial dependence is realized in 
spin-singlet $d_{xy}$-wave and spin-triplet $p_{x}$-wave 
superconductor for low transparent limit. \par

In \S \ref{sec:2.2.1} and \S \ref{sec:ABSinUSC}, 
we illustrate these results by numerical calculations. 
In order to understand the angular momentum dependence of 
pair amplitudes, we define 
$\hat{f}_{1}$, $\hat{f}_{2}$ 
as
\begin{eqnarray}
\hat{f}_{1(2)}=
\left\{
\begin{array}{ll}
f_{1(2)+}(i\omega_n,\theta,x), & \mbox{for $-\pi/2<\theta<\pi/2$}, \\
f_{1(2)-}(i\omega_n,\pi-\theta,x), & \mbox{for $\pi/2<\theta<3\pi/2$}. 
\end{array}
\right. 
\label{eq:hatf}
\end{eqnarray}
We decompose 
$\hat{f}_{1(2)}$ into various angular momentum components. 
%
\subsubsection{Proximity effect in spin-singlet $s$-wave superconductors} 
\label{sec:2.2.1}

\begin{figure}[tb]
\begin{center}
\scalebox{0.8}{
\includegraphics[width=10cm,clip]{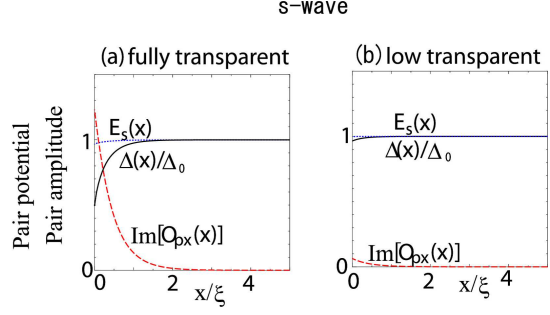}
}
\end{center}
\caption{(Color online) Spatial dependence of the pair potential for 
spin-singlet $s$-wave superconductor. 
Distance $x$ is normalized by $\xi=v_{F}/\Delta_{0}$. 
Pair potential normalized by its value in the 
bulk (solid line). 
Even frequency pair amplitude $E_{s}(x)$ (dotted line), 
and odd frequency pair amplitude $O_{p_x}(x)$ (dashed line). 
(a)Fully transparent case with $Z=0$ and 
(b)Low transparent case with $Z=5$. 
$\omega _{n}=\pi T$ and  $T=0.05T_{C}$. 
$T_{C}$ is the 
transition temperature of bulk superconductor. 
$R=Z^{2}/(Z^{2}+4\cos^{2}\theta)$. }
\label{fig:2}
\end{figure}

We choose conventional spin-singlet 
$s$-wave superconductor as a typical example of 
ESE pairing. As seen from Table \ref{table:2}(1-1), 
both ESE and OSO pairing exist at the N/S interface. 
We plot in Fig. \ref{fig:2} 
the $s$-wave component of $\hat{f}_{2}$ 
$E_{s}(x)$ 
and the $p_{x}$-wave component of  $\hat{f}_{1}$ $O_{p_x}(x)$. 

As seen from Fig. \ref{fig:2}(a), 
the pair potential $\Delta(x)$ is suppressed for 
fully transparent case (a) ($T_m=1$, $R=0$) due to 
the penetration of Cooper pair into normal metal. 
At the same time, odd-frequency pair amplitude 
$O_{p_x}(x)$ is enhanced at the interface. 
On the other hand, for low transparent case (b) ($T_m < 1$), 
$O_{p_x}(x)$ is almost zero and 
$\Delta (x)$ remains  constant up to the interface. 

\begin{figure}[tb]
\begin{center}
\scalebox{0.8}{
\includegraphics[width=10cm,clip]{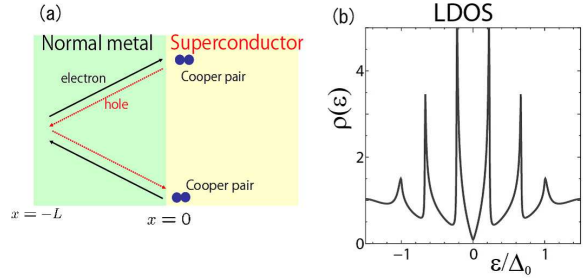}
}
\end{center}
\caption{(Color online) (a)Formation of standing wave by the closed loop of 
electron and hole. 
(b)Local density of state of quasiparticle $\rho(\varepsilon)$,
}
\label{fig:3}
\end{figure}

In order to highlight the effect of the induced odd-frequency pairing 
more clearly, here we consider N/S junction with 
finite length of N \cite{odd3b} [see Fig.\ref{fig:3} (a)].
We assume that the transparency at the interface is unity. 
In the present system, it is known that 
ABS with nonzero energy is formed \cite{McMillan}
as a standing wave which is generated by the 
interference between electron and hole in N.
%
As seen in Fig.\ref{fig:3} (b), there are many peaks in local density of state
(LDOS), which have been  observed as 
Rowell-McMillan oscillation in tunneling spectroscopy
\cite{McMillan,Rowell}. 
Here, we focus on this well-known phenomena in the light of 
odd-frequency pairing. 
Let us consider the pair amplitudes in the normal 
metal which are decomposed into 
components with different injection angle $\theta$. 
We will define the following spatially averaged 
odd-frequency pair amplitude
$F_{1+}^{(N)}(\varepsilon,\theta)$
and even-frequency pair amplitude
$F_{2+}^{(N)}(\varepsilon,\theta)$
\begin{equation}
F_{1(2)+}^{(N)}(\varepsilon,\theta) =\frac{1}{L}
\int^{0}_{-L} 
f^{R}_{1(2)+}(\varepsilon,\theta,x) dx.
\end{equation}
Then we introduce the ratio of 
$F_{1+}^{(N)}(\varepsilon,\theta)$ to 
$F_{2+}^{(N)}(\varepsilon,\theta)$
as 
\begin{equation}
R_{{\rm odd/even}} = 
\frac{\left\vert F_{1+}^{(N)}(\varepsilon,\theta)\right\vert }{\left\vert
F_{2+}^{(N)}(\varepsilon,\theta)\right\vert }=\left\vert \tan 
\left( \frac{2\varepsilon L }{v_{Fx}} \right) 
\right\vert.
\label{f12_E}
\end{equation}
\begin{figure}[tb]
\begin{center}
\scalebox{0.8}{
\includegraphics[width=7cm,clip]{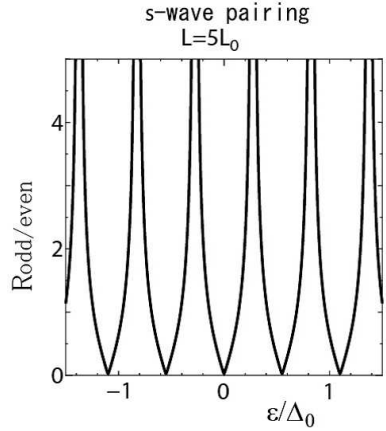}
}
\end{center}
\caption{
Ratio of the spatially averaged 
odd-frequency pair amplitude to 
that of even-frequency one  $R_{{\rm odd/even}}$ is 
plotted as a function of $\varepsilon$ for 
$\theta=0$. 
$Z=0$, $L=5L_{0}$, and $L_{0}=v_{F}/(\pi T_{C})$. }
\label{fig:4}
\end{figure}

As seen in Fig. \ref{fig:4}, 
in a certain energy regime, 
the magnitude of the odd-frequency pair amplitude 
exceeds over that of even-frequency one. 
It is remarkable that odd-frequency pairing can become 
dominant even in  simple normal metal / spin-singlet 
$s$-wave superconductor junctions for high transparency at the 
interface. 
In the extreme case with $L \gg L_{0}$ for fully transparent 
N/S interface, ABS levels 
with nonzero energy are given by \cite{McMillan,Rowell}
\begin{equation}
\varepsilon _{n}=\frac{\pi v_{Fx}}{2L}(n+1/2),\text{ \ \ }n=0,1,2,...
\end{equation}
It should be noted that just at $\varepsilon=\varepsilon_{n}$, 
$F_{2+}^{(N)}(\varepsilon,\theta)$ vanishes and 
$R_{{\rm odd/even}}$ diverges. 
Thus, the well-known quasiparticle bound states in N/S junction 
with conventional spin-singlet $s$-wave superconductor 
can be expressed by the generation of odd-frequency pairing. 

For low transparency limit, {\it i.e.}, 
$T_{m} \rightarrow 0$, the present ABS disappears since 
the coupling between N and S is weakened. 
Thus, the present OSO state is sensitive to the transparency at the interface. 
(We represent such a state in the braket in Table \ref{table:2}.)

In the next subsection, we will focus on 
spin-triplet $p_{x}$-wave superconductor, 
where odd-frequency pairing exceeds over 
even-frequency one at the interface independent of the 
transparency.

\par

\subsubsection{Andreev bound state in unconventional superconductor}
\label{sec:ABSinUSC}

\begin{figure}[tb]
\begin{center}
\scalebox{0.8}{
\includegraphics[width=10cm,clip]{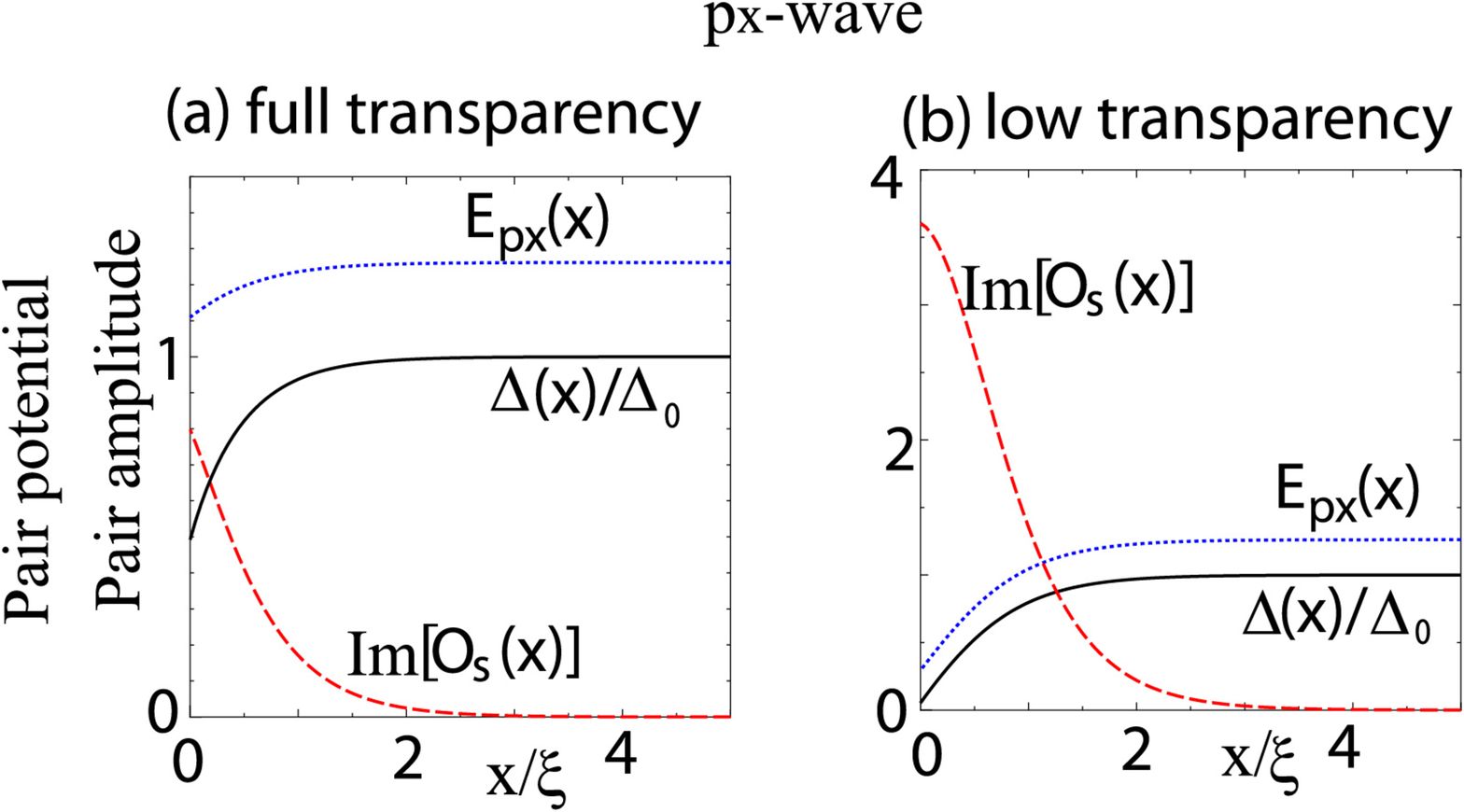}
}
\end{center}
\caption{(Color online) Spatial dependence of the pair potential for 
spin-triplet $p_{x}$-wave superconductor similar to Fig. \ref{fig:2}. 
Distance $x$ is normalized by $\xi=v_{F}/\Delta_{0}$. 
Pair potential normalized by its value in the 
bulk (solid line), 
Even frequency pair amplitude $E_{p_x}(x)$ (dotted line), 
and odd frequency pair amplitude $O_{s}(x)$ (dashed line).  
(a)Fully transparent case with $Z=0$, 　
(b)Low transparent case with $Z=5$. 
$\omega _{n}=\pi T$ and  $T=0.05T_{C}$. 
}
\label{fig:5}
\end{figure}

In this subsection, we show that the mid gap ZEABS 
specific to unconventional (anisotropic) superconductor 
is expressed by odd-frequency pairing \cite{TGKU07}. 
We focus on the case with $\Phi_{+}\Phi_{-}<0$ 
where mid gap ABS is generated. 
In this case, as discussed in \S \ref{sec:2b}, 
eq.(\ref{eq:Riccatib}) is satisfied by using Riccati parameters. 
In the low transparency limit with $T_{m}\rightarrow 0$ ($R\rightarrow 1$),
only the $f_{1\pm }$ is nonzero. 
Namely,
only the odd-frequency pair amplitude 
exists at the interface for $T_{m}\rightarrow 0$. 
First, we focus on the 
spatial dependence of the pair potential and 
pair amplitudes for spin-triplet $p_{x}$-wave superconductor. 
We plot in Fig. \ref{fig:5} 
the $p_{x}$-wave component of ETO pairing 
$E_{p_x}(x)$ 
and 
the $s$-wave component of OTE pairing $O_{s}(x)$. 

\begin{figure}[tb]
\begin{center}
\scalebox{0.8}{
\includegraphics[width=7cm,clip]{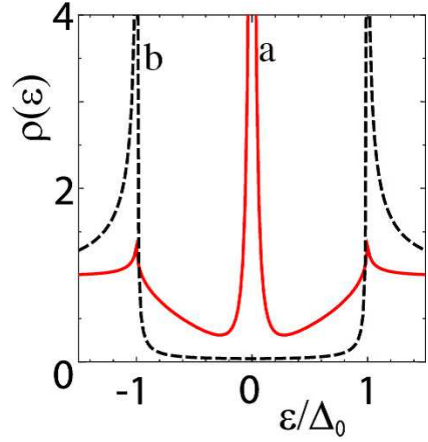}
}
\end{center}
\caption{(Color online) Quasiparticle density of state $\rho(\varepsilon)$ 
normalized by its value in the normal state at the 
N/S interface for $Z=5$.  
(a)Spin-triplet $p_{x}$-wave superconductor junction. 
(b)Spin-singlet $s$-wave 
superconductor junction. }
\label{fig:6}
\end{figure}

For fully transparent junctions, 
similar to spin-singlet $s$-wave junction case 
(Fig. \ref{fig:2}(a)), 
odd-frequency component of the pair amplitude $O_{s}(x)$ 
becomes of the same order of even-frequency one $E_{p_x}(x)$ 
as shown in Fig. \ref{fig:5}(a).  
For low transparent junction, $O_{s}(x)$ largely 
exceeds over $E_{p_x}(x)$ near the interface 
while $\Delta(x)$ is suppressed [Fig. \ref{fig:5}(b)]. 
The corresponding LDOS $\rho(\varepsilon)$ at $x=0$ has a ZEP for 
spin-triplet $p_{x}$-wave superconductor junction
(curve (a) in Fig. \ref{fig:6}), while $\rho(\varepsilon)$ has a 
gapped line shape for spin-singlet $s$-wave superconductor junction.  
Similar ZEP also appears for spin-singlet $d_{xy}$-wave 
superconductor. ZEP of $\rho(\varepsilon)$ 
ubiquitously appears 
for unconventional superconductor junctions, 
where the pair potential changes sign on the Fermi surface
\cite{TK95,ABSR1,Hu}.
Experimentally, ZBCP by this ZEP  has been observed in many unconventional 
superconductors including cuprate
\cite{TK95,Experiment1,Experiment3,Experiment4,Experiment5,Iguchi00,Experiment6,Experiment7,Ichimura,Laube00,Mao,Kashiwaya11,Walti,Turel,Rourke}. 
It is noted that in the presence of the ZEP  
originating from ABS,  
the magnitude of the odd-frequency pairing is enhanced. \par

Hereafter, we discuss more details about the odd-frequency pairing 
in the low transparency limit. As shown in the last paragraph, 
only $f_{1\pm}$ exists  for $R \rightarrow 1$ with 
$\Phi_{+}\Phi_{-} < 0$. 
In this limit, $f_{2\pm}$ becomes zero and 
$f_{1\pm}$ is given by 
$f_{1\pm}=2iD_{+}/(1-D_{+}^{2})$. 
Furthermore, ABS is 
expressed only by the induced odd-frequency pairing. 
Then, $\hat{f}_{1}$ in eq.(\ref{eq:hatf}) is given by 
\begin{eqnarray}
\hat{f}_{1}=
\left\{
\begin{array}{ll}
\frac{2i D_{+}(\theta)}{1-D^{2}_{+}(\theta)}, & -\pi/2< \theta < \pi/2\\ 
\frac{-2i D_{-}(\pi-\theta)}{1-D^{2}_{-}(\pi-\theta)}, & 
\pi/2< \theta < 3\pi/2
\end{array}
\right. .
\end{eqnarray}
with $D_{+}(\theta)=-D_{-}(\theta)$ and $D_{-}(\pi-\theta)=D_{+}(\theta)$. 
Here, we explicitly write the $\theta$ dependence of $D_{+}$. 
When the parity of bulk superconductor is even [odd], 
$D_{+}(\theta)=D_{+}(\pi+\theta)=D_{-}(-\theta)$ 
[$D_{+}(\theta)=-D_{+}(\pi+\theta)=-D_{-}(-\theta)]$ is satisfied. 
Thus, $\hat{f}_{1}$ has an odd-parity (even-parity), when 
the bulk paring symmetry is even-parity(odd-parity). 
The difference in the parity of the 
induced odd-frequency pairing results in a
serious difference when we consider proximity effects 
into DN attached to 
superconductor \cite{TG07,odd3}. 
In DN, only $s$-wave even parity 
pairing is possible. 
Thus, pair amplitudes with 
angular momentum larger than 0 cannot penetrate into DN. 
This implies that odd-frequency pairing generated from 
$d_{xy}$-wave superconductor cannot enter into DN 
\cite{Proximityd,Proximityd2}
since it is
expressed by OSO pairing. 
On the other hand, for $p_{x}$-wave 
superconductor, 
ABS can enter into DN \cite{Proximityp,Proximityp2,Proximityp3} 
since it is expressed by 
 OTE including $s$-wave channel 
\cite{TG07,Tshape,odd3b}. \par
If we neglect the spatial change of the pair potential near the 
interface, it is possible to obtain $D_{+}$ and $\hat{f}$ analytically. 
Since $D_{+}$ is give by $\Delta_{+}/(\omega_{n} 
+ \sqrt{\omega_{n}^{2} + \Delta_{+}^{2}})$, 
$\hat{f}_{1}$ is given by 
\begin{eqnarray}
\hat{f}_{1}=
\left\{
\begin{array}{ll}
\frac{i \Delta_{+}(\theta)}{\omega_{n}}, & -\pi/2< \theta < \pi/2\\ 
\frac{-i \Delta_{-}(\pi-\theta)}{\omega_{n}}, & 
\pi/2< \theta < 3\pi/2
\end{array}
\right.
\end{eqnarray}
For spin-triplet $p_{x}$-wave bulk superconductor, 
the resulting $\hat{f}_{1}$ is 
\[
\hat{f}_{1}=\frac{i \Delta_{0}}{\omega_{n}}\mid \cos \theta \mid, 
\]
while for spin-singlet $d_{xy}$-wave bulk superconductor, 
the resulting $\hat{f}_{1}$ is 
\[
\hat{f}_{1}=\frac{i \Delta_{0}}{\omega_{n}}\mid \sin 2\theta \mid
\rm{sgn}(\sin \theta). 
\]
Reflecting on the ZEABS, 
$\hat{f}_{1}$ is proportional to the inverse of $\omega_{n}$ for 
both cases. 
The difference of the parity of $\hat{f}_{1}$ between two cases 
appears in the topological features of wave function of 
two ABSs as will be discussed in \S \ref{sec:3c}.

\subsection{Odd-frequency pairing in vortex core}
\label{sec:2c}

In this subsection, we discuss the pairing 
symmetry in vortex core \cite{Yokoyamavortex}. 
The study of the mixed state in type-II superconductors, where
magnetic flux enters a sample in the form of quantized vortices, 
has a long history.\cite{Abrikosov,Blatter} 
It is known that the 
ABS is generated in
the Abrikosov vortex core due to the spatial structure of the 
pair potential. \cite{Caroli,Hess,Makhlin,LO,Kopnin97,Volovik,Gygi91,Hayashi98}. 
One of 
the manifestations of the bound states is the enhancement of 
LDOS in the core, observable as a ZBCP by scanning
tunneling spectroscopy (STM)
\cite{Hess,Fischer}. 
Since an Abrikosov vortex breaks translational
symmetry in a superconductor, 
we can expect the emergence of an
odd-frequency pairing state around the vortex even in a conventional 
spin-singlet $s$-wave superconductor.

Here, based on the quasiclassical theory, 
we clarify a pairing symmetry in the vortex core. 
The electronic structure at the core of a single Abrikosov vortex in a
ballistic superconductor is described by the quasiclassical Eilenberger
equations \cite{Eilenberger,Schopohl,Schopohl98}.  
We assume that the pairing symmetry of bulk superconductor is 
ESE or ETO. 
 Along a trajectory
${\bm r}(x') = {\bm r}_0 + x' \; \hat{\bm {v}}_F $ with unit vector
$ \hat{\bm v}_F$  parallel to the Fermi velocity 
$\bm {v}_F$, the Eilenberger equations
are generally represented in a 4$\times4$ matrix.\cite{Eschrig00}
For a singlet (triplet) superconductor with $S_{z}=0$, 
these equations are reduced to the set of two decoupled differential equations of the Riccati type for the functions $a(x')$ and $b(x')$ 
\cite{Schopohl,Schopohl98}
\begin{eqnarray}
 v_F \partial_{x'}  a(x') + \left[ 2 \omega_n + \bar{\Delta}^* a(x') \right] a(x')
- \bar{\Delta} & = & 0, \nonumber \\
 v_F \partial_{x'} b(x') - \left[ 2 \omega_n + \bar{\Delta} b(x') \right] b(x') + \bar{\Delta}^{*} & = & 0 \label{Riccati}
\end{eqnarray}
where $\omega_n$ is the Matsubara frequency. 
Quasiparticle Green's function $g$ and anomalous Green's function 
$f$ are expressed by $a$ and $b$ 
as $g=(1 -ab)/(1+ab)$ and $f=2a/(1+ab)$, respectively. 
For simple case of a cylindrical Fermi surface, 
the Fermi velocity can be written as
${\bm v}_F = v_F (\cos \theta,  \sin \theta)$. 
We choose the following form of the pair potential:
\begin{equation}
\bar{\Delta}=
\bar{\Delta} ({\bf r},\theta) 
= \Delta _0 \Phi(\theta) 
F(r)
\exp(im \varphi) \nonumber 
\end{equation}
with $r=\sqrt {x^2  + y^2 }$, $\exp(i\varphi)=(x + iy)/r$  
and integer $m$. 
The center of a vortex is situated at $x=y=0$ and $\exp(im\varphi)$ is
the phase factor which originates from the vortex.
We consider axially symmetric cores.
For the calculation of the normalized local 
DOS by its value in the normal state, 
 the quasiclassical propagator has to be integrated over the
angle $\theta$ which defines the direction of the Fermi velocity.
The normalized LDOS in terms of functions $a$ and $b$ is given by
\begin{equation}
\rho({\bm r}_0,\varepsilon) = \int_0^{2\pi} \frac{d \theta}{2 \pi}  {\rm Re} \;
\left[ \frac{1-a  b}{1+ a b} \right]_{i \omega_n \rightarrow \varepsilon + i
\delta}, \label{angularaverage}
\end{equation}
where $\varepsilon$ 
denotes the quasiparticle energy with respect to the Fermi
level and $\delta$ is an effective broadening parameter of quasiparticle 
energy level. In the actual numerical calculations, we will fix this value as
$\delta=0.1\Delta_0$.

First, we show the general property of the 
symmetry at the vortex center independent of the detailed 
spatial dependence of $\bar{\Delta}$. 
Let us consider a trajectory passing through the center of the vortex. 
By setting $x'=0$ at the center of the vortex, 
we obtain $b(x',i\omega_n ) =  - 1/a( - x', -i \omega _n )$ from the Eilenberger equations with odd integer $m$. 
 Similarly, we obtain $b(x',i\omega _n ) =  1/a( - x', - i\omega _n )$ for 
even integer $m$. Thus, at the vortex center $x'=0$,  we get 
$ f(i \omega _n ) =  - f( -i \omega _n )$ in the former case, while 
$ f(i \omega _n ) =   f( -i \omega _n )$ in the latter. 
Since we do not consider the Zeeman effect, 
spin flip does not occur in the present system. 
Thus, we can summarize 
pairing symmetry at the vortex center in Table \ref{table:3} 
based on the broken inversion symmetry and Fermi-Dirac statistics 
\cite{Yokoyamavortex}. \par
\begin{center}
\begin{table}[h]
\begin{tabular}{|c|p{1.4cm}|p{1.4cm}|p{1.9cm}|p{1.9cm}|}
\hline
& bulk state & vorticity  m  & parity of the bulk states  & symmetry at the center \\ \hline
(1) & ESE  & odd & even & OSO \\ \hline
(2) & ESE  & even & even & ESE \\ \hline
(3) & ETO  & odd & odd & OTE \\ \hline
(4) & ETO & even & odd & ETO \\ \hline
\end{tabular}%
\caption{Pairing symmetry in the vortex state. }
\label{table:3}
\end{table}
\end{center}
 For conventional ESE $s$-wave case, there have been
several studies of multi-flux state with $m \ge 1$
\cite{SSC,Melnikov} in the context of superconducting quantum dot 
where normal nano-scale region is surrounded by superconductor. 
It has been shown that ZEP in the DOS only appears
for odd number $m$ at the vortex center \cite{SSC,Melnikov}. 
This statement is consistent with our present results 
for the conventional spin-singlet $s$-wave case, 
where odd-frequency pairing  is generated only for odd integer $m$. 

\begin{figure}[htb]
\begin{center}
\scalebox{0.4}{
\includegraphics[width=16.0cm,clip]{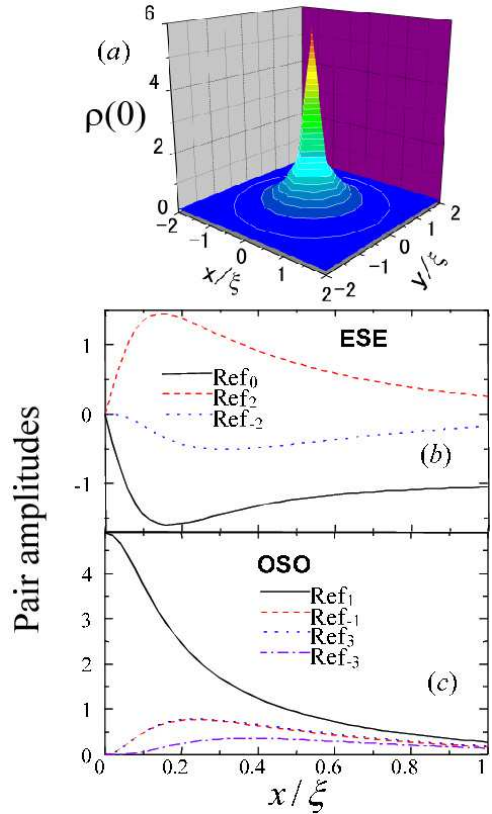}}
\end{center}
\caption{(Color online) (a) Normalized LDOS 
around the vortex at $\varepsilon=0$. The center
of the vortex is situated at $x=y=0$. (b) Spatial dependencies of various 
pairing components at $\varepsilon=0$.  Spatial dependencies of (b) ESE component and (c) OSO component.
[Reproduced from Fig. 1 of 
Phys. Rev. B 78 012508 (2008) by T. Yokoyama, $et$. $al$.] 
} 
\label{vortex1}
\end{figure}
In general, the most stable vorticity is 
$\mid m \mid=1$ realized in Abrikosov 
vortex. As a symmetry of the bulk superconductor, 
we choose ESE $s$-wave pairing. 
Also, spatial dependence of the gap is chosen as $F(r)=\tanh (r/\xi)$, 
where  $\xi = v_F/\Delta_0$ 
is the coherence length. 
Due to the broken translational symmetry of the system,
various pairing symmetries  are
mixed near the vortex within the spin-singlet pairings. 
We decompose anomalous 
Green's function $f^{R}(\varepsilon,\theta)$ 
into various angular momentum components 
at the center of the core, and $f^{R}(\varepsilon,\theta)$ can be given by 
\begin{eqnarray}
f^{R}(\varepsilon,\theta)
 = \sum_{l_{z}=0,\pm1,\pm2,...} f_{l_{z}}\exp(i l_{z} \theta). 
\end{eqnarray}
All the above pairing components $f_{l_{z}}$ are spin singlet. 
Spatial dependence of 
the LDOS around the vortex at 
$\varepsilon=0$ is shown in Fig. \ref{vortex1} (a). 
It has a peak just at the core center. 
The spatial dependencies of decomposed anomalous
Green's function $f_{l_{z}}$ 
at $\varepsilon=0$ are shown in Figs. \ref{vortex1} 
(b) and (c).
Interestingly, only OSO pairing component ${\rm Re}(f_{1})$ 
is nonzero at the center of the vortex core. 
Thus, we see that anomalous Green's function at the core center has chiral 
$p$-wave symmetry. \par
\begin{figure}[htb]
\begin{center}
\scalebox{0.4}{
\includegraphics{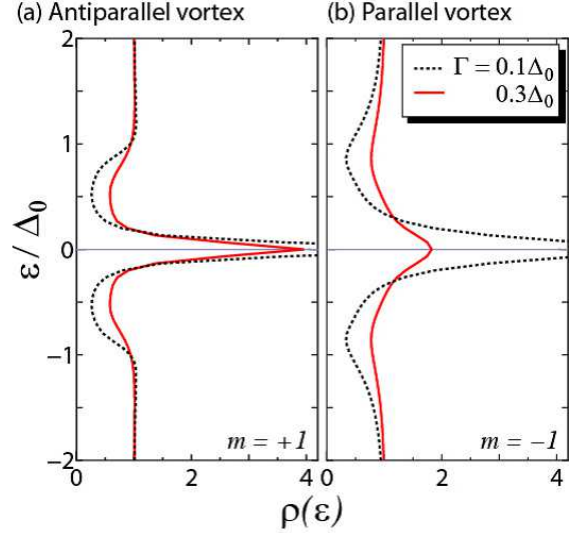}
}
\caption{(Color online)
The LDOS at the vortex center 
for the antiparallel (a) and (b) the parallel
vortex.
The dashed and solid lines are plots for
$\Gamma = 0.1\Delta_0$ and $0.3\Delta_0$,
respectively.
%
%
[Reproduced from Fig. 3 of Phys. Rev. Lett. 102 117003(2009) by 
Y. Tanuma $et.$ $al$. ]
\label{vortex:02}}
\end{center}
\end{figure}

It is interesting to consider the bulk chiral superconductor with 
the angular momentum $l$. 
The corresponding pair potential is given by 
$\bar{\Delta}=\Delta_{0}\Phi(\theta)F(r)\exp(il_{z}\theta)\exp(im\varphi)$. 
At the center of the core, it is shown analytically that 
the angular momentum of the pair amplitude is 
$l_{z} + m$. 
Here, we choose Abrikosov vortex in bulk chiral $p$-wave superconductor 
with $l_{z}=1$. 
As seen from Table \ref{table:3}(3), symmetry at the center of core 
is OTE. The angular momentum of the pair amplitude depends on 
whether chirality and vorticity is antiparallel or parallel. 
For the antiparallel case ($m=-1$), the OTE $s$-wave pairing  is 
generated at the vortex core \cite{Tanuma09},
while for the parallel case ($m=1$), the OTE $d$-wave 
symmetry is  realized at the core. 
Since  $s$-wave pair amplitude is robust against impurity scattering 
\cite{Anderson}, 
it is useful to consider impurity scattering effect on the 
vortex core state (odd-frequency pairing) to distinguish two states. 
We introduce this effect within the Born approximation, 
where we denote the impurity scattering rate in the normal state 
$\Gamma$ with $\Gamma =1/(2\tau)$ with mean free path $v_{F}\tau$. 
The resulting LDOS of 
anti-parallel and parallel vortex case is shown in 
Fig. \ref{vortex:02}. 
The ZEP is robust with the increase of $\Gamma$ for anti-parallel 
vortex case while it is fragile against $\Gamma$ for parallel vortex 
case \cite{Kato1,Kato2,Hayashi,Hayashi03}. 

\begin{figure}[htb]
\begin{center}
\scalebox{0.37}{
\includegraphics{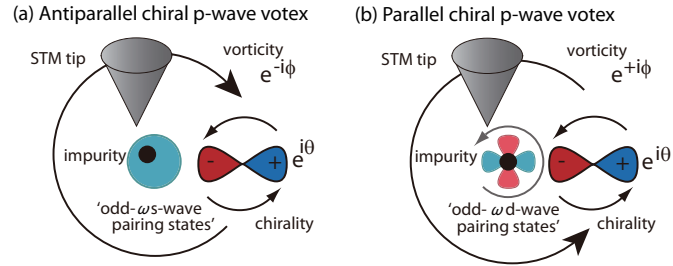}
}
\caption{(Color online)
Schematic illustration of 
odd-$\omega$ pair amplitude
for (a) the antiparallel and (b) the parallel vortex
in chiral $p$-wave SCs.
The arrows represent the phase rotation.
[Reproduced from Fig. 4 of Phys. Rev. Lett. 102 117003(2009) by 
Y. Tanuma $et.$ $al$. ]
\label{vortex:03}}
\end{center}
\end{figure}

The robustness of the ZEP at the vortex center against nonmagnetic 
impurities for antiparallel case originates from the odd-frequency 
$s$-wave pair amplitude. 
In the actual sample of spin-triplet 
chiral $p$-wave superconductor, 
degenerated chiral state, {\it i.e.}, $l_{z}=1$ and $l_{z}=-1$, 
form a domain structure.  
Near the antiparallel vortex core, strong ZEP of LDOS is expected, 
while near the parallel vortex core, weak ZEP appears. 
Accordingly, measurements of the ZEP of LDOS 
in the presence of impurities reflects the 
detection of the symmetry of odd-frequency pair amplitude 
at the center of the core as shown in Fig. \ref{vortex:03}. 

Extension of the calculation in the Abrikosov vortex lattice has been done by 
Yokoyama $et.$ $al$ \cite{Yokoyama10}. 
They have found that only odd-frequency pairing 
exists at core centers. \par

\subsection{Anomalous proximity effect in spin-triplet superconductor junctions}\label{sec:2d}

\begin{figure}[tb]
\begin{center}
\scalebox{0.8}{
\includegraphics[width=13cm,clip]{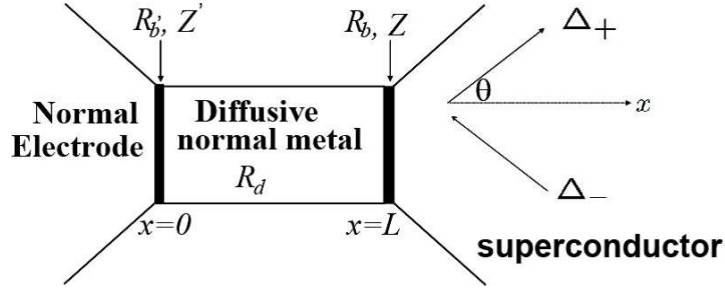}
}
\end{center}
\caption{Schematic illustration of the model of 
normal electrode attached to 
diffusive normal metal (DN) 
/ superconductor (semi-infinite）junction. 
[Revised from Fig.1 of Phys. Rev. B 72 140503(R) (2005) by 
Y. Tanaka $et.$ $al$.]  
}
\label{Anomalous0}
\end{figure}

In this subsection, we discuss the anomalous proximity effect 
specific to spin-triplet superconductor junctions. 
We consider diffusive normal metal(DN) ($0<x<L$) / 
superconductor (S) ($x>L$) 
junction where the length of DN is $L$ as shown in 
Fig. \ref{Anomalous0}. 
We assume that the DN is attached to the normal electrode  
at $x=0$. 
The interface
between the DN and the superconductor (S) at $x=L$ has a resistance
$R_{b}$ and the N/DN interface at $x=0$ has a resistance
$R_{b^{\prime }}$. We also denote resistance in DN as $R_{d}$. 
For $R_{b^{\prime }}=\infty$, the present model
is reduced to the DN/S bilayer with vacuum at the DN free surface.

We assume that 
pairing symmetry in S is ESE or ETO in the ballistic regime. 
We choose $S_{z}=0$ for ETO case. 
In the DN region, 
only the $s$-wave pairing which is robust against 
impurity can survive in the limit of strong disorder. 
Since there is no spin-flip in the present junctions, 
spin structure of pair amplitude in DN is the same as that in bulk. 
Then, the pairing symmetry in the DN can be derived 
in accordance with Fermi-Dirac statistics. 
If the pairing symmetry in the bulk is ESE, 
the resulting symmetry in DN is ESE $s$-wave. 
On the other hand, if the pairing symmetry in the bulk is ETO, 
the resulting symmetry in DN is OTE $s$-wave. 
In DN, it is convenient to use Usadel equation 
which is derived from Eilenberger equation in the 
diffusive limit after angular average of the direction of motion. 
The Green's functions in Usadel equation do not have a 
$\theta$ dependence anymore due to the isotropization by impurity scattering. 
Here, in order to calculate LDOS in the DN region, 
we use retarded Green's function. 
Due to the fermion's commutation relation and 
complex conjugation of Green's function, 
anomalous Green's function should satisfy 
$f^{R}_{1(2)}(\varepsilon,x) = [f^{A}_{1(2)}(\varepsilon,x)]^{*}$ 
where the suffix $1$ and $2$ denote odd and even-frequency component.  
$f^{R}_{1}(\varepsilon,x)=-[f^{R}_{1}(-\varepsilon,x)]^{*}$ is satisfied 
by the relation $f^{R}_{1}(\varepsilon,x)=-f^{A}_{1}(-\varepsilon,x)$.  
On the other hand, 
$f^{R}_{2}(\varepsilon,x)=[f^{R}_{2}(-\varepsilon,x)]^{*}$ is satisfied 
by the relation $f^{R}_{1}(\varepsilon,x)=f^{A}_{2}(-\varepsilon,x)$.  
Usadel equation \cite{Usadel} in DN is given by 
\begin{equation}
D\frac{\partial ^{2}\zeta }{\partial x^{2}}+2i\varepsilon \sin \zeta =0
\label{eq.1}, 
\end{equation}%
with diffusion constant $D$. 
$g^{R}(\varepsilon,x)$ and $f^{R}_{2}(\varepsilon,x)$ 
are parameterized by 
$g^{R}(\varepsilon,x)=\cos \zeta$, $f^{R}_{2}(\varepsilon,x)=\sin \zeta$, 
and  $f^{R}_{1}(\varepsilon,x)=0$, 
when the bulk superconductor has an ESE symmetry. 
On the other hand, 
$g^{R}(\varepsilon,x)=\cos \zeta$, $f^{R}_{1}(\varepsilon,x)=\sin \zeta$, 
and  $f^{R}_{2}(\varepsilon,x)=0$ are satisfied 
when the bulk superconductor has an ETO symmetry. 
The boundary condition of $\zeta$ is given by 
\cite{Proximityd,Proximityd2,Proximityp,Proximityp2}
\begin{equation}
\frac{L}{R_{d}}\left. \left( \frac{\partial \zeta }{\partial x}\right) \right|_{x=L}=
\frac{\langle F_{1}\rangle }{R_{b}},  
\label{eq.2}
\end{equation}%
\begin{equation}
F_{1}=\frac{2T_{1}(f_{S}\cos \zeta _{L}-g_{S}\sin \zeta _{L})}{%
2-T_{1}+T_{1}(\cos \zeta _{L}g_{S}+\sin \zeta _{L}f_{S})}
\label{boundary}
\end{equation}
at DN/S interface ($x=L$) 
and 
\begin{equation}
\frac{L}{R_{d}}\left.\left(\frac{\partial \zeta }{\partial x}\right)\right|_{x=0}=\frac{%
\langle F_{2}\rangle }{R_{b^{\prime }}},\ \ F_{2}=\frac{2T_{2}\sin \zeta
_{0}}{2-T_{2}+T_{2}\cos \zeta _{0}},  
\label{eq.3}
\end{equation}%
at normal electrode /DN interface ($x=0$). 
Here, $\zeta _{L}=\zeta \mid _{x=L}$,  
and $\zeta _{0}=\zeta\mid _{x=0}$.
$\langle \ldots \rangle $ denotes the 
angular average of the injection angle 
$\theta$ of quasiparticles, 
\begin{equation}
\langle F_{1(2)}(\theta )\rangle =\frac{ \int_{-\pi /2}^{\pi /2}d\theta 
\cos \theta
F_{1(2)}(\theta )}{
\int_{-\pi /2}^{\pi /2}d\theta T_{1(2)}\cos \theta} ,
\label{average}
\end{equation}%
\begin{equation}
T_{1}=\frac{4\cos ^{2}\theta }{Z^{2}+4\cos ^{2}\theta },
\;\;T_{2}=\frac{4\cos
^{2}\theta }{Z^{\prime }{}^{2}+4\cos ^{2}\theta },
\end{equation}%
$T_{1,2}$ denotes the transmission coefficient at the interface 
with barrier parameters 
$Z$ and $Z^{\prime }$. 
$g_{S}$ and $f_{S}$ are determined by the 
pair potential in bulk superconductor with 
$g_{S}=(g_{+}+g_{-})/(1+g_{+}g_{-}+f_{+}f_{-})$, 
$g_{\pm }=\varepsilon /\sqrt{\varepsilon ^{2}-\Delta _{\pm }^{2}}$,  
$f_{\pm }=\Delta _{\pm }/\sqrt{\Delta _{\pm }^{2}-\varepsilon ^{2}}$, 
and $\Delta _{\pm }=\Delta_{0} \Phi_{\pm}$.  
For even-parity bulk superconductor, 
$f_{S}=(f_{+}+f_{-})/(1+g_{+}g_{-}+f_{+}f_{-})$ and 
for odd-parity superconductor 
$f_{S}=i(f_{+}g_{-}-f_{-}g_{+})/(1+g_{+}g_{-}+f_{+}f_{-})$ \cite{Proximityp,Proximityp2}.  
Here, we neglect the spatial dependence of $\Delta_{\pm}$. 
By solving, above Usadel equation with proper boundary 
condition we can determine the pair amplitude in DN  and the 
resulting density of state. 

\begin{figure}[tb]
\begin{center}
\scalebox{0.8}{
\includegraphics[width=7cm,clip]{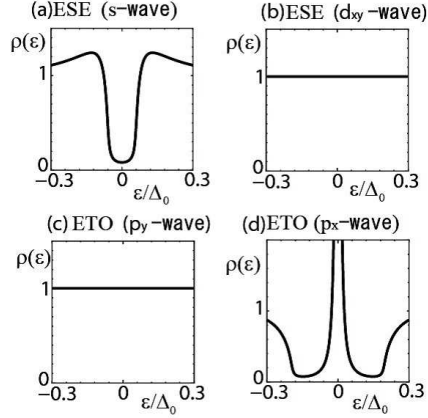}
}
\end{center}
\caption{LDOS $\rho(\varepsilon)$ of quasiparticle inside 
the diffusive normal metal at $x=L/2$. 
Symmetry in the bulk superconductor is 
(a)ESE $s$-wave, 
(b)ESE $d_{xy}$-wave, 
(c)ETO $p_{y}$-wave, and 
(d)ETO $p_{x}$-wave. 
}
\label{anomalous1}
\end{figure}

\begin{center}
\begin{table}[h]
\begin{tabular}{|c|p{2.8cm}|p{1.8cm}|p{1.8cm}|}
\hline
& Bulk & interface & DN \\ \hline
(1-1) & ESE ($s$, $d_{x^{2}-y^{2}}$-wave) & ESE (OSO) & ESE($s$-wave) \\ \hline
(1-2) & ESE ($d_{xy}$-wave) & OSO (ESE) & No \\ \hline
(2-1) & ETO ($p_{y}$-wave) & ETO (OTE) & No \\ \hline
(2-2) & ETO ($p_{x}$-wave) & OTE (ETO) & OTE($s$-wave)  \\ \hline
\end{tabular}%
\caption{Paring symmetry in the bulk, 
at the interface, and in DN. 
}
\label{table:4}
\end{table}
\end{center}
The obtained results are summarized in Table \ref{table:4}. 
LDOS $\rho(\varepsilon)$ normalized by its value in the 
normal state  is plotted in Fig. \ref{anomalous1}. 
The pair amplitude $\hat{f}_{1}$ and $\hat{f}_{2}$ at the 
interface of S side discussed in \S \ref{sec:2b} 
can not enter into DN if their angular momentum $l$ 
is not zero. 
When the symmetry of the bulk superconductor is ESE $s$-wave (1-1), 
$f^{R}_{1}(\varepsilon,\theta)=0$ and $f^{R}_{2}(\varepsilon,\theta) \neq 0$ 
are satisfied. This means that 
ESE $s$-wave pair is induced in DN. The OSO pair amplitude induced 
at the interface can not enter into DN due to the odd-parity. 
The LDOS has a gap like structure as shown in Fig. \ref{anomalous1}(a). 
This is a standard proximity effect known for the mesoscopic superconducting 
systems \cite{Volkov,Golubov88,Belzig96,TGK}. 
The order of the magnitude of the energy gap is 
Thouless energy $E_{Th}=D/L^{2}$ in DN. 
When the symmetry of the bulk superconductor is ESE $d_{xy}$-wave (1-2), 
the resulting $\zeta$ is zero and $f^{R}_{1}(\varepsilon,\theta)=f^{R}_{2}(\varepsilon,x)=0$ inside 
DN. Then the resulting $\rho(\varepsilon)$  is unity as shown in Fig. \ref{anomalous1}(b). 
The OSO pair amplitude at the interface can not enter into DN. Also, 
the subdominant ESE state can not enter since its angular momentum $l$ is 
$l \geq 2$. This is the reason for the absence of the proximity effect into DN. In other words, mid gap ABS generated at the interface can not 
penetrate into DN for bulk spin-singlet $d_{xy}$-wave superconductor. 
For ETO state with $p_{y}$-wave pairing (2-1),  
$\zeta$ is zero  and $f^{R}_{1}(\varepsilon,\theta)=f^{R}_{2}(\varepsilon,\theta)=0$ 
inside DN. Then the resulting $\rho(\varepsilon)$  is unity as shown in Fig. \ref{anomalous1}(c). In this case, pair amplitudes at the S side of the interface 
have angular momenta with $l>0$. 
Proximity effect specific to spin-triplet superconductor 
appears for   ETO $p_{x}$-wave bulk superconductor case (2-2). 
The resulting $\zeta$ and $f^{R}_{1}(\varepsilon,x)$ are nonzero in DN 
with $f^{R}_{2}(\varepsilon,x)=0$. 
It is remarkable that $\rho(\varepsilon)$  has a sharp ZEP. 
This is a new type of proximity effect. The present anomalous proximity effect 
is generated  by odd-frequency pairing \cite{TG07}. 
The mid gap ABS at the interface can penetrate into DN \cite{Proximityp,Proximityp2}. 
In this case, mid gap ABS can be interpreted as an OTE pairing 
which  has an $s$-wave component \cite{TG07}.  
Then, the proximity effect into DN becomes possible. \par
\begin{figure}[tb]
\begin{center}
\scalebox{0.8}{
\includegraphics[width=5cm,clip]{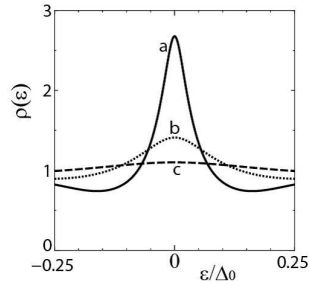}
}
\end{center}
\caption{Similar plot to 
Fig. \ref{anomalous1} of $\rho(\varepsilon)$ 
inside the diffusive normal metal at $x=L/2$ for 
ETO chiral $p$-wave superconductor. 
$Z=Z'=1$, $R_{d}/R_{b}=1$ and $E_{Th}=0.25\Delta_{0}$. 
(a)$R_{d}/R_{b'}=0.01$, (b)$R_{d}/R_{b'}=1$, and 
(c)$R_{d}/R_{b'}=100$. 
[Revised from Fig.2(a) of Phys. Rev. B 72 140503(R) (2005) by 
Y. Tanaka $et.$ $al$.]  

}
\label{anomalous2}
\end{figure}

Although we have shown anomalous proximity effect 
for ETO $p_{x}$-wave superconductor, 
it appears for generic ETO superconductor as far as the 
the interface induced OTE pair amplitude  has an $s$-wave component 
\cite{Proximityp,Proximityp2}. 
ETO $p_{y}$-wave case is exceptional since OTE $s$-wave 
component is canceled by the angular integral of $\theta$ 
in the boundary condition. 
One of the strong candidate material of ETO $p$-wave 
superconductor is realized in chiral $p$-wave 
superconductor Sr$_{2}$RuO$_{4}$ 
where ZBCP originating from ABS \cite{YTK97,HS98} has been observed 
\cite{Laube00,Mao,Kashiwaya11}. In this case, 
ZEP of $\rho(\varepsilon)$ in DN by the anomalous proximity effect appears. 
In the actual calculation, 
we must extend boundary condition at DN/S interface 
\cite{Meissner3}. In Fig. \ref{anomalous2}, we plot LDOS 
$\rho(\varepsilon)$ for various parameters. 
The height of the ZEP of LDOS increases with the decrease of $R_{d}/R_{b'}$ 
since the odd pair amplitude $f^{R}_{1}(\varepsilon,x)$ is more 
strongly confined within DN for large magnitude of $R_{b'}$. 

\begin{figure}[tb]
\begin{center}
\scalebox{0.8}{
\includegraphics[width=10cm,clip]{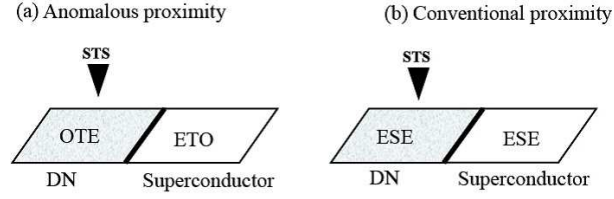}
}
\end{center}
\caption{Schematic illustration of an experimental setup by 
scanning tunneling microscope (STM) in diffusive normal metal (DN) 
attached to a superconductor. (a)Anomalous proximity effect in DN/ETO superconductor junction.  
(b)Conventional proximity effect in DN/ETO superconductor junction.  
The pair amplitude has an OTE and ESE symmetry for 
anomalous case (a) and conventional  case (b), 
respectively. Only for  anomalous proximity case (a), 
the LDOS  has a zero energy peak. }
\label{Anomalous3}
\end{figure}

\begin{figure}[tb]
\begin{center}
\scalebox{0.8}{
\includegraphics[width=7cm,clip]{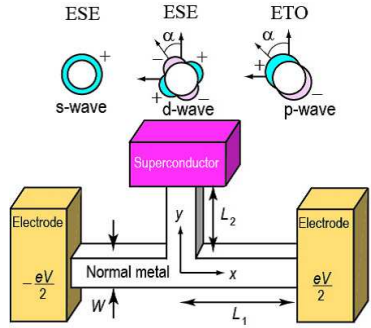}
}
\end{center}
\caption{(Color online) Schematic illustration of an experimental setup by 
$T$-shaped proximity structure.
We consider spin-singlet $s$-wave, $d$-wave 
and spin-triplet $p$-wave pair superconductors. 
[Reproduced from Fig.1 of Phys. Rev. Lett. 99 067005 (2007) by Y. Asano $et.$ $al$.]  
}
\label{Anomalous4}
\end{figure}

In order to detect anomalous proximity effect 
by odd-frequency pairing, 
we propose an experimental setup 
using scanning tunneling spectroscopy (STS) as shown in 
Fig. \ref{Anomalous3}. 
The ZEP of LDOS can be detected by scanning tunneling 
spectroscopy (STS) and we can 
distinguish anomalous proximity effect by OTE pairing from 
conventional proximity effect by ESE pairing. 
However, to fabricate well-controlled 
surface of junctions available for STS is not so easy at present. 
In order to detect anomalous proximity effect, 
Asano $et.$ $al$ proposed a $T$-shaed junction \cite{Tshape} 
which is more accessible for the fabrication as shown in Fig. \ref{Anomalous4}. 
In this system, the proximity effect from the superconductor 
modifies the conductance between two electrodes depending 
remarkably on the pairing symmetry: spin-singlet or spin-triplet.
Only for spin-triplet pairing, OTE state can be generated and 
conductance has a zero bias peak. \par
\begin{figure}[bh]
\begin{center}
\scalebox{0.4}{
\includegraphics[width=10.0cm,clip]{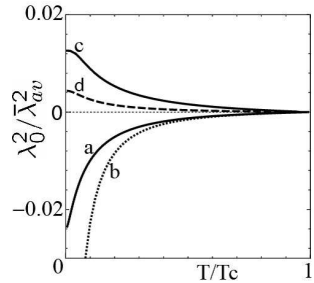}
}
\end{center}
\caption{ Averaged value of the Meissner screening length 
$\bar{\protect\lambda}_{av}^{2}$ is plotted for 
various superconductor with $Z=1$, $%
R_{d}/R_{b}=1$, $Z^{\prime }=1$, $R_{d}/R_{b^{\prime }}=1$ and $%
E_{Th}=0.25\Delta _{0}$, respectively. 
(a)ETO chiral $p$-wave, (b)ETO $p_{x}$
-wave, (c)ESE $s$-wave and 
(d)ESE chiral $d$-wave superconductors. 
[Revised from Fig.3 of Phys. Rev. B 72 140503(R) (2005) by 
Y. Tanaka $et.$ $al$.]  
}
\label{Anomalous5}
\end{figure}
The anomalous proximity effect 
induces an peculiar response of DN to external magnetic
field. When a magnetic field is applied parallel to the interface,  the magnetic field in the DN behaves as $%
H(x)\sim \exp (-x/\lambda (x))$ with the local penetration depth $\lambda
(x) $, which is given by
\begin{equation}
\frac{1}{\lambda ^{2}(x)}=
\frac{T\sum_{\omega_{n}}f^{2}_{1}(i\omega_{n},x)}{%
\lambda _{0}^{2}},
\end{equation}%
using Matsubara representation 
as shown in eq. (18) of ref. \cite{Belzig} or eq. (2.5) of 
ref. \cite{Narikiyo} with $\lambda _{0}^{-2}=32\pi ^{2}e^{2}N(0)D$. 
In Fig. \ref{Anomalous5}, the averaged
value of $\lambda ^{2}$, ($\bar{\lambda}_{av}^{2}=L/\int_{0}^{L}\frac{dx}{%
\lambda ^{2}(x)}$) is plotted as a function of temperatures ($T$), where 
$T_{C}$ is the transition temperature of bulk superconductor. 
As is shown in Fig. \ref{Anomalous5} by curves c and d, $\bar{%
\lambda}_{av}^{2}>0$ for corresponding ESE superconductor 
junctions. Thus $\bar{\lambda}_{av}$ is a
real number and a magnetic field is screened by the usual Meissner effect in
the DN irrespective of the fact whether bulk superconductor keeps 
$T$ symmetry or not.  
On the other hand, in ETO superconductor junctions, 
we find $\bar{\lambda}_{av}^{2}<0$ as shown by curves a and b. 
Therefore, the $\bar{\lambda}_{av}$
becomes a purely imaginary number in the anomalous proximity case. 
This is the consequence of the fact that the pair
amplitude $f_{1}(i\omega _{n},x)$ is purely imaginary. 
The negative value of $\bar{\lambda}_{av}^{2}$ means the generation of 
negative superfluid density locally. 
It is a novel feature 
of the anomalous proximity effect  that the
applied magnetic field is not screened in the DN region and is screened only
by the S region. 
The magnetic field can spatially oscillate
in DN and paramagnetic state becomes possible.
The paramagnetic Meissner effect in the surface of 
high Tc cuprate has been studied \cite{Higashitani97} 
and anomalous temperature dependence of the 
penetration depth has been reported 
\cite{BKK00,WPSKAHBRS98}. 
Recently, Asano $et.$  $al$ have calculated dynamical 
response of OTE pairing to the electromagnetic field \cite{Asano2011}. 
It has been found that the surface impedance $(Z = R - iX)$ of a DN has  
anomalous features where $R$ and $Z$ are resistance and reactance, 
respectively. 
In contrast to the standard relation $(R << X)$, 
an anomalous relation $R > X$ is satisfied  at low temperatures. \par
Before closing this subsection, it is noted that 
purely odd-frequency pairing
is realized in DN for DN/ETO superconductor junctions. 
The present odd-frequency pairing in DN arouses plenty of 
anomalous quantum phenomena including enhanced 
Josephson current in S/DN/S junctions  \cite{Proximityp3}. 
Since there are many new quantum phenomena  relevant to anomalous proximity 
effect via odd-frequency pairing, we hope it will be verified by experiments 
in Sr$_{2}$RuO$_{4}$ junctions \cite{Meissner3,Tshape,Asano2011}.

\subsection{odd-frequency pairing in ferromagnet  / superconductor junctions}
\label{sec:2e}

In this subsection, we discuss proximity effect in 
diffusive ferromagnet (DF)/ superconductor (S) junctions.  
The presence of odd-frequency pairing in 
DF/S junction has been 
originally proposed by Bergeret Volkov and Efetov \cite{Efetov1,Efetov2}. 
In their work, since inhomogeneous ferromagnet is assumed,  
equal-spin-triplet odd-frequency pairing 
is generated  by the spin flip of an electron \cite{Efetov1,Efetov2,BVE2003}. 
Then, so called long range proximity effect appears \cite{Efetov1,Efetov2,BVE2003,VFE2005,KSJ2001}, where the characteristic length of proximity effect 
is not $\sqrt{D/h}$ but $\sqrt{D/T}$ similar to the conventional 
proximity effect without exchange energy $h$, 
where DF becomes diffusive non-magnetic metal 
denoted as DN in previous subsections. 
It is noted that there have been 
several relevant theoretical 
\cite{Golubov,Buzdinrev,Linder08,Linder10,Lofwander2005} 
and experimental \cite{Keizer,Sosnin,Birge,Robinson,Millo} 
works in DF/S junctions up to now. 
Although original idea by Bergeret $et.$ $al$ has assumed inhomogeneous 
magnetization, it is possible to generate 
odd-frequency pairing even if we assume uniform ferromagnet with constant 
 $h$  in DF. 
Here, we discuss the symmetry of pair amplitudes and 
resulting LDOS \cite{Yokoyama2007}. 
We assume ESE superconductor and ETO one with $S_{z}=0$. 
In this case, straightforward extension of the previous subsection becomes 
possible. 

\par
\begin{figure}[htb]
\begin{center}
\scalebox{0.4}{
\includegraphics[width=22.0cm,clip]{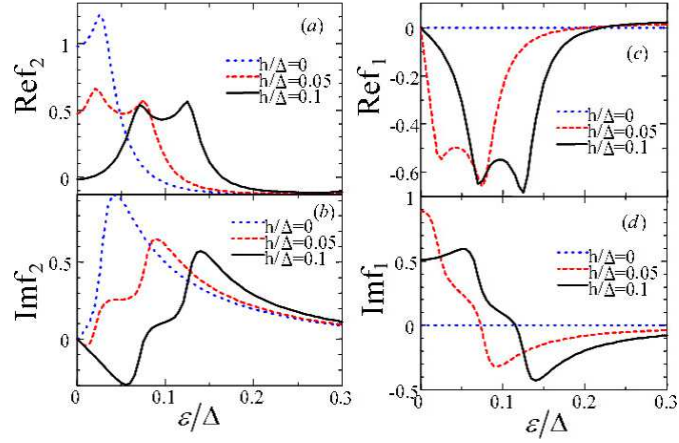}
}
\end{center}
\caption{(Color online) Real and Imaginary parts of 
pair amplitudes $f_{2}=f^{R}_{2}(\varepsilon,x)$ and $f_{1}=f^{R}_{1}(\varepsilon,x)$ at $x=0$ 
are plotted for DF/S junctions  with
$R_d/R_b = 1$ for various exchange energy $h$. 
The pairing symmetry of 
$f_{2}$ and $f_{1}$ are ESE and OTE, respectively. 
[Reproduced from Fig. 2  of Phys. Rev. B, \textbf{75},134510,(2007) by Yokoyama $et.$ $al$.]
}
\label{ferro1}
\end{figure}

Before calculation of LDOS by Usadel equation, we discuss general properties. 
In DF, only even parity $s$-wave pairing symmetry is possible due to the 
impurity scattering. 
When the symmetry of bulk superconductor is ESE, original pairing symmetry in 
DF without $h$ is  ESE as discussed in previous subsection. 
By the nonzero $h$, OTE pairing is induced by the explicit spin 
rotational symmetry breaking. 
On the other hand, when the bulk symmetry is ETO, only OTE pairing 
is possible in DF without $h$.
Then, ESE pairing is induced by $h$. 
Thus for generic case,  both ESE and OTE pair amplitudes exist in DF. 
Hereafter, we assume that the magnitude of $h$ is much smaller than 
Fermi energy in S and DF. In this case, it is possible to apply 
Usadel equation. 
The boundary condition of the 
Green's function discussed in previous section is 
also available now. 
In the present case, we must pay attention to the 
direction of $S_{z}$ in the Green's function 
which we have not explicitly written in the 
previous sections. 
As in the case of \S \ref{sec:2d}, we neglect the 
spatial dependence of pair potential in the bulk. 
Then the bulk Green's function $g_{\pm}$ 
and $f_{\pm}$ are given as follows 
\[
g_{+} \equiv g^{R}_{\uparrow,\uparrow }(\varepsilon,\theta) 
= g^{R}_{\downarrow,\downarrow }(\varepsilon,\theta)
\]
\begin{equation}
g_{-} \equiv g^{R}_{\uparrow,\uparrow }(\varepsilon,\pi-\theta) 
= g^{R}_{\downarrow,\downarrow }(\varepsilon,\pi-\theta)
\end{equation}

\begin{equation}
f_{+} \equiv f^{R}_{\uparrow,\downarrow }(\varepsilon,\theta) 
\ \ 
f_{-} \equiv f^{R}_{\uparrow,\downarrow }(\varepsilon,\pi-\theta).  
\end{equation}
For ESE superconductor, 
$f^{R}_{\uparrow,\downarrow }(\varepsilon,\theta) 
=-f^{R}_{\downarrow,\uparrow }(\varepsilon,\theta)$ is satisfied  
while 
$f^{R}_{\uparrow,\downarrow }(\varepsilon,\theta) 
=f^{R}_{\downarrow,\uparrow }(\varepsilon,\theta)$ is satisfied  
for ETO one. 
$g_{\pm}$ and $f_{\pm}$ are given by 
$g_{\pm }=\varepsilon /\sqrt{\varepsilon ^{2}-\Delta _{\pm }^{2}}$  and  
$f_{\pm }=\Delta _{\pm }/\sqrt{\Delta _{\pm }^{2}-\varepsilon ^{2}}$, respectively, with $\Delta _{\pm }=\Delta_{0} \Phi_{\pm }$ as 
in the last subsection. 
In DF, $\zeta$ in eq. (\ref{eq.1}) follows
\begin{equation}
D\frac{\partial ^{2}\zeta }{\partial x^{2}}+2i(\varepsilon +h)\sin \zeta =0. 
\label{eq.1nn}
\end{equation}
with $\sin \zeta =f^{R}_{\uparrow,\downarrow }(\varepsilon,x)$. 
On the other hand, $\sin \bar{\zeta}$ with  
$\sin \bar{\zeta} =f^{R}_{\downarrow,\uparrow }(\varepsilon,x)$
satisfies 
\begin{equation}
D\frac{\partial ^{2}\bar{\zeta} }{\partial x^{2}} +2i(\varepsilon - h) \sin
\bar{\zeta} =0  \label{eq.1n}
\end{equation}%
The boundary condition of 
$\zeta$ is given by 
eqs. ~(\ref{boundary})~(\ref{eq.3})\cite{Proximityd,Proximityp}. 
On the other hand, the boundary condition of 
$\bar{\zeta}$ is given by

\begin{equation}
\frac{L}{R_{d}}\left.\left(\frac{\partial \bar{\zeta} }{\partial x}\right)\right|_{x=L}
=\frac{
\langle \bar{F}_{1}\rangle }{R_{b}},  \label{eq.2n}
\end{equation}
at the DF/S interface. 
$\bar{F}_{1}$ is given by
\begin{equation}
\bar{F}_{1}=\frac{2T_{1}(f_{S}\cos \bar{\zeta}_{L}-g_{S}\sin \bar{\zeta}
_{L})}{2-T_{1}+T_{1}(\cos \bar{\zeta} _{L}g_{S}+\sin \bar{\zeta} _{L}f_{S})%
}
\end{equation}%
for ETO superconductor 
and 
\begin{equation}
\bar{F}_{1}=\frac{2T_{1}(-f_{S}\cos \bar{\zeta} _{L}-g_{S}\sin \bar{\zeta}
_{L})}{2-T_{1}+T_{1}(\cos \bar{\zeta} _{L}g_{S} - \sin \bar{\zeta}
_{L}f_{S})}
\end{equation}%
for ESE superconductor. 
At the interface between normal electrode and DF, 

\begin{equation}
\frac{L}{R_{d}}\left.\left(\frac{\partial \bar{\zeta} }{\partial x}\right)\right|_{x=0}
= \frac{
\langle \bar{F}_{2}\rangle }{R^{\prime }_{b}}, \bar{F}_{2}=\frac{2T_{2} \sin
\bar{\zeta} _{0}}{2-T_{2}+T_{2}\cos \bar{\zeta} _{0}}.  \label{eq.3n}
\end{equation}%
Here,  
$\bar{\zeta} _{L}=\bar{\zeta} \mid _{x=L}$, 
$\zeta_{0}=\bar{\zeta} \mid _{x=0}$. 
Even-frequency and odd-frequency pair amplitudes 
are given by 
\begin{equation}
f^{R}_{2}(\varepsilon,x )=(\sin \zeta -\sin \bar{\zeta})/2
\end{equation}%
and 
\begin{equation}
f^{R}_{1}(\varepsilon,x)=(\sin \zeta +\sin \bar{\zeta})/2, 
\end{equation}
respectively. 
After solving Usadel equation, we can determine the 
symmetry of pair amplitudes in DF as shown in Table \ref{table:5}. 
Proximity effect is absent for cases (1-2) and (2-1). 
For (1-1) and (2-2) cases, pairing symmetries in DF are
ESE and OTE, respectively. 
\begin{center}
\begin{table}
\begin{tabular}{|c|p{2.2cm}|p{2.2cm}|p{2.2cm}|}
\hline
& Bulk  & DN ($h=0$) & DF \\ \hline
(1-1) & ESE ($s$, $d_{x^{2}-y^{2}}$-wave) & ESE & ESE + \textbf{OTE} \\ \hline
(1-2) & ESE ($d_{xy}$-wave) & No & No
\\ \hline
(2-1) &　ETO ($p_{y}$-wave)  & No & No
\\ \hline
(2-2) & ETO ($p_{x}$-wave) & OTE & OTE + \textbf{ESE}
\\ \hline
\end{tabular}
\caption{
Symmetries of Cooper pair in 
Ferromagnet (diffusive) / superconductor junctions. 
The bold letter expresses the symmetry induced by the exchange field 
$h$ in DF. }
\label{table:5}
\end{table}
\end{center}
In the following, we calculate  
$f^{R}_{2}(\varepsilon,x)$,  
$f^{R}_{1}(\varepsilon,x)$,  
and LDOS $\rho(\varepsilon)$.  
$\rho(\varepsilon)$ in DF is given by 
\begin{equation}
\rho(\varepsilon)
=\frac{1}{2}(\mathrm{Re}\cos \zeta +\mathrm{Re}\cos \bar{\zeta})
\end{equation}%
We consider a model similar to Fig. \ref{Anomalous0}, 
where DN is replaced by DF. 
We assume ESE $s$-wave superconductor as a bulk state and 
choose 
$Z=3$, $Z^{\prime }=3$, $E_{Th}\equiv D/L^{2}=0.1\Delta$, 
$R_{d}/R_{b}^{\prime }=0.1$. 
Pair amplitudes at $x=0$, $i.e.$,  the interface between normal 
electrode and DF, are plotted in Fig. ~\ref{ferro1}. 

\begin{figure}[tbh]
\begin{center}
\scalebox{0.4}{
\includegraphics[width=18.0cm,clip]{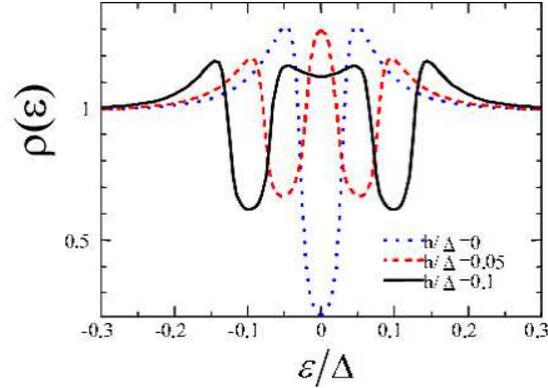}
}
\end{center}
\caption{(Color online) 
Normalized quasiparticle density of state 
$\rho(\varepsilon)$ at $x=0$ for 
DF/S junctions 
is plotted as a function of 
$\varepsilon$ for $R_{d}/R_{b}=1$.
[Reproduced from Fig. 3 of Phys. Rev. B, \textbf{75},134510,(2007) by Yokoyama 
$et$ $al$. ] 
 }
\label{ferro2}
\end{figure}

From the definition of Green's function, 
imaginary part of 
$f^{R}_{2}(\varepsilon,x)$ and real part of 
$f^{R}_{1}(\varepsilon,x)$ become zero at $\varepsilon=0$ for any $h$.  
For $h=0$, $f^{R}_{1}(\varepsilon,x)$ is zero and only 
$f^{R}_{2}(\varepsilon,x)$ has a nonzero value~\cite{TGK,Volkov}. 
With the increase of $h$, 
the magnitude of $f_{2}=f^{R}_{2}(\varepsilon,x)$ around $\varepsilon=0$ 
is suppressed. On the other hand, the magnitude of 
the imaginary part of $f_{1}=f^{R}_{1}(\varepsilon,x)$ is enhanced by $h$.  
The corresponding $\rho(\varepsilon)$ is shown in Fig. \ref{ferro2}. 
For $h=0$, $\rho(\varepsilon)$ has a mini  gap 
in consistent with previous theory ~\cite{Volkov}. 
$\rho(\varepsilon)$ is sensitive to $h$. 
For $h/\Delta=0.05$, 
$\rho(\varepsilon)$ has a ZEP. 
The presence of ZEP of 
$\rho(\varepsilon)$ in F/S junction has been discussed in 
previous works 
\cite{Golubov,Ferrozero1,Ferrozero2,Ferrozero3,Ferrozero4}.
It has been clarified that 
for $E_{Th}\sim 2hR_{b}/R_{d}$,  
$\rho(\varepsilon)$ has a ZEP \cite{Yokoferro1,Yokoferro2}. 
This condition is consistent with the present choice of the 
parameters. 
As seen from Figs. \ref{ferro1} and \ref{ferro2}, 
$\rho(\varepsilon=0)$ is enhanced when the imaginary part of $f^{R}_{1}(\varepsilon,x)$ has a 
large value (Fig. \ref{ferro1}(d)). 
It is noted that odd-frequency pair amplitude plays a 
pivotal role for the generation of ZEP \cite{Yokoferro1,Yokoferro2}. 

\begin{figure}[htb]
\begin{center}
\scalebox{0.4}{
\includegraphics[width=12.0cm,clip]{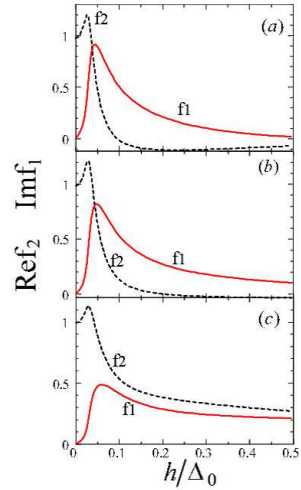}
}
\end{center}
\caption{(Color online) Pair amplitudes with ESE symmetry 
$f_{2}=f^{R}_{2}(\varepsilon,x)$ and 
that with OTE symmetry 
$f_{1}=f^{R}_{1}(\varepsilon,x)$ at $\protect\varepsilon=0$ 
are plotted as a function of $h$. 
(a) $x=0$ (normal electrode /DF interface). (b) $x=L/2$ (middle of DF). 
(c) $x=L$ (DF/S interface). 
[Reproduced from Fig. 5 of Phys. Rev. B, \textbf{75},134510,(2007) by Yokoyama 
$et.$ $al$.]）
}
\label{ferro3}
\end{figure}

In Fig.~\ref{ferro3}, we  study the crossover between 
ESE  and OTE pairings. We plot ${\rm Re}f_{2}$ and ${\rm Im}f_{1}$ as a function of $h$ for $\varepsilon =0$ at (a) $%
x=0$, (b) $x=L/2$ and (c) $x=L$ in Fig. \ref{ferro3}. 
${\rm Re}f_{1}$ increases from 
zero with $h$ and it has a maximum 
 at a certain value of $h$. 
As shown in Fig. \ref{ferro3}(a) and Fig. \ref{ferro3}(b), 
if the value of $h$ is larger than this value, 
OTE pairing  becomes dominant. 
If we use 
$\sin \bar{\zeta%
}(\varepsilon )=-\sin \zeta ^{\ast }(-\varepsilon )$ and 
$\cos \bar{\zeta}(\varepsilon )=\cos \zeta ^{\ast }(-\varepsilon )$ 
with $\zeta=\zeta(\varepsilon)$ and $\bar{\zeta}=\bar{\zeta}(\varepsilon)$, 
following relations are satisfied in pair amplitudes in DF,  
\begin{eqnarray}
f_{1}=f^{R}_{1}(\varepsilon,x ) &=&[\sin \zeta (\varepsilon )+\sin \zeta ^{\ast
}(-\varepsilon )]/2, \\
f_{2}=f^{R}_{2}(\varepsilon,x ) &=&[\sin \zeta (\varepsilon )-\sin \zeta ^{\ast
}(-\varepsilon )]/2.
\end{eqnarray}%
The ratio of $f_{2}$ to $f_{1}$ at 
$\varepsilon =0$ is given by 
\begin{equation}
\frac{{f_{2}}}{{f_{1}}}=\frac{{\tan {\mathop{\rm Re}\nolimits}\zeta
(0)}}{{i\tanh {\mathop{\rm Im}\nolimits}\zeta (0)}}.
\end{equation}
If $\left\vert {{\mathop{\rm Re}\nolimits}\zeta (0)}%
\right\vert <\left\vert {{\mathop{\rm Im}\nolimits}\zeta (0)}\right\vert $
is satisfied, 
the crossover occurs and the OTE pair amplitude becomes dominant. 
The threshold $h$, where the present crossover occurs,  
is given by 
$h\sim(R_{d}/R_{b})(E_{Th}/2)$ \cite{Yokoyama2007}.

The generation of OTE state is possible without using 
DF region \cite{Linder09L,Linder10B,Yaroslav}. 
Linder $et$ $al.$ 
has clarified that even in DN/S junctions, if the interface is spin active,
OTE pairing can be induced in DN where the bulk symmetry is 
ESE $s$-wave \cite{Linder09L,Linder10B}. In this case, also ZEP of 
LDOS can be expected. 
Recently, Yokoyama $et.$ $al$ has predicted the 
presence of anomalous Meissner effect is generated by the spin active 
interface. The magnetic susceptibility has a nonmonotonic 
temperature dependence accompanied by its sign change. 
Correspondingly, magnetic field and current density can
spatially oscillate in DN \cite{Meissner10}. 
We hope these features will be observed by experiments with 
$\mu$SR or microwave resonance. \par
\begin{figure}[htb]
\begin{center}
\scalebox{0.4}{
\includegraphics[width=12.0cm,clip]{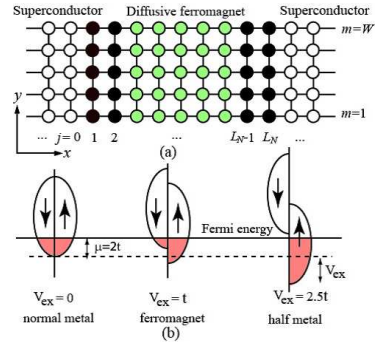}
}
\end{center}
\caption{(Color online) (a) A schematic figure of a SFS junction
on the tight-binding lattice. (b) The density of states for each
spin direction. The Josephson junction is of the S/DN/S, S/DF/S, and S/
HM/S type for $V_{ex}/t=0$,$1$ and $2.5$, respectively. 
$V_{ex}$ is the exchange potential in DF and HM. 
[Reproduced from Fig. 1 of Phys. Rev. Lett, \textbf{98},107002,(2007) by Asano 
$et.$ $al$.] }
\label{ferro4}
\end{figure}
\begin{figure}[htb]
\begin{center}
\scalebox{0.4}{
\includegraphics[width=20.0cm,clip]{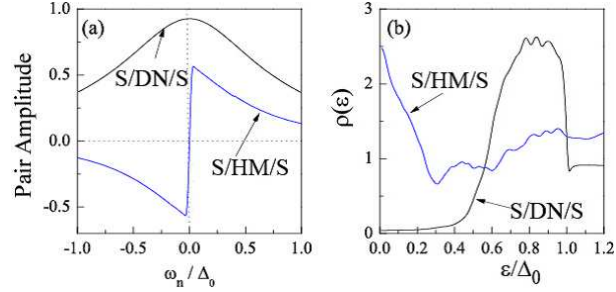}
}
\end{center}
\caption{(Color online) (a) $\omega_{n}$ dependences of 
pair amplitudes in the middle of HM(half metal)  and in a DN 
(diffusive normal metal). 
(b)Corresponding  local density of state $\rho(\varepsilon)$ 
[Reproduced from Fig. 4 of Phys. Rev. Lett, \textbf{98},107002,(2007) by Asano 
$et.$ $al$.] }
\label{ferro5}
\end{figure}

Proximity effect in fully polarized ferromagnet, $i.e.$, half metal(HM),  
has become a hot topic  now. 
Keizer $et.$ $al$ reported the existence of Josephson coupling
in superconductor/half metal/superconductor (S/HM/S)
junctions \cite{Keizer}, where symmetry of superconductor is conventional 
ESE $s$-wave. One of the possible pairing  carrying a Josephson current 
in HM is a ETO $p$-wave pairing generated by the spin flip scattering 
at the interface \cite{Eschrig2003}. 
In real S/HM/S junctions, however, half metals are close to the dirty
limit in the diffusive transport regime. 
Then more promising symmetry which can carry Josephson current is 
OTE $s$-wave pairing. 
Stimulated by the experiment by Keizer $et.$ $al$, there have been 
several theoretical proposals \cite{Asano2007PRL,Braude,Eschrig2008,
Asano2007PRB,Takahashi} and an experimental report 
\cite{Anwar}. Since the magnitude of the exchange energy is large, 
theoretical treatment is not so straightforward. 
Asano $et.$ $al$ have used a recursive Green's function in the lattice model 
(see Fig. \ref{ferro4})
and calculated pair amplitudes, LDOS and Josephson current. 
In the actual calculation, by changing the exchange potential 
$V_{ex}$, they have studied Josephson current in 
S/DN/S, S/DF/S and S/HM/S junctions.  
At the interface, spin flip scattering is introduced. 
On-site site scattering potentials are given randomly. 
The spin-flip scattering at the junction interfaces 
opens the Josephson coupling via 
odd-frequency spin-triplet Cooper pairs. 
In the middle of HM, the pairing symmetry is purely OTE state with 
equal-spin-triplet pairing. 
In Fig. \ref{ferro5}(a), the pair amplitudes in the middle of the 
DF and DN are plotted for S/HM/S and S/DN/S junctions, respectively . 
For S/HM/S junction, the pair amplitude is an odd-function of $\omega_{n}$  
and has an OTE symmetry.
On the other hand, it is an even function for S/DN/S reflecting on the 
ESE symmetry. 
In Fig. \ref{ferro5}(b), corresponding $\rho(\omega)$ in the middle of the 
DF and DN are plotted for 
S/HM/S and S/DN/S junctions, respectively . 
The $\rho(\omega)$  in the S/HM/S junction  
has a large peak at the Fermi energy 
by contrast to that in 
S/DN/S junctions. Therefore, the odd-frequency 
pairs can be detected experimentally by using the 
scanning tunneling spectroscopy \cite{Asano2007PRL,Asano2007PRB}. 


\section {Topology and bulk-edge correspondence }
\label{sec:3}

In \S \ref{sec:2}, we discussed various properties of
superconductors in terms of symmetry.
In this section, we argue topological properties of superconductors.

\subsection{Quantum Hall effect and TKNN number }

Before going to discuss topological properties of superconductors, we
would like to see more details of the integer quantum Hall states, which are a
representative example of topological order. 


The quantum Hall states are realized in two-dimensional
electron systems under uniform magnetic field perpendicular to the plane. 
%
In a commensurate periodic potential of
the crystal field, each level $n$ is described by the Bloch wave
function $\psi_{n,{\bm k}}$ with the crystal momentum ${\bm k}$ in the magnetic
Brillouin zone (BZ).
The integer quantum Hall effects occur
when the Fermi energy is located in a band gap, and 
all the levels below it are fully occupied.

To see the topological characterization, 
consider the first BZ with 
$- \pi < k_x,k_y <\pi$.
Because of the periodicity, we can identify the edges of the 1st BZ, {\it i.e.},
$k_x= -\pi$ ($k_y= - \pi$) and  $k_x= \pi$  ($k_y= \pi$)  are equivalent.
Therefore, the 1st BZ can be regarded as the torus $T^2$, and the 
occupied Bloch wave functions $\psi_{n, {\bm k}}$  defines the 
mapping from this $T^2$ to the $U(1)$ phase of the wave functions.
The topological index characterizing 
this mapping is so called the Chern number (or the TKNN integer) defined by
\cite{TKNN82,Kohmoto85}
\begin{equation}
C_1 = \frac{1}{2\pi} \int_{T^2} d^2 k
\epsilon^{ij}\partial_{k_i}{\cal A}_j({\bm k})
=\frac{1}{2\pi} \int_{T^2} d^2 k
{\cal B}_z({\bm k})
\label{eq:C1}
\end{equation}
Here, ${\cal A}_i({\bm k})$ is the ``vector potential'' of the $U(1)$ phase 
in the momentum 
space defined as
\begin{equation}
{\cal A}_i ({\bm k}) =i \sum_{E_n<E_F} \langle u_{n, {\bm k}}|\partial_{k_i} 
u_{n, {\bm k}} \rangle 
\end{equation}
in terms of the periodic part   $|u_{n, {\bm k}} \rangle $
of the Bloch wave function, and  ${\cal B}_z= \partial_{k_x}{\cal A}_y
-\partial_{k_y}{\cal A}_x$ is the
"magnetic flux density".
It can been seen that $C_1$
is an integer
corresponding to the winding number
for the mapping from $T^2$ to the phase of the Bloch wave function. 
From the Kubo formula,  the Hall conductance $\sigma_{H}$ is found to satisfy
eq.(\ref{eq:Hall1}).
Thus the quantization of the Hall conductance for IQHE can be naturally
explained as the quantization of the Chern number itself.

Recently, the topological characterization has been generalized to
superconductors in the name of ``topological superconductor''
\cite{RG00,Volovikbook,QHRZ09,SRFL08,RSFL10,Roy08,SF09,QZ10}.
The key observation is that there
exists a close similarity between the quantum Hall states and
superconducting states, summarized in Table \ref{table:analogy}:
Both states are gapped in
the bulk, so we need a finite energy to create bulk excitations. 
On the other hand, we may have gapless states on their boundaries, {\it i.e.}
chiral edge state in quantum Hall states and ABS in superconducting state.
The analogy between them enables us to apply the topological idea of the
quantum Hall states to
superconductors, while they show originally different physical phenomena
from each other.

As discussed in \S\ref{sec:TOandTQN}, for quantum Hall states, there is a
relation between the chiral gapless edge states and the bulk
Chern number $C_1$. 
So one can naturally expect that a similar bulk-edge
correspondence also holds for superconductors. 
In the following sections, 
we will show that this is the case, and gapless ABSs can be related
to bulk topological numbers of superconductors.

\begin{table}
\begin{tabular}{c||c|c}
\hline
 &IQH state & superconducting state \\ \hline
\mbox{bulk} & gapped (Landau level) & gapped (Cooper pair) \\ \hline
\mbox{edge} & chiral gapless edge state  & gapless ABS\\
\hline
\end{tabular}
\caption{Similarity between integer quantum Hall state and
 superconducting state.}
\label{table:analogy}
\end{table}

\subsection{Bulk-edge correspondence}
\label{sec:bec}
Here we will generalize the idea of the bulk-edge correspondence, or
bulk-boundary correspondence \cite{Sato10b,STYY11}.
Except for quantum Hall states, we may not use the Hall conductance
to discuss the bulk-edge correspondence, however, we will show in the
following that the bulk-edge correspondence can be obtained from a
general argument of topology. 

\begin{figure}[h]
\begin{center}
\includegraphics[width=5cm]{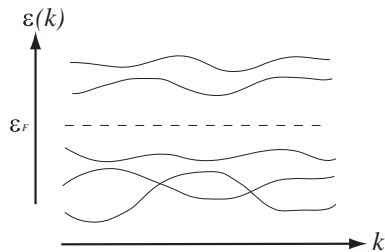}
\caption{Schematic picture for the spectrum of a band insulator. The Fermi
 energy is located in the band gap. All the states below the Fermi
 energy is fully occupied.
[Reproduced from Fig.2 of Bussei Kenkyu {\bf
 94} 311 (2010) by Sato.]
}
\label{fig:band}
\end{center}
\end{figure}

First, we discuss how a bulk-topological number can be
defined generally. 
We suppose a band description of theory, and the bulk wave functions of
electrons or quasiparticles are given by Bloch wave functions $|u_n({\bm
k})\rangle$ in the first BZ.
As is illustrated in Fig.\ref{fig:band}, we assume that the Fermi energy
is located in a band gap so the system is a band insulator in the bulk.

In this situation, we can introduce a bulk topological number by using
occupied Bloch wave functions as follows:
Let us first note that a Bloch wave function generally maps a point
${\bm k}$ in the momentum space to a point in the Hilbert space. 
Thus, using the occupied Bloch wave functions, in which all the momentum
space is filled with electrons or quasiparticles,   
we can map the whole of the first BZ to the Hilbert space.
Then, depending on a type of wave functions, the
image of the BZ may ``wind the Hilbert space'' in some way so
that it can not be deformed into a point smoothly.
If this happens, a bulk topological number is defined as a ``winding
number'' of the image of the BZ.

The above definition of topological number is rather abstract, however it is
sufficient to derive a general property of bulk topological numbers.
The definition infers that the topological number takes only discrete
integer values since it counts a``winding number'' of the image of the
BZ.
However, at the same time, the topological number could change
only continuously since it is defined by using the wave functions which can
change only continuously.
From the consistency, we can conclude that the topological
number cannot change actually as long as no singularity arises.

\begin{figure}[h]
\begin{center}
\includegraphics[width=5cm]{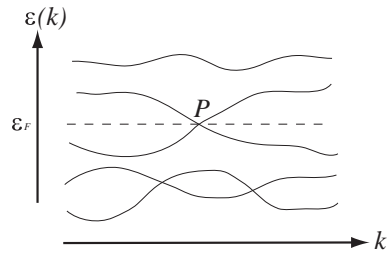}
\caption{When the bulk gap closes, a gap closing
 point $P$ appears. We cannot distinguish an occupied sate from an
 empty state at $P$.
[Reproduced from Fig.4 of Bussei Kenkyu {\bf
 94} 311 (2010) by Sato.]
}
\label{fig:band2}
\end{center}
\end{figure}

Then a question is when a singularity arises. 
The answer is when the bulk gap closes.
If the bulk gap closes, we have a gap closing point $P$. (See
Fig.\ref{fig:band2}.)
Since we cannot distinguish an occupied state from an empty state at
$P$, the concept of occupied states becomes ill-defined, so is the
topological number itself. 
Thus the topological number may change discontinuously in this case.

\begin{figure}[h]
\begin{center}
\includegraphics[width=6.5cm]{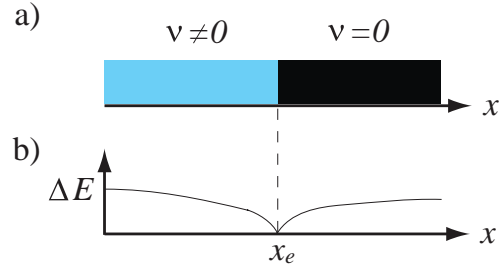}
\caption{(Color online) (a) An interface between the insulator with a nonzero
 topological number $\nu\neq 0$ and that with $\nu=0$. (b) Gap $\Delta
 E$ of the system. When the topological number $\nu$ changes at $x=x_{\rm
 e}$, the gap closes.
[Reproduced from Fig.5 of Bussei Kenkyu {\bf
 94} 311 (2010) by Sato.]
}
\label{fig:edge}
\end{center}
\end{figure}

From the above properties of topological numbers, we can
derive a generalized bulk-edge correspondence.
Consider an interface between two different band insulators illustrated
in Fig.\ref{fig:edge}(a).
We suppose that a bulk topological number $\nu$ is defined in both sides of
insulators, and $\nu\neq 0$ in the left hand side while $\nu=0$ in the
right hand side. 
Let us see what happens to the topological number when we go from the
left to the right.
We immediately find that the nonzero value of the topological number changes
abruptly to zero near the interface at $x_{\rm e}$.
Thus, the above discussion of bulk topological numbers implies that the
gap of the system must close near the interface.
In other words, we have a gapless state near the interface.
Regarding the vacuum as the topologically trivial insulator in right
hand side, we have a generalized bulk-edge correspondence:
{\it On a boundary (edge or surface) of an insulator with a nonzero bulk
topological number, there exist gapless states corresponding to the
topological number.}

While the above argument is rather intuitive, we can make it rigorous by
introducing the confining potential separating the topological phases
\cite{STYY11}.
The quantitative arguments confirm the robustness of the bulk-edge
correspondence. 

In the following sections, we will see how the bulk-edge correspondence
works for superconductors.
In the ground state of a fully gapped superconductor,
the negative energy states are fully occupied while the positive energy
states are empty. So  
we can regards a superconductor as a kind of ``insulator''.
This identification enables us to apply the bulk-edge correspondence to
superconductors.

\subsection{Topology of Andreev bound state with flat dispersion}
\label{sec:3c}

First, we consider a class of superconductors in which 
the gap function is a single component real function \cite{Sato10b,RH02,STYY11}.
It includes conventional $s$-wave
superconductors, $p_x$-wave (or $p_y$-wave) superconductors, high $T_c$
cuprates, and so on.
When the spin-orbit interaction is negligible, and 
the Cooper pairs preserve the time-reversal invariance and a spin 
in a certain direction, say $S_z$, this class of superconductors is realized.

The Hamiltonian is given by
\begin{eqnarray}
{\cal H}=\sum_{{\bm k}}
\left(c^{\dagger}_{{\bm k}\uparrow}, c_{-{\bm k}\downarrow}\right)
{\cal H}({\bm k})
\left(
\begin{array}{c}
c_{{\bm k}\uparrow} \\
c^{\dagger}_{-{\bm k}\downarrow}
\end{array}
\right) 
\end{eqnarray}
where ${\cal H}({\bm k})$ is a $2\times 2$ Bogoliubov de Gennes (BdG)
Hamiltonian
\begin{eqnarray}
{\cal H}({\bm k})=\left(
\begin{array}{cc}
\varepsilon({\bm k}) & \Delta({\bm k})\\
\Delta({\bm k}) & -\varepsilon({\bm k})
\end{array}
\right), 
\label{eq:2x2BdG}
\end{eqnarray}
with 
\begin{eqnarray}
\Delta({\bm k})=\left\{
\begin{array}{ll}
\psi({\bm k})=\psi(-{\bm k}) & \mbox{for spin-singlet}\\
d_z({\bm k})=-d_z(-{\bm k}) & \mbox{for spin-triplet}
\end{array}
\right..
\label{eq:gapfunction}
\end{eqnarray}
Here
$c_{{\bm k}\sigma}$ ($c_{{\bm k}\sigma}^{\dagger}$) is the
annihilation (creation) operator of electron with momentum ${\bm k}$ and
spin $\sigma$, and $\varepsilon({\bm k})$ the energy dispersion in the
normal state. ($\varepsilon({\bm k})=\varepsilon(-{\bm k})$.)
The time-reversal invariance implies that  
the gap function $\Delta({\bm k})$ can be chosen to be real. 
Diagonalizing the $2\times 2$ BdG Hamiltonian ${\cal H}({\bm k})$, we
find that the quasiparticle spectrum  $E({\bm k})$ is given by $E({\bm
k})=\pm \sqrt{\varepsilon({\bm k})^2+\Delta({\bm k})^2}$.
The gap of the system closes when the following condition is satisfied,
\begin{eqnarray}
\varepsilon({\bm k})=0, \quad \Delta({\bm k})=0.
\label{eq:gapclosing}
\end{eqnarray}
The negative energy state is fully occupied in the ground state.

We will identify the Hilbert space of this model.
Since the BdG Hamiltonian is a $2\times 2$ real symmetric matrix, the
occupied state  $|u({\bm k})\rangle$ is a two component real
vector with unit norm, which is given by
\begin{eqnarray}
|u({\bm k})\rangle=
\left(
\begin{array}{c}
\cos \alpha({\bm k}) \\
\sin \alpha({\bm k})
\end{array}
\right), 
\label{eq:occupied}
\end{eqnarray}  
with an angle variable $\alpha({\bm k})$.
Because of an sign ambiguity in the occupied state, 
$
|u({\bm k})\rangle \rightarrow -|u({\bm k})\rangle,
$
the angle variable $\alpha({\bm k})$ should be
identified with $\alpha({\bm k})+\pi$,  
\begin{eqnarray}
\alpha({\bm k})\sim \alpha({\bm k})+\pi. 
\end{eqnarray}
In other words, the state with $\alpha({\bm k})$ is physically the same
as that with $\alpha({\bm k})+\pi$.
Thus, instead of the eigenstate (\ref{eq:occupied}) itself, 
the physical state should be  
rather characterized by the unit vector, 
\begin{eqnarray}
\left(
\begin{array}{c}
\cos2\alpha({\bm k}) \\
\sin2\alpha({\bm k})
\end{array}
\right). 
\label{eq:hilbert}
\end{eqnarray}
The Hilbert space of the model is a one-dimensional sphere $S^1$
parameterized by the vector (\ref{eq:hilbert}). 

\begin{figure}[h]
\begin{center}
\includegraphics[width=3cm]{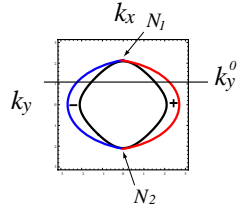}
\caption{(Color online) Two-dimensional $p_x$-wave superconductor. Since nodes exist 
at $N_1$ and $N_2$, the system is not gapful in the whole of the BZ. 
However, if we restrict ourself to the momentum space with a fixed
 $k_y=k_y^0$, we obtain a fully gapped one-dimensional system.
[Reproduced from Fig.6 of Bussei Kenkyu {\bf
 94} 311 (2010) by Sato.]
}
\label{fig:mostsimple}
\end{center}
\end{figure}

Now let us see how to define a topological number in this class of
superconductors. 
We first note a subtlety of the system.
As was discussed in \S\ref{sec:bec},  a fully gapped system is needed to
obtain a well-defined bulk topological number. 
However, for superconducting states described by eq.(\ref{eq:2x2BdG}),
we often have a nodal superconductor such as two-dimensional $p_x$-wave
superconductor or a high $T_c$ cuprate.  
This is because the gap closing condition (\ref{eq:gapclosing}) is rather easily met 
in the two- or three-dimensional momentum space.
To resolve this problem, we regard the momenta in certain directions as
parameters of the system.
Fixing them to certain values, 
we effectively have a ``fully gapped one-dimensional
system'', in which a bulk topological number can be defined.
For concreteness, in the following arguments, 
we fix the momenta in the $y$ and $z$-directions, and consider the
system as a one-dimensional system extended in the $x$-direction.  
(See Fig.\ref{fig:mostsimple}.)

In the one-dimensional gapped system, the BZ is
$-\pi<k_x\le \pi$. It is essentially a one-dimensional sphere (circle) $S^1$
since $k_x=\pi$ is identified with $k_x=-\pi$.
Thus the occupied state $|u({\bm k})\rangle$ maps $S^1$ of the Brillouin
zone to $S^1$ of the Hilbert space.  
Counting how many times the image of the BZ winds around the
Hilbert space, we can define a topological number. 
Every time the image winds around the Hilbert space, $2\alpha({\bm k})$ in
eq.(\ref{eq:hilbert}) changes by $2\pi$.
So the winding number $w_{\rm 1d}$ is given  by
\begin{eqnarray}
w_{\rm 1d}(k_y,k_z)=\frac{1}{2\pi}\int_{-\pi}^{\pi} dk_x 
\partial_{k_x} (2\alpha({\bm k})). 
\end{eqnarray}

Using $\theta({\bm k})$ defined by
\begin{eqnarray}
\cos\theta({\bm k})=\frac{\varepsilon({\bm k})}{\sqrt{\varepsilon({\bm
 k})^2+\Delta({\bm k})^2}},
\nonumber\\
\sin\theta({\bm k})=\frac{\Delta({\bm k})}{\sqrt{\varepsilon({\bm
 k})^2+\Delta({\bm k})^2}},
\end{eqnarray}
we can show that the occupied state $|u({\bm k})\rangle$ is given by
\begin{eqnarray}
|u({\bm k})\rangle
=\left(
\begin{array}{c}
\cos[(\theta({\bm k})-\pi)/2] \\
\sin[(\theta({\bm k})-\pi)/2]
\end{array}
\right).
\end{eqnarray}
Thus, we obtain $\alpha({\bm k})=\theta({\bm k})/2-\pi/2$.
From this relation, the topological number $w_{\rm 1d}$ can be
rewritten in terms of the parameters of BdG Hamiltonian,
\begin{eqnarray}
w_{\rm 1d}=\frac{1}{2\pi}\int_{-\pi}^{\pi}dk_x 
\epsilon^{ab}
m_a({\bm k})\partial_{k_x}m_b({\bm k}), 
\end{eqnarray}
where $m_a({\bm k})$ is given by
\begin{eqnarray}
m_1({\bm k})&=&
\frac{\varepsilon({\bm k})}
{\sqrt{\varepsilon({\bm  k})^2+\Delta({\bm k})^2}},
\nonumber\\
m_2({\bm k})&=& \frac{\Delta({\bm k})}
{\sqrt{\varepsilon({\bm  k})^2+\Delta({\bm k})^2}}.
\end{eqnarray}
Then, the above integral is evaluated as a simple sum
\cite{Sato09,Sato10b, STYY11}  
\begin{eqnarray}
w_{\rm 1d}=-\frac{1}{2}\sum_{k_x;\varepsilon({\bm k})=0}{\rm
 sgn}[\Delta({\bm k})]\cdot{\rm sgn}[\partial_{k_x}\varepsilon({\bm k})], 
\label{eq:formula}
\end{eqnarray}
where the summation is taken for $k_x$ with $\varepsilon({\bm k})=0$.
From the bulk-edge correspondence, there exist the gapless 
states on the boundary when $w_{\rm 1d}$ is nonzero.  

We note here that the resultant ABS has flat dispersion.
This is because the topological number $w_{\rm 1d}(k_y,k_z)$ is nonzero 
in a finite region of $(k_y,k_z)$ since it cannot change unless the
integration path intersects a gap node.
From the bulk-edge correspondence, this implies that the zero energy
state also exists in a finite region of $(k_y,k_z)$. 
In other words, the ABS corresponding to nonzero $w_{\rm 1d}$ has a flat
dispersion.

For the flat dispersion ABS, the bulk-edge correspondence is nicely
summarized in the form of the index theorem \cite{STYY11,index}.
Since the BdG Hamiltonian (\ref{eq:2x2BdG}) has the so called chiral symmetry
\begin{eqnarray}
\left\{{\cal H}({\bm k}), \sigma_y\right\}=0,
\end{eqnarray}
it can be shown that the ZEABS is an eigenstate of the
chirality operator $\sigma_y$.
Then, denoting the number of the zero energy states with the eigenvalue
$\sigma_y=\pm 1$ as $n_0^{(\pm)}$, 
we can relate the index $n_0^{(+)}-n_0^{(-)}$ to the winding number
$w_{\rm 1d}$ in the form of the index theorem,
\begin{eqnarray}
w_{\rm 1d}=n_0^{(+)}-n_0^{(-)} 
\label{eq:index2x2}
\end{eqnarray}
or
\begin{eqnarray}
w_{\rm 1d}=n_0^{(-)}-n_0^{(+)}, 
\label{eq:index2x2_2}
\end{eqnarray}
where eq.(\ref{eq:index2x2}) (eq.(\ref{eq:index2x2_2})) holds for the ABS on the
surface of the semi-infinite superconductor on $x>0$ $(x<0)$ \cite{winding1d}.

\begin{figure}[h]
\begin{center}
\includegraphics[width=8cm]{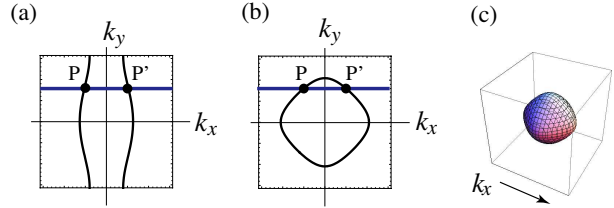}
\caption{(Color online) Fermi surfaces with simple topology in (a) 
quasi-one-dimensional system, (b) quasi-two-dimensional one, and (c) three
 dimensional one. The thick blue lines denote the integral path of
 $w_{\rm 1d}$. 
For simplicity, we illustrate the integral path only in (a) and (b). For
 each case, the integral path gets across the Fermi surface only twice at
 $k_x=\pm k_x^0$.  In (a) and (b), P and P’ denote the intersection
 point $(-k_x^0,k_y)$ and $(k_x^0,k_y)$, respectively. [Reproduced from
 Fig.4 of arXiv:1102.1322 to appear in Phys. Rev B by Sato et al.]}
\label{fig:simpleFS}
\end{center}
\end{figure}

Now we will see that the bulk-edge correspondence
reproduces the criterion of the ZEABS 
proposed previously \cite{Sato10b, STYY11}. 
In the case where the topology of the Fermi surface is simple as
illustrated in Fig.\ref{fig:simpleFS}, it has been known that
if the gap function satisfies 
\begin{eqnarray}
\Delta(k_x,k_y,k_z)\Delta(-k_x,k_y,k_z)<0, 
\label{eq:signchange}
\end{eqnarray} 
then a ZEABS exists on the boundary
perpendicular to $x$-direction \cite{ABSR1,Hu}.
In other words, 
a sign change of the gap function with respect to $k_x \rightarrow -k_x$
implies the existence of the ZEABS on a 
surface perpendicular to the $x$-direction.
The bulk-edge correspondence reproduces this result exactly:
Equation (\ref{eq:formula}) leads to
\begin{eqnarray}
w_{\rm 1d}&=&-\frac{1}{2}\left[
{\rm sgn}[\partial_{k_x}\varepsilon(-k_x^0,k_y,k_z)]
{\rm sgn}[\Delta(-k_x^0,k_y,k_z)] \right.
\nonumber\\
&&\hspace{-5ex}+\left.{\rm sgn}[\partial_{k_x}\varepsilon(k_x^0,k_y,k_z)]
{\rm sgn}[\Delta(k_x^0,k_y,k_z)] 
\right],
\end{eqnarray}
where $(\pm k_x^0, k_y,k_z)$ denotes the intersection points
between the integral path of $w_{\rm 1d}(k_y)$ and the Fermi surface. See
Fig. \ref{fig:simpleFS}. 
Noticing that ${\rm sgn}[\partial_{k_x}\varepsilon(k_x^0,k_y,k_z)]
=-{\rm sgn}[\partial_{k_x}\varepsilon(-k_x^0,k_y,k_z)]$, 
we can rewrite this as
\begin{eqnarray}
w_{\rm 1d}&=&-\frac{1}{2}
{\rm sgn}[\partial_{k_x}\varepsilon(-k_x^0,k_y,k_z)]
\nonumber\\
&&\hspace{-11ex}\times
\left[
{\rm sgn}[\Delta(-k_x^0,k_y,k_z)]-
{\rm sgn}[\Delta(k_x^0,k_y,k_z)] 
\right].
\end{eqnarray}
Thus the topological number $w_{\rm 1d}$ becomes nonzero only when the gap
function satisfies eq.(\ref{eq:signchange}), which means that the bulk-edge
correspondence reproduces the previous one in this particular simple case.

It should be noted here that the bulk-edge correspondence does not
merely reproduce the known criterion, but is more informative.
It is also applicable to more
complicated cases in which the previous criterion does not work. 
Such examples were presented in refs.\cite{Sato10b} and 
\cite{STYY11}.

As a concrete example, 
we first consider the two-dimensional $d_{xy}$-wave superconductor where
$\varepsilon({\bm k})$ and  $\Delta({\bm k})$ in eq.(\ref{eq:2x2BdG}) are given by
\begin{eqnarray}
\varepsilon({\bm k})=\frac{{\bm k}^2}{2m}-\mu,
\quad 
\Delta({\bm k})=\Delta_0 \frac{k_x k_y}{{\bm k}^2}.
\label{eq:epsilon}
\end{eqnarray}
Here $\Delta_0$ is a positive constant.
From eq.(\ref{eq:formula}), the topological number $w_{\rm 1d}(k_y)$ is
evaluated as
\begin{eqnarray}
w_{\rm 1d}(k_y)=
\left\{
\begin{array}{rl}
-1, & \mbox{for $0<k_y< k_{\rm F}$}\\
1,& \mbox{for $0>k_y> -k_{\rm F}$}\\
0, & \mbox{for $|k_y|>k_{\rm F}$}
\end{array}
\right., 
\label{eq:wdxy}
\end{eqnarray}
where $k_{\rm F}=\sqrt{2m\mu}$ is the Fermi momentum. 
Thus the index theorems (\ref{eq:index2x2}) and (\ref{eq:index2x2_2})
imply the existence of the corresponding ZEABS. 

The ZEABS is obtained by solving the BdG equation 
directly. 
For the semi-infinite $d_{xy}$ superconductor 
on $x>0$ with the boundary condition 
$|u(x=0,k_y)\rangle=0$, 
the ZEABS on $x=0$ is given by
\cite{Hu}
\begin{eqnarray}
|u_0(x)\rangle=C
\left(
\begin{array}{c}
1 \\
-i{\rm sgn}k_y
\end{array}
\right)e^{ik_y y}\sin (k_x x) e^{-x/\xi},
\end{eqnarray}
where $C$ is a normalization constant, $|k_y|<k_{\rm F}$,
$k_x=\sqrt{k_{\rm F}^2-k_y^2}$
and $\xi^{-1}=m\Delta_0 k_y/k^2_{\rm F}$. 
Since the ABS
is an eigenstate of $\sigma_y$ with eigenvalue $\sigma_y=-1$
($\sigma_{y}=1$) for $0<k_y<k_{\rm F}$ ($0>k_y>-k_{\rm F}$), 
it is found that $n_0^{(+)}=0$ and $n_0^{(-)}=1$ for $0<k_y<k_{\rm F}$ 
($n_0^{(+)}=1$ and
$n_0^{(-)}=0$ for $0>k_y>-k_{\rm F}$).
When $|k_y|>k_{\rm F}$, no
ZEABS is found, thus $n_0^{(+)}=n_0^{(-)}=0$.
As summarized in Table \ref{table:dxy} (a), the index theorem
(\ref{eq:index2x2}) holds.

On the other hand, for the
semi-infinite $d_{xy}$ superconductor on $x<0$, the ZEABS on
the surface at $x=0$
is given by
\begin{eqnarray}
|u_0(x)\rangle=C
\left(
\begin{array}{c}
1 \\
i{\rm sgn}k_y
\end{array}
\right)e^{ik_y y}\sin (k_x x) e^{x/\xi}.
\end{eqnarray}
Thus, the index theorem (\ref{eq:index2x2_2}) holds in this case.
(See also Table \ref{table:dxy} (b).)

\begin{table}
\begin{center} 
\begin{tabular}[t]{|c|c|c|c|c|}
\hline
\hline
\multicolumn{5}{c}{(a) $d_{xy}$-wave superconductor on $x>0$} \\ 
 \hline
$k_y$ 
& $n_0^{(+)}$& $n_0^{(-)}$& $n_0^{(+)}-n_0^{(-)}
$ & $w_{\rm 1d}(k_y)$ \\ 
\hline 
$0<k_y<k_{\rm F}$ 
&0 &1 &-1 &-1 \\ 
$0>k_y>-k_{\rm F}$
&1 &0& 1 & 1 \\
$|k_y|>k_{\rm F}$
&0 &0 &0 & 0 \\
\hline
\multicolumn{5}{c}{}\\
\multicolumn{5}{c}{(b) $d_{xy}$-wave superconductor on $x<0$} \\ 
 \hline
$k_y$ 
& $n_0^{(+)}$& $n_0^{(-)}$& $n_0^{(+)}-n_0^{(-)}
$ & $w_{\rm 1d}(k_y)$ \\ 
\hline 
$0<k_y<k_{\rm F}$ 
&1 &0 &1 &-1 \\ 
$0>k_y>-k_{\rm F}$
&0 &1& -1 & 1 \\
$|k_y|>k_{\rm F}$
&0 &0 &0 & 0 \\
\hline
\multicolumn{5}{c}{}\\
\hline
\hline
\end{tabular} 
\end{center}
\caption{The number $n_0^{(\pm)}$ of the zero energy ABSs with the
 $\sigma_y=\pm 1$ for (a) the semi-infinite $d_{xy}$-wave
superconductor on $x>0$ and (b) that on $x<0$. 
For comparison, we also show the topological number $w_{\rm 1d}(k_y)$ given in
 eq.(\ref{eq:wdxy}). The index theorem (\ref{eq:index2x2}) and
 (\ref{eq:index2x2_2}) hold in (a) and (b), respectively.}
\label{table:dxy}
\end{table}

Now consider the two-dimensional $p_x$-wave superconductor.
The gap function is $\Delta({\bm k})=\Delta_0 k_x/k$ with $k=\sqrt{{\bm
k}^2}$ and $\varepsilon({\bm k})$ is the same as that in eq.(\ref{eq:epsilon}).
For the $p_x$-wave superconductor, we have
\begin{eqnarray}
w_{\rm 1d}(k_y)=
\left\{
\begin{array}{rl}
-1, & \mbox{for $|k_y|<k_{\rm F}$}\\
0, & \mbox{for $|k_y|>k_{\rm F}$}
\end{array}
\right., 
\label{eq:wpx}
\end{eqnarray}
from eq.(\ref{eq:formula}).
Correspondingly, if $|k_y|<k_{\rm F}$, we obtain the following ZEABS on $x=0$ 
\begin{eqnarray}
|u(x,k_y)\rangle=
C\left(
\begin{array}{c}
1\\
-i
\end{array}
\right)e^{ik_y y}\sin(k_x x)e^{-x/\xi_{p}}, 
\end{eqnarray}
for the semi-infinite $p_x$-wave superconductor on $x>0$, and 
\begin{eqnarray}
|u(x,k_y)\rangle=
C\left(
\begin{array}{c}
1\\
i
\end{array}
\right)e^{ik_y y}\sin(k_x x)e^{x/\xi_{p}}, 
\end{eqnarray}
for the semi-infinite $p_x$-wave  superconductor on $x<0$.
Here $C$ is a normalization constant, $k_x=\sqrt{k_{\rm F}^2-k_y^2}$ and $\xi_p^{-1}=m\Delta_0 /k_{\rm
F}$.
It is also found that these solutions are the eigenstates of $\sigma_y$ with
the eigenvalue $\sigma_y=-1$ and $\sigma_y=1$, respectively. 
Thus $n_0^{(+)}$ and $n_0^{(-)}$ are summarized as Table \ref{table:px}
(a) and (b).
We confirm the relations (\ref{eq:index2x2}) and (\ref{eq:index2x2_2}),
respectively again. 

\begin{table}
\begin{center} 
\begin{tabular}[t]{|c|c|c|c|c|}
\hline
\hline
\multicolumn{5}{c}{(a) $p_{x}$-wave superconductor on $x>0$} \\ 
 \hline
$k_y$ 
& $n_0^{(+)}$& $n_0^{(-)}$& $n_0^{(+)}-n_0^{(-)}
$ & $w_{\rm 1d}(k_y)$ \\ 
\hline 
$|k_y|<k_{\rm F}$ 
&0 &1 &-1 &-1 \\ 
$|k_y|>k_{\rm F}$
&0 &0 &0 & 0 \\
\hline
\multicolumn{5}{c}{}\\
\multicolumn{5}{c}{(b) $p_{x}$-wave superconductor on $x<0$} \\ 
 \hline
$k_y$ 
& $n_0^{(+)}$& $n_0^{(-)}$& $n_0^{(+)}-n_0^{(-)}
$ & $w_{\rm 1d}(k_y)$ \\ 
\hline 
$|k_y|<k_{\rm F}$ 
&1 &0 &1 &-1 \\ 
$|k_y|>k_{\rm F}$
&0 &0 &0 & 0 \\
\hline
\multicolumn{5}{c}{}\\
\hline
\hline
 \end{tabular} 
\end{center}
\caption{The number $n_0^{(\pm)}$ of the zero energy ABSs with the
$\sigma_y=\pm 1$ for (a) the semi-infinite $p_{x}$-wave
 superconductor on $x>0$ and (b) that on $x<0$, respectively.
For comparison, we also show the topological number $w_{\rm 1d}(k_y)$ given in
 eq.(\ref{eq:wpx}). The index theorem (\ref{eq:index2x2}) and
 (\ref{eq:index2x2_2}) hold in (a) and (b), respectively. }
\label{table:px}
\end{table}

Finally, we would like to point out the relevance between the
topological structures of the ABS and the odd-frequency pairing.
The surface odd-frequency pair amplitude discussed in \S \ref{sec:ABSinUSC}
is rewritten as
\begin{equation}
\hat{f}_{1}=
\frac{i{\rm sgn}(k_{y}) \Delta_{0}}{\omega_{n}}
\frac{\mid k_{x} \mid \mid k_{y} \mid}{{\bm k}^{2}}
\label{zeroodd1}
\end{equation}
for spin-singlet $d_{xy}$-wave superconductor and
\begin{equation}
\hat{f}_{1}=\frac{i \Delta_{0}}{\omega_{n}}
\frac{\mid k_{x} \mid}{\sqrt{ {\bm k}^{2}}},
\label{zeroodd2}
\end{equation}
for spin-triplet $p$-wave superconductor
by using $k_{x}=k_{F}\cos\theta$ and $k_{y}=k_{F}\sin\theta$.
As well as the wave function of ABS of spin-singlet $d_{xy}$ and
spin-triplet $p_{x}$-wave superconductor
derived in this section, the
factor ${\rm{sgn}}k_{y}$ exists only for spin-singlet $d_{xy}$-wave case.
This factor decides the difference of the parity
of induced Cooper pair.
Next, we consider a two-dimensional semi-infinite superconductor
in $x<0$.
The corresponding pair amplitude at surface ($x=0$) is
given by
\begin{equation}
\hat{f}_{1}=
-\frac{i{\rm sgn}(k_{y}) \Delta_{0}}{\omega_{n}}
\frac{\mid k_{x} \mid \mid k_{y} \mid}{{\bm k}^{2}}
\label{zeroodd3}
\end{equation}
for spin-singlet $d_{xy}$-wave superconductor and
\begin{equation}
\hat{f}_{1}=-\frac{i \Delta_{0}}{\omega_{n}}
\frac{\mid k_{x} \mid}{\sqrt{ {\bm k}^{2}}}
\label{zeroodd4}
\end{equation}
for spin-triplet $p_{x}$-wave one, respectively.
Comparing
eq. (\ref{zeroodd1}) [(\ref{zeroodd2})] with
(\ref{zeroodd3}) [(\ref{zeroodd4})],
it is evident that the
difference between them is
the presence of $-$ sign.
The present $-$ sign exactly corresponds to the different values
of $n_{0}^{(+)}$ and $n_{0}^{(-)}$ between case (a) and
case (b) in Tables \ref{table:dxy} and \ref{table:px}. \par

\subsection{Time-reversal breaking superconductors and Majorana fermion}
\label{sec:TRBSandMF}

Next, consider time-reversal symmetry breaking superconducting states.
The analogy between the quantum
Hall states and superconducting states is direct in this
case\cite{Volovikbook,GI99,FMS01,RG00}.
The simplest one is a superconductor with a single complex gap function. 
As well as the previous subsection, 
it is  realized when the spin-orbit interaction is negligible, and the
Cooper pairs preserve $S_z$, but break the time-reversal
invariance in this time.
Also, it includes a spinless superconductor where the Cooper pairs
are formed by fully spin polarized electrons. 
We consider the latter case first.

For the spinless superconductor, 
the Hamiltonian is given by
\begin{eqnarray}
{\cal H}=\frac{1}{2}\sum_{{\bm k}}
\left(c^{\dagger}_{{\bm k}\uparrow}, c_{-{\bm k}\uparrow}\right)
{\cal H}({\bm k})
\left(
\begin{array}{c}
c_{{\bm k}\uparrow} \\
c^{\dagger}_{-{\bm k}\uparrow}
\end{array}
\right), 
\label{eq:2x2Hamiltonianspinless}
\end{eqnarray}
with 
\begin{eqnarray}
{\cal H}({\bm k})=\left(
\begin{array}{cc}
\varepsilon({\bm k}) & \Delta({\bm k})\\
\Delta^*({\bm k}) & -\varepsilon({\bm k})
\end{array}
\right). 
\label{eq:2x2BdGspinless}
\end{eqnarray}
Here we assume that the electron is fully polarized with up spin.
From the Fermi statistics, $\Delta({\bm k})$ is an odd function of ${\bm k}$.
Thus the superconducting state is spin-triplet.

The above BdG Hamiltonian ${\cal H}({\bm k})$ is a $2\times 2$ hermitian
matrix, so the occupied state $|u({\bm k})\rangle$ is a two-dimensional
complex vector with unit norm, 
\begin{eqnarray}
|u({\bm k})\rangle=
\left(
\begin{array}{c}
\cos\alpha({\bm k})e^{-i\beta({\bm k})} \\
\sin\alpha({\bm k})e^{-i\gamma({\bm k})}
\end{array}
\right), 
\label{eq:state2}
\end{eqnarray}
where $\alpha({\bm k})$, $\beta({\bm k})$ and $\gamma({\bm k})$ are
angle variables with $0\le \alpha({\bm k})<\pi/2$, $0\le\beta({\bm
k})<2\pi$ and $0\le \gamma({\bm k})<2\pi$.
Using a phase ambiguity of the eigenstate,
\begin{eqnarray}
|u({\bm k})\rangle\rightarrow e^{i\theta({\bm k})}|u({\bm k})\rangle, 
\end{eqnarray}
we can set $\gamma({\bm k})=0$ in eq.(\ref{eq:state2}).
To identify the Hilbert space, we calculate the expectation value of the
Pauli matrices $\sigma_i$, 
\begin{eqnarray}
(\langle u({\bm k})|\sigma_x|u({\bm k})\rangle, \langle u({\bm
 k})|\sigma_y|u({\bm k})\rangle, \langle u({\bm k})|\sigma_z|u({\bm
 k})\rangle)
\nonumber\\
=(\sin2\alpha({\bm k})\cos\beta({\bm k}), 
\sin2\alpha({\bm  k})\sin\beta({\bm k}),
\cos2\alpha({\bm k})). 
\label{eq:s2}
\end{eqnarray}
The vector (\ref{eq:s2}) parameterizes the two-dimensional sphere
$S^2$.
Thus we have a one-to one correspondence between the occupied
state $|u({\bm k})\rangle$ and a point on $S^2$.
%
In other words, the Hilbert space in this model is $S^2$.

In two dimensions, the BZ is topologically equivalent to the
two-dimensional torus $T^2$. Thus the occupied state $|u({\bm k})\rangle$
maps $T^2$ of the BZ into $S^2$ of the Hilbert space.
Since the surface element of $S^2$ of the Hilbert space is
\begin{eqnarray}
\sin 2\alpha d(2\alpha)d\beta, 
\end{eqnarray}
the winding number of the image of the BZ is evaluated as
\begin{eqnarray}
w_{\rm 2d}=\frac{1}{4\pi}\int_{-\pi}^{\pi}\int_{-\pi}^{\pi}dk_xdk_y 
\sin[ 2\alpha({\bm k})] 
\epsilon^{ij}\partial_{k_i}2\alpha({\bm k})\partial_{k_j}\beta({\bm k}). 
\nonumber\\
\end{eqnarray}
Introducing $\theta({\bm k})$ and $\varphi({\bm
k})$ as 
\begin{eqnarray}
\cos\theta({\bm k})=\frac{\varepsilon({\bm k})}
{\sqrt{\varepsilon({\bm k})^2+|\Delta({\bm k})|^2}},
\nonumber\\
\sin\theta({\bm k})e^{i\varphi({\bm k})}=\frac{\Delta({\bm k})}
{\sqrt{\varepsilon({\bm k})^2+|\Delta({\bm k})|^2}},
\end{eqnarray}
the occupied state is given by
\begin{eqnarray}
|u({\bm k})\rangle=
\left(
\begin{array}{c}
\cos(\theta({\bm k})/2-\pi/2)e^{i\varphi({\bm k})} \\
\sin(\theta({\bm k})/2-\pi/2)
\end{array}
\right),
\end{eqnarray}
which leads to
\begin{eqnarray}
\alpha({\bm k})=\theta({\bm k})/2-\pi/2,
\quad
\beta({\bm k})=-\varphi({\bm k}). 
\end{eqnarray}
From this, the winding number is recast into
\begin{eqnarray}
&&w_{\rm 2d}
\nonumber\\
&&=-\frac{1}{8\pi}\int_{-\pi}^{\pi}\int_{-\pi}^{\pi}dk_xdk_y 
\epsilon^{ij}\epsilon^{abc}m_a({\bm k})
\partial_{k_i}m_b({\bm k})
\partial_{k_j}m_c({\bm k}),
\nonumber\\
\end{eqnarray}
where $m_a({\bm k})$ is given by
\begin{eqnarray}
m_1({\bm k})=\frac{{\rm Re} \Delta({\bm k})}
{\sqrt{\varepsilon({\bm k})^2+|\Delta({\bm k})|^2}},
\nonumber\\
m_2({\bm k})=\frac{{\rm Im} \Delta({\bm k})}
{\sqrt{\varepsilon({\bm k})^2+|\Delta({\bm k})|^2}},
\nonumber\\
m_3({\bm k})=\frac{{\varepsilon({\bm k})}}
{\sqrt{\varepsilon({\bm k})^2+|\Delta({\bm k})|^2}}.
\end{eqnarray}
Finally, this can be evaluated as a simple sum \cite{Sato09, Sato10b}
\begin{eqnarray}
w_{\rm 2d}=-\frac{1}{2}\sum_{\Delta({\bm k})=0}{\rm
 sgn}[\varepsilon({\bm k})]{\rm sgn}[\det \partial_{k_i}\Delta_j({\bm k})], 
\end{eqnarray}
with $\Delta_1({\bm k})={\rm Re} \Delta({\bm k})$ and $\Delta_2({\bm
k})={\rm Im}
\Delta({\bm k})$ \cite{winding2d}. 
Here the summation is taken for $(k_x,k_y)$ with $\Delta({\bm k})=0$.
When $w_{\rm 2d}$ is not zero, we have a chiral edge state on the boundary. 

\begin{figure}[h]
\begin{center}
\includegraphics[width=6.5cm]{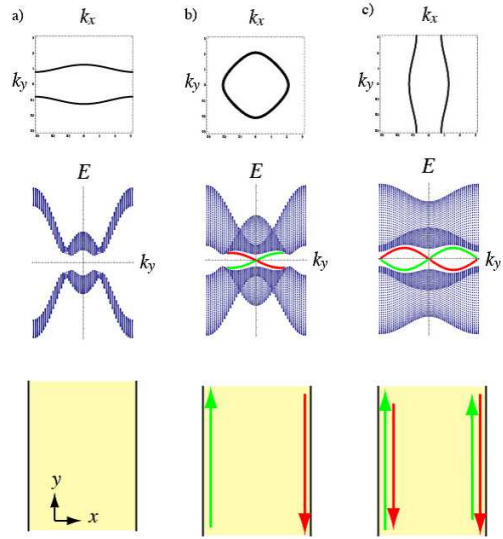}
\caption{(Color online) Edge states in spinless chiral $p+ip$-wave
 superconductors given by eq.(\ref{eq:latticemodel}). 
We take (a) $t_x=0.2$, $t_y=1$, $\mu=-1$, $\Delta_0=0.5$, 
(b) $t_x=1$, $t_y=1$, $\mu=-1$, $\Delta_0=0.5$, and
(c) $t_x=1$, $t_y=0.2$, $\mu=-1$, $\Delta_0=0.5$. 
(Top panels) Fermi surfaces in the normal states.
(Middle panels) 
Quasiparticle spectra of
the corresponding superconducting state with edges at $x=0$ and $x=L$.
The gapless modes are localized on the edges. 
The winding number $w_{\rm 2d}$ is (a) $w_{\rm 2d}=0$, (b) $w_{\rm
 2d}=1$, and (c) $w_{\rm 2d}=0$, respectively. 
(bottom panels) Schematic
 illustration of the corresponding edge state. The arrows indicate the
 direction of the group velocity of the edge modes.
}
\label{fig:edgestate_chiral}
\end{center}
\end{figure}

In Fig \ref{fig:edgestate_chiral}, we illustrate the quasiparticle spectra 
of spinless chiral $p$-wave superconductors \cite{Sato10b}.
Here we have used the lattice model with 
\begin{eqnarray}
\varepsilon({\bm k})&=&-2t_x\cos k_x-2t_y\cos k_y-\mu,
\nonumber\\
\Delta({\bm k})&=&\Delta_0(\sin k_x+i\sin k_y), 
\label{eq:latticemodel}
\end{eqnarray}
and the quasiparticle spectra are calculated under the open boundary
condition at $x=0$ and $x=L$ in the $x$-direction.
The resultant winding number $w_{\rm 2d}$ depends on the topology of the
Fermi surface, and correspondingly, the gapless edge modes also depend
on the topology of the Fermi surface.
For a two-dimensional Fermi surface, we find that $w_{\rm 2d}=1$,  so we
have a single chiral edge mode on each side of boundary.
The relation between the Fermi surface topology and the winding number
$w_{\rm 2d}$ (and the corresponding edge state) is a general property of
spin-triplet superconductors \cite{Sato09}, as will be explained in
\S\ref{sec:tos}.

As well as the edge state, we can apply the bulk-edge correspondence to
the vortex in superconductors \cite{RG00,Roy10}.
A vortex in a superconductor can be regarded as a hole in the bulk.
Then, applying the bulk-edge correspondence to the edge of the hole, we
find that there is a zero energy state in the vortex core when
$w_{\rm 2d}\neq 0$.
Some zero modes become non-zero mode when deforming the hole into
a real vortex.
When $w_{\rm 2d}$ is odd, however, at least a single
zero mode survives 
since the zero modes become massive in a pair due to the
particle-hole symmetry.  
In particular, there is a single zero mode in a vortex when $w_{\rm
2d}=\pm 1$ \cite{KS91,Roy10,FF10,FF10b,SNCM11}.

Here we would like to mention that the analogy between the quantum
Hall states and superconducting states is direct in this case. 
The quantum Hall states are
topologically characterized by nonzero Chern number.
For comparison, let us evaluate the Chern number in this model.
By using the occupied state (\ref{eq:state2}), the Chern number is
given as
\begin{eqnarray}
C_1=\frac{1}{2\pi}\int_{-\pi}^{\pi}\int_{-\pi}^{\pi}dk_xdk_y
\epsilon^{ij}\partial_{k_i}{\cal A}_j({\bm k}).  
\end{eqnarray}
where
$
{\cal A}_i({\bm k})=i\langle u({\bm k})|\partial_{k_i}u({\bm k})\rangle.
$
From a straightforward calculation, we obtain the relation
\begin{eqnarray}
\epsilon^{ij}\partial_{k_i}{\cal A}_{j}=-\frac{1}{4}\epsilon^{ij}\epsilon^{abc}
m_a\partial_{k_i}m_b\partial_{k_j}m_c.  
\end{eqnarray} 
This implies that winding number $w_{\rm 2d}$ is nothing but the
Chern number, $w_{\rm 2d}=C_1$.

In spite of the analogy above, there are differences between them. 
The $U(1)$ electromagnet gauge symmetry
is spontaneously broken in the superconductor
while it is not in the quantum Hall states. 
Thus no quantized Hall current is carried by the edge state in the former.
Furthermore, the quasiparticle $\Psi_{\bm k}$ in the superconductor
\begin{eqnarray}
\Psi_{\bm k}=\left(
\begin{array}{c}
c_{{\bm k}\uparrow} \\
c^{\dagger}_{-{\bm k}\uparrow}
\end{array}
\right), 
\end{eqnarray}
satisfies the additional condition,
\begin{eqnarray}
\Psi^{\dagger}_{-{\bm k}}=\Gamma \Psi_{\bm k},
\quad
\Gamma=\left(
\begin{array}{cc}
0 &1 \\
1 & 0
\end{array}
\right),  
\label{eq:Majorana}
\end{eqnarray}
The condition (\ref{eq:Majorana}) is called the Majorana condition.
It means that the quasiparticle $\Psi_{\bm k}$ is essentially the same as its
antiparticle $\Psi^{\dagger}_{-{\bm k}}$. 
As a result, the edge state in the superconductor
is a $1+1$ dimensional Majorana chiral fermion, not a 
chiral fermion in the quantum Hall state.

An important consequence of the Majorana condition is 
that it gives rise to the non-Abelian statistics of the vortices
\cite{RG00,Ivanov01}.
As shown in the above, for the spinless chiral $p$-wave superconductor, 
there is a single zero mode $\gamma_0$ in the vortex core.
The zero mode satisfies the anticommutation relation
\begin{eqnarray}
\{\gamma_0^{\dagger}, \gamma_0\}=1.
\end{eqnarray}
At the same time, the Majorana condition reads
\begin{eqnarray}
\gamma_0=\gamma_0^{\dagger}. 
\end{eqnarray}
Thus $\gamma_0$ is neither the annihilation operator nor the creation
operator. 
To obtain the well-defined creation operator, we need a pair of vortices. 
Using the zero mode $\gamma_0^{(i)}$ of the vortex $i$ $(i=1,2)$,
we can construct the creation operator $\gamma^{\dagger}$ as
\begin{eqnarray}
\gamma^{\dagger}=\frac{\gamma_0^{(1)}+i\gamma_0^{(2)}}{\sqrt{2}},
\end{eqnarray}
which satisfies 
\begin{eqnarray}
\gamma\neq \gamma^{\dagger},
\quad
\{\gamma,\gamma^{\dagger}\}=1. 
\end{eqnarray}
Since the vortex 1 and 2 are separated from each other, the creation
operator is defined non-locally.
This non-locality changes the statistics of vortex drastically.
Indeed, the vortices obey the non-Abelian statistics.

\begin{figure}[h]
\begin{center}
\includegraphics[width=6cm]{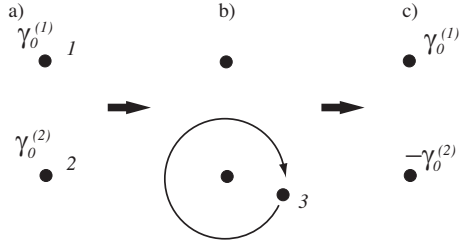}
\caption{Movement of the vortex 3 around the vortex 2.
}
\label{fig:loop2}
\end{center}
\end{figure}

To see the non-Abelian statistics of the vortices, consider 
the vortex 1 and 2 illustrated in Fig.\ref{fig:loop2}. Then, the  
vortex 3 encircles the vortex 2 and it goes far away.
In classical theory, the final configuration is completely the same as
the initial one.
In quantum theory, however, the final state can be completely different
from the initial one. 

Suppose that the initial state $|0\rangle$ is
annihilated by $\gamma$, $\gamma|0\rangle=0$. 
When the vortex 3 encircles the vortex 2, the vortex 2 also encircles
the vortex 3 in the rest frame of the vortex 2. 
Thus the zero mode $\gamma_0^{(2)}$ in the vortex 2 gets the
Aharanov-Bohm phase in this process.
If an electron (a hole) moves around the vortex 3, the Aharanov-Bohm
phase is $e^{ie\Phi_0}=-1$ ($e^{-ie\Phi_0}=-1$), where $\Phi_0$ is the
unit magnetic flux of the vortex, $\Phi_0=\pi/e$. 
Thus the zero mode $\gamma_0^{(2)}$ gets the same factor, 
\begin{eqnarray}
\gamma_0^{(2)}\rightarrow -\gamma_0^{(2)}, 
\end{eqnarray}
since it is a superposition of an electron and a hole.
This means that $\gamma$ changes as
\begin{eqnarray}
\gamma=\frac{\gamma_0^{(1)}+i\gamma_0^{(2)}}{\sqrt{2}}
\rightarrow 
\gamma^{\dagger}=\frac{\gamma_0^{(1)}-i\gamma_0^{(2)}}{\sqrt{2}}.
\end{eqnarray}
Therefore, the final state is annihilated by $\gamma^{\dagger}$, 
$\gamma^{\dagger}|1\rangle=0$, not by
$\gamma$.
The final state $|1\rangle$ is completely different from the initial
one $|0\rangle$. Indeed, there is no overlap between them,
{\it i.e.} $\langle 0|1\rangle=0$.

Note that such an exotic phenomenon never occurs if the vortices is
boson, fermion, or Abelian anyon. 
For these cases, the encircling process considered here results in a
overall phase at most, so the final state is essentially the
same as the initial one.   
In other words, the above phenomenon is a result of the non-Abelian
statistics of vortices.
A particle obeying the non-Abelian statistics is called non-Abelian
anyon, and a topological phase with non-Abelian anyons is called non-Abelian
topological phase.

Up to now, we have focused on the spinless superconductor.
Now consider the spinful time-reversal breaking superconductor. 
An example is a  chiral $p$-wave
superconductor in Sr$_2$RuO$_4$ or in a $^3$He A-phase
superfluid thin film.
When the spin-orbit interaction is negligible and the
Cooper pairs preserve a spin in a certain direction, say $S_z$,  
the Hamiltonian is given by
\begin{eqnarray}
{\cal H}=\sum_{{\bm k}}
\left(c^{\dagger}_{{\bm k}\uparrow}, c_{-{\bm k}\downarrow}\right)
{\cal H}({\bm k})
\left(
\begin{array}{c}
c_{{\bm k}\uparrow} \\
c^{\dagger}_{-{\bm k}\downarrow}
\end{array}
\right),
\label{eq:2x2BdGspinful} 
\end{eqnarray}
with ${\cal H}({\bm k})$ in eq.(\ref{eq:2x2BdGspinless}). 
The gap function
$\Delta({\bm k})$ is given  by
eq.(\ref{eq:gapfunction}).
Since ${\cal H}({\bm k})$ is the same as that for the spinless
superconductor, the Hilbert space is also the same, that is $S^2$.
Thus the topological number characterizing the state is the same winding number
$w_{\rm 2d}$.
The difference between the spinless superconductor and the spinful one
is the Majorana condition. 
For the spinful superconductor, the quasiparticle operator $\Psi_{{\bm k}}$
\begin{eqnarray}
\Psi_{{\bm k}}=\left(
\begin{array}{c}
c_{{\bm k}\uparrow} \\
c^{\dagger}_{-{\bm k}\downarrow}
\end{array}
\right) 
\end{eqnarray}
in eq.(\ref{eq:2x2BdGspinful}) does not satisfy the Majorana condition
(\ref{eq:Majorana}). 
Nevertheless, this does not imply that the Majorana fermion cannot appear
in the spinful superconductor.
Indeed, in a special kind of vortex called half-quantum vortex, 
the Majorana fermion appears.

The half-quantum vortex is realized in spinful spin-triplet
superconductors, where
the spin components of the gap
function can rotate around vortex. 
Using this degrees of freedom, 
the half-quantum vortex has the following configuration, 
\begin{eqnarray}
{\bm d}({\bm k})=\Delta({\bm k},r)e^{i\theta/2}
\left(\cos \theta/2, \sin \theta /2, 0\right), 
\end{eqnarray}
where ${\bm d}({\bm k})$ is the ${\bm d}$-vector of the spin-triplet gap
function,
$r$ the distance from the vortex core, and $\theta$ the angle around the
vortex. 
Since this configuration does not preserve $S_z$, its BdG Hamiltonian
is not given in the form of eq.(\ref{eq:2x2BdGspinful}). 
Instead, we can treat the system as two copies of spinless superconductors.
Indeed, the matrix representation of the gap function leads to
\begin{eqnarray}
\Delta={\bm d}({\bm k},{\bm r}){\bm \sigma}\sigma_y=
\left(
\begin{array}{cc}
-\Delta({\bm k},r) & 0\\
0 & \Delta({\bm k},r)e^{i\theta}
\end{array}
\right),
\label{eq:hqv}
\end{eqnarray}
so there is no mixing between the different spin sectors.
Then, each spin sector is the spinless superconductor
described by eq.(\ref{eq:2x2Hamiltonianspinless}).
(For the down spin sector, $c_{{\bm k} \uparrow}$ and $c^{\dagger}_{{\bm k}
\uparrow}$ should be replaced by $c_{{\bm k} \downarrow}$ and
$c^{\dagger}_{{\bm k} \downarrow}$, respectively.)
Equation (\ref{eq:hqv}) implies that only the down-spin sector has a vortex.
Therefore, we can regard as the half quantum vortex as the vortex in spinless
superconductor with down spin. 
This means that if ${w_{\rm 2d}}$ is odd, there is a Majorana zero mode
in the half-quantum vortex.
Thus the half-quantum vortex can obey the non-Abelian statistics.
Recently, 
an experimental evidence of the half-quantum vortex in Sr$_2$RuO$_4$ has been
reported \cite{JFVBCGM11}.
Also, a half-quantum vortex is expected to exist in a thin film of
$^3$He A phase \cite{Kawakami09,Kawakami11}. 
 
\subsection{Helical edge modes in TI/QSHE}
\label{sec:3e}

In this subsection, we consider the band insulators without the
superconductivity from the viewpoint of the topology.
For time-reversal invariant systems, 
new topological phases arise when the spin-orbit interaction is strong enough.
Here we review briefly such states, the quantum spin Hall
states \cite{KM05a, KM05b, BZ06, BHZ06} and its three dimensional
generalization, topological insulators \cite{FKM07, MB07, Roy09}.

The quantum spin Hall state is easily understood as
a pair of the integer quantum Hall states if $S_z$ is a good quantum number.
Consider the following two-dimensional system preserving $S_z$,
\begin{eqnarray}
{\cal H}({\bm k})=
\left(
\begin{array}{cc}
{\cal H}_{+}({\bm k}) &0 \\
0 & {\cal H}_{-}({\bm k})
\end{array}
\right),
\end{eqnarray}
where ${\cal H}_{+}({\bm k})$ (${\cal
H}_{-}({\bm k})$) is the Hamiltonian for
up-spin (down-spin) electron. 
Here we suppose that the spin-orbit interaction works as a kind of
magnetic field,  
and each spin sector realizes an integer quantum
Hall state with the Hall conductance
$\sigma^{(\sigma)}_{H}=-C_1^{(\sigma)}e^2/h$ ($\sigma=\pm$).
Due to the time-reversal invariance, the Chern numbers of the two spin
sectors are opposite from each other, $C_1^{(+)}=-C_1^{(-)}$.
Thus, the spin Hall conductance
$\sigma_{sH}=\sigma_H^{(+)}-\sigma_H^{(-)}$ is quantized,
while the total Hall conductance,
$\sigma_H=\sigma_H^{(+)}+\sigma_H^{(-)}$, is zero.
Correspondingly, there exist $C_1^{(+)}$ ($C_1^{(-)}$)
gapless edge states with up-spin (down-spin) carrying the spin Hall conductance.  


Now consider a time-reversal invariant perturbation which does not
preserve $S_z$.
The spin indices $\sigma=\pm$ turns into that of
pseudo-spin $\alpha, \beta$, and the Chern numbers $C_1^{(+)}$ and
$C_1^{(-)}$ are replaced by $C_1^{(\alpha)}$ and $C_1^{(\beta)}$.
Then, due to the spin mixing, most of the gapless edge states become
massive.
Nevertheless, if $C_1^{(\alpha)}(=-C_1^{(\beta)})$ is an odd number, at least
one pair of gapless edge states survives:
Because of the time-reversal invariance, the gapless states form a set
of Kramers doublets.
Since there is no mixing in the same Kramers doublet,
the Kramers doublets become massive in a pair by the mixing with a
different Kramers doublet.
Thus, if there are an odd number of the
Kramers doublets, which is realized when $C_1^{(\alpha)}$ is odd, 
at least one Kramers doublet of the gapless edge
states remains massless without a partner.

As indicated by the above argument, for a generic time-reversal invariant
system, where $S_z$ is not a good quantum number,
the quantum spin Hall state is characterized by the parity of
$C_1^{(\alpha)}$, $(-1)^{\nu_{\rm 2dTI}}\equiv (-1)^{C_1^{(\alpha)}}$, which
is called the ${\bm Z}_2$ invariant in the literature.
Then the resultant gapless edge state is called the helical edge mode.

The topological insulator is a three dimensional generalization of the
quantum spin Hall state \cite{FKM07, MB07, Roy09}.
It is characterized by the four ${\bm Z}_2$ invariants
$(\nu_0;\nu_1,\nu_2,\nu_3)$, which can take the value 0 or 1. 
When, $\nu_0 =1$,
the system is called strong topological insulator which supports an odd
number of 2D Dirac fermions on the surface. When some of $\nu_1$,
$\nu_2$ and $\nu_3$ are 1, while $\nu_0=0$, the system is classified
into weak topological insulator. In this case,
an even number of Dirac fermions are on the surface, which can in principle 
be paired and gapped due to the disorder scattering etc., and are not stable.
However, the weak topological insulator is still distinct from the conventional
insulator with all $\nu$'s being zero, as has been evidenced by 
one-dimensional channels appearing along the dislocation \cite{RZV09}.

\subsection{Helical superconductors and helical Majorana modes}
\label{sec:3f}

There exist
a topological state of superconductors which is analogous to the quantum
spin Hall state. Such a superconductor is called the helical
superconductor \cite{QHRZ09,Roy08}.
A representative example of the helical superconductor is
two-dimensional Rashba
noncentrosymmetric superconductors \cite{SF09,TYBN08}. 
Let us see its basic property.

The model Hamiltonian of the Rashba superconductor in two dimensions is 
\begin{eqnarray}
&&{\cal H}={\cal H}_{\rm kin}+{\cal H}_{\rm SO}+{\cal H}_{\rm pairing},
\nonumber\\
&&{\cal H}_{\rm kin}=\sum_{{\bm k},\sigma}\varepsilon_{\bm k}
 c_{{\bm k}\sigma}^{\dagger}c_{{\bm k}\sigma}
-\mu_{\rm B}H_z \sum_{{\bm k},\sigma}(\sigma_z)_{\sigma\sigma'}
c_{{\bm k}\sigma}^{\dagger}c_{{\bm k}\sigma},
\nonumber\\
&&{\cal H}_{\rm SO}=
\alpha\sum_{{\bm k},\sigma,\sigma'}{\bm {\mathcal L}}_{\bm k}
\cdot{\bm \sigma}_{\sigma\sigma'}c_{{\bm k}\sigma}^{\dagger}c_{{\bm k}\sigma'}, 
\nonumber\\
&&{\cal H}_{\rm pairing}=
\frac{1}{2}\sum_{{\bm k}\sigma\sigma'}
\Delta_{\sigma\sigma'}({\bm k})c_{{\bm k}\sigma}^{\dagger}
c_{-{\bm k}\sigma'}^{\dagger}
+{\rm h.c},
\label{eq:Hamiltonian}
\end{eqnarray}
where $c_{{\bm k}\sigma}^{\dagger}$ ($c_{{\bm k}\sigma}$) is a creation (an
annihilation) operator for an electron with momentum ${\bm
k}=(k_x,k_y)$, spin $\sigma$.
The energy band dispersion is $\varepsilon_{\bm k}=-2t(\cos k_x+\cos
k_y)-\mu$ with the hopping parameter $t$ and the chemical potential
$\mu$, and the Rashba spin-orbit coupling is ${\bm {\mathcal L}}_{\bm
k}=(\sin k_y, -\sin k_x)$.
Because of parity mixing of Cooper pairs, 
the gap function $\Delta({\bm k})$ has both a spin-triplet component
${\bm d}({\bm k})$ and a spin-singlet one $\psi({\bm k})$ at the same time,
$\Delta({\bm k})=i\psi({\bm k})\sigma_y+i{\bm d}({\bm k}){\bm \sigma}\sigma_y$. 
Due to the strong spin-orbit coupling, the spin-triplet component ${\bm d}({\bm k})$ is
aligned with the Rashba coupling, ${\bm d}({\bm k})=\Delta_{\rm t}{\bm {\mathcal
L}}_{\bm k}$ \cite{FAKS04}.
For the spin-singlet component $\psi({\bm k})$, we assume an $s$-wave pairing,
$\psi({\bm k})=\Delta_{\rm s}$. 
The amplitudes $\Delta_{\rm t,s}$ are chosen as real.
The Zeeman coupling
$\mu_{\rm B}H_z\sum_{\bm k}
(c^{\dagger}_{{\bm k}\uparrow}c_{{\bm k}\uparrow}
-c^{\dagger}_{{\bm k}\downarrow}c_{{\bm k}\downarrow})$ 
with $H_z$ a magnetic field in the $z$ direction
has been also introduced for later use. 

Before going to study topological properties of the system, we first
examine the bulk spectrum of the system. 
Topological nature of the system changes only when
the gap of the bulk spectrum closes.
The bulk spectrum $E({\bm k})$ of the system is obtained by diagonalizing the
following $4\times 4$ matrix, 
\begin{eqnarray}
&&{\cal H}({\bm k})=
\nonumber\\
&& \left(
\begin{array}{cc}
\varepsilon_{\bm k}-h\sigma_z+\alpha{\bm {\mathcal L}}_{\bm k}
\cdot
{\bm \sigma} &
i\Delta_{\rm s}\sigma_y+i\Delta_{\rm t}{\bm {\mathcal L}}_{\bm k}
\cdot{\bm \sigma}\sigma_y\\
-i\Delta_{\rm s}\sigma_y
-i\Delta_{\rm t}{\bm {\mathcal L}}_{\bm k}\sigma_y\cdot{\bm \sigma}&
- \varepsilon_{\bm k}+h\sigma_z+\alpha{\bm {\mathcal
L}}_{\bm k}
\cdot{\bm \sigma}^{*}
\end{array}
 \right),
\nonumber\\
\end{eqnarray}
with $h=\mu_{\rm B}H_z$, and we have
\begin{eqnarray}
E({\bm k})&=&\pm\Biggl[
\varepsilon_{\bm k}^2+(\alpha^2+\Delta_{\rm
 t}^2)
{\bm {\mathcal L}}_{\bm k}^2+h^2+\Delta_{\rm s}^2
\nonumber\\
&&\pm2\sqrt{
(\varepsilon_{\bm k}\alpha
+\Delta_{\rm s}\Delta_{\rm t})^2{\bm {\mathcal L}}_{\bm k}^2
+(\varepsilon_{\bm k}^2+\Delta_{\rm s}^2)h^2}
\Biggr]^{1/2}
\end{eqnarray}
The gap of the system closes only when 
\begin{eqnarray}
&&\varepsilon_{\bm k}^2+(\alpha^2
+\Delta_{\rm t}^2){\bm {\mathcal L}}_{\bm k}^2
+h^2+\Delta_{\rm s}^2
\nonumber\\
&&=2\sqrt{(\varepsilon_{\bm k}\alpha
+\Delta_{\rm s}\Delta_{\rm t})^2{\bm {\mathcal L}}_{\bm k}^2
 +(\varepsilon_{\bm k}^2+\Delta_{\rm s}^2)h^2}, 
\end{eqnarray}
which is equivalent to
\begin{eqnarray}
\varepsilon_{\bm k}^2+\Delta_{\rm s}^2
=h^2+(\alpha^2+\Delta_{\rm t}^2){\bm {\mathcal L}}_{\bm
k}^2,
\nonumber\\
\varepsilon_{\bm k}\Delta_{\rm t}{\bm {\mathcal L}}_{\bm k}
=\Delta_{\rm s}\alpha{\bm {\mathcal L}}_{\bm k}.
\label{eq:gap-closing}
\end{eqnarray}
When $\Delta_{\rm t}\neq 0$, eq.(\ref{eq:gap-closing}) is met either
when  
\begin{eqnarray}
\varepsilon_{\bm k}=\frac{\Delta_{\rm s}}{\Delta_{\rm t}}\alpha,
\quad 
\left(1+\frac{\alpha^2}{\Delta_{\rm t}^2}\right)
\left(\Delta_{\rm t}^2{\bm {\mathcal L}}_{\bm k}^2-\Delta_{\rm
 s}^2\right)+h^2=0.
\label{eq:gap-closing1-1}
\end{eqnarray}
or
\begin{eqnarray}
\varepsilon_{\bm k}^2+\Delta_s^2=h^2,
\quad
{\bm {\mathcal L}}_{\bm k}=0.
\label{eq:gap-closing1-2}
\end{eqnarray}
In the absence of the magnetic field, only eqs.(\ref{eq:gap-closing1-1}) can be
met and they are rewritten in simpler forms,
\begin{eqnarray}
\varepsilon_{\bm k}^2=\alpha^2{\bm {\mathcal L}}_{\bm k}^2,
\quad
\Delta_{\rm t}^2{\bm {\mathcal L}}_{\bm k}^2=\Delta_{\rm s}^2.
\label{eq:gap-closing2}
\end{eqnarray}
Topological nature of the system does not change unless
eq.(\ref{eq:gap-closing1-1}) or eq.(\ref{eq:gap-closing1-2}) (or
eq.(\ref{eq:gap-closing2}) when $H_z=0$) is satisfied. 
  

When $H_z=0$, the system is time-reversal invariant, and the
topological property is characterized by the ${\bm Z}_2$ invariant
like the quantum spin Hall state.
Below, we will show that if the spin-triplet pairing is stronger than
the spin-singlet one, the ${\bm Z}_2$ number is non-trivial.

To see this, we adiabatically deform the
Hamiltonian of the system without gap closing. 
This process does not change the ${\bm Z}_2$ topological number, 
since it changes only when the gap closes \cite{SF09,TYBN08}.
From eq.(\ref{eq:gap-closing2}), it is found that if
the spin-triplet amplitude $\Delta_{\rm t}{\bm {\mathcal L}}_{\bm k}$
is larger than the spin-singlet one $\Delta_{\rm s}$ on the Fermi
surface given by $\varepsilon_{\bm k}=\alpha{\bm {\mathcal L}}_0({\bm
k})$, we can  take $\Delta_{\rm s}\rightarrow 0$, then
$\alpha\rightarrow 0$ without gap closing.
(If $\varepsilon_{\bm k}=0$ at one of the time-reversal momenta ${\bm k}=(0,0),
(\pi,0), (0,\pi), (\pi,\pi)$, the gap closes when $\Delta_{\rm
s}=0$. However, this undesirable gap closing can be avoided by
changing $\mu$ or $t$ slightly.)
Thus, its ${\bm Z}_2$ number is the same as that of the
pure spin-triplet SC with ${\bm d}({\bm k})=\Delta_{\rm t}{\bm {\mathcal
L}}_{\bm k}$.
The resultant system preserves $S_z$, and 
the BdG Hamiltonian is
decomposed into two spinless chiral superconductors in up-spin sector and
the down-spin sector,
${\cal
H}_{+}({\bm k})$ and ${\cal H}_{-}({\bm k})$, respectively,
\begin{eqnarray}
&&{\cal H}_{\pm}({\bm k})= 
\nonumber\\
&&\left(
\begin{array}{cc}
\varepsilon_{\bm k}& -\Delta_{\rm t}(\sin k_x \pm i\sin k_y)\\
-\Delta_{\rm t}(\sin k_x \mp i\sin k_y) & -\varepsilon_{\bm k}
\end{array}
\right).
\nonumber\\
\end{eqnarray}
If the Fermi surface is two-dimensional and electron-like, their Chern numbers
is $C_1^{(\pm)}=\mp 1$ for $\Delta_{\rm t}>0$. (see
Fig.\ref{fig:edgestate_chiral}).
Thus the ${\bm Z}_2$ number is non-trivial.

\begin{figure}
\begin{center}
\includegraphics[width=6cm]{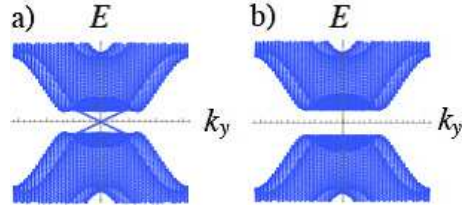}
\caption{(Color online) The energy spectra of the 2D noncentrosymmetric superconductor with edges in the absence of
 magnetic field.
We take $t=1$, $\mu=-3$, $\alpha=0.6$.
(a) noncentrosymmetric superconductor with dominating $p$-wave paring.
 $\Delta_{\rm t}=0.6$ and  $\Delta_{\rm s}=0.1$. (b) noncentrosymmetric superconductor with purely $s$-wave
 paring. $\Delta_{\rm t}=0$ and $\Delta_{\rm s}=0.6$.}
\label{fig:edgestatewithouth} 
\end{center}
\end{figure}

From the bulk-edge correspondence, 
there should exist gapless edges if the spin-triplet pairs
dominate the superconductivity.
In Fig.\ref{fig:edgestatewithouth} (a), we show the energy spectrum of
the 2D noncentrosymmetric superconductor with edges.
It is found that there exist gapless edge states in the bulk gap.
The gapless edges states form a Kramers pair.

For comparison, we also illustrate the energy spectrum for
the 2D noncentrosymmetric superconductor with purely $s$-wave paring in Fig.\ref{fig:edgestatewithouth} (b).
As is seen clearly, no edge state is obtained.
This is also consistent with the trivial ${\bm Z}_2$ number of the
purely $s$-wave paring.

\begin{figure}
\begin{center}
\includegraphics[width=8cm]{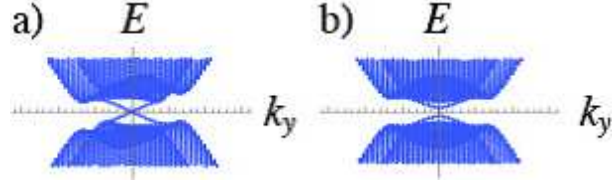}
\caption{(Color online) The dependence of gapless edge states on the
 direction of the magnetic field. 
We take $t=1$, $\mu=-3$, $\alpha=0.6$, 
$\Delta_{\rm t}=0.6$ and $\Delta_{\rm s}=0.1$. 
a) $\mu_{\rm B}H_x=0.15$, $\mu_{\rm B}H_y=0$ and $\mu_{\rm B}H_z=0$. b)
 $\mu_{\rm B}H_x=0$, $\mu_{\rm B}H_y=0.15$ and $\mu_{\rm B}H_z=0$.}
\label{fig:edgestatehy} 
\end{center}
\end{figure}

We also notice that 
the helical Majorana gapless edge states
are very sensitive to the direction of the applied magnetic field.
As seen in Fig.\ref{fig:edgestatehy},
they become unstable under a small magnetic field in the
$y$-direction, while the gapless edge states are stable under a magnetic field 
in the $x$- and $z$-direction.
As a result, the magnetic field $H_y$ along the edge causes a tiny gap
of the order
$O(\mu_{\rm B}H_y)$ for the edge states.
The experimental detection of the edge states have been proposed 
theoretically \cite{TYBN08,SF09,YTJ05,IHSYMTS07,EIT10,VVE08,
LY09,SBMT10,SRL09}. 

\subsubsection{Non-Abelian topological order and Majorana fermion
   induced by spin-orbit interaction and Zeeman field}
\label{subsec:withh}

The spin-orbit interaction enables us to realize Majorana fermion and
non-Abelian topological order in the presence of the Zeeman field
\cite{SF09,STF09,STF10}. 
In particular, the Majorana fermion can be realized even if the symmetry
of the gap function is spin-singlet dominant
\cite{STF09,STF10,SLTD10,SF10}.
Here we see this interesting phase of noncentrosymmetric
superconductors.

\begin{figure}
\begin{center}
\includegraphics[width=8cm]{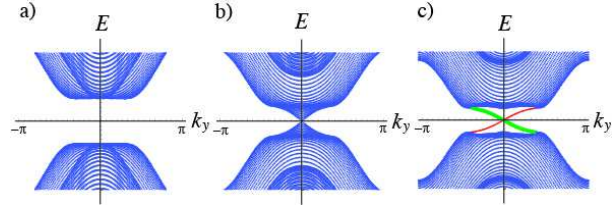}
\caption{(Color online) Majorana edge modes in an $s$-wave
 superconducting state with Rashba spin-orbit interaction under the
 Zeeman magnetic field $h$ in the $z$-direction. 
Energy spectra of the system with open boundary in the $x$-direction
 and the periodic boundary condition in the $y$-direction. (a) $h=0$ (b)
$h=h_c\equiv\sqrt{\varepsilon_{{\bm k}=0}^2+\Delta_{\rm s}^2} $ (c) $h>h_c$.
When the Zeeman field $h$ is larger than the critical value
 $h_c$, 
the Majorana zero mode appears.
[Reproduced from Fig.1 of 
Phys. Rev. Lett.{\bf 103}, 020401 (2009) by Sato {\it et al.}] }
\label{fig:edgestate_with_h} 
\end{center}
\end{figure}

An intuitive understanding on this mechanism of non-Abelian topological
phase is obtained if we perform the dual transformation of the BdG
Hamiltonian \cite{STF09,STF10}.
For the two-dimensional Rashba superconductor (\ref{eq:Hamiltonian}), 
the dual Hamiltonian is obtained by the following unitary transformation,
\begin{eqnarray}
{\cal H}^{\rm D}({\bm k})=D{\cal H}({\bm k})D^{\dagger},
\quad
D=\frac{1}{\sqrt{2}}
\left(
\begin{array}{cc}
1 & i\sigma_y\\
i\sigma_y & 1
\end{array}
\right),
\end{eqnarray}
which leads to
\begin{eqnarray}
&&{\cal H}^{\rm D}({\bm k})
\nonumber\\
&&=
\left(
\begin{array}{cc}
\Delta_{\rm s}+\Delta_{\rm t}{\bm {\mathcal L}}_{\bm k}\cdot{\bm \sigma}
-h\sigma_z &
 -i\varepsilon_{\bm k}\sigma_y
-i\alpha{\bm {\mathcal L}}_{\bm k}\cdot{\bm \sigma}\sigma_y \\
i\varepsilon_{\bm k}\sigma_y+i\alpha{\bm {\mathcal L}}_{\bm
 k}\sigma_y{\bm \sigma}&  -\Delta_{\rm s}+\Delta_{\rm t}{\bm {\mathcal
 L}}_{\bm k}\cdot{\bm \sigma}^{*}+
h\sigma_z
\end{array}
\right).
\label{eq:dual}
\end{eqnarray}
From eq. (\ref{eq:dual}), it is found that the Rashba spin-orbit interaction
in the original Hamiltonian is formally transformed
into a ``$p$-wave pairing interaction''  with the ${\bm d}$ vector,
${\bm d}({\bm k})=-\alpha{\bm {\mathcal L}}_{\bm k}$, in the dual
Hamiltonian ${\cal H}^{\rm D}({\bm k})$. 
Since ${\cal H}^{\rm D}({\bm k})$ has a nonstandard constant kinetic
term, this does not necessarily mean that the topological properties of
${\cal H}({\bm k})$ are the same as those of a usual $p$-wave superconductor.
However, we find that that the topological order emerges when $h$ satisfies
\begin{eqnarray}
h^2>\Delta_{\rm s}^2+\varepsilon_{{\bm k}={\bm 0}}^2, 
\label{eq:condition}
\end{eqnarray}
due to the Rashba spin-orbit interaction.
Indeed, one can prove that the Chern number $C_1=-1$ if the condition
(\ref{eq:condition}) is met \cite{SF09,STF09,STF10}. 
Also, the same condition (\ref{eq:condition}) is obtained 
from the analysis of Majorana zero mode in a vortex \cite{SLTD10}.  
In Fig.\ref{fig:edgestate_with_h}, we illustrate the energy bands of
our Hamiltonian (\ref{eq:Hamiltonian}) with edges. 
This figure clearly shows the existence
of a chiral gapless state if the Zeeman field $h$ satisfies
eq.(\ref{eq:condition}), even when the spin-triplet superconductor
is absent.

From the argument in \S\ref{sec:TRBSandMF}, it is evident that a vortex
supports a Majorana zero mode in this case. 
Indeed, we can demonstrate the existence of the Majorana zero mode
explicitly by solving the BdG equation for a single vortex \cite{SF09,STF09,STF10,SLTD10,STLSD10}.
As explained in \S\ref{sec:TRBSandMF},
if there exists a single Majorana fermion zero mode for
each vortex, vortices obey the non-Abelian statistics.
Following ref.\cite{STF09}, we use here the dual Hamiltonian ${\cal H}^{\rm
D}({\bm k})$ to solve the BdG equation, 
then construct a solution in the original Hamiltonian
${\cal H}({\bm k})$ by using the duality transformation
(\ref{eq:dual}). 
For simplicity,
we assume $\varepsilon_{{\bm k}=0}=0$ and $\Delta_{\rm t}=0$.
Then, if $h$ satisfies $|\Delta_{\rm s}|<|h|<2|\Delta_{\rm s}|$, 
the low energy properties are governed by quasiparticles on the smaller Fermi
surface, which is split from the larger one by the spin-orbit
interaction \cite{STF10}. 
The larger Fermi surface can be neglected to construct the zero mode. Thus, we
concentrate on
fermions with $|{\bm k}|\sim 0$ for which ${\cal H}^{\rm D}({\bm k})$ 
is decomposed
\begin{eqnarray}
{\cal H}_{\pm}^{\rm D}({\bm k})=
\left(
\begin{array}{cc}
\Delta_{\rm s}\mp h & \alpha(\pm k_y+ik_x)\\
\alpha(\pm k_y-ik_x) & -\Delta_{\rm s}\pm h
\end{array}
\right). 
\end{eqnarray}
The BdG equations for ${\cal H}^{\rm D}_{\pm}$
 with a single vortex can be solved \cite{RG00}, then, we find a unique
 zero energy solution with a quasiparticle field 
$\gamma^{\dagger}=\int d{\bm r}[u_0\psi_{+}^{\dagger}+v_0\psi_{+}]$,
where $u_0=i(re^{i\theta})^{-1/2}e^{-(h-\Delta_{\rm s})r/\alpha}$,
$v_0=-i(re^{-i\theta})^{-1/2}e^{-(h-\Delta_{\rm s})r/\alpha}$. 
The solution is normalizable when eq.(\ref{eq:condition})
is satisfied. 
This is the Majorana zero energy mode; {\it i.e.},
$\gamma^{\dagger}=\gamma$. 
Using the duality transformation
 (\ref{eq:dual}),  we also found
that a vortex in the original Hamiltonian has a single
Majorana zero mode, which implies
that the vortex is a non-Abelian anyon.

Here note that the condition
(\ref{eq:condition}) implies the Zeeman energy larger than
the $s$-wave BCS gap $\Delta_{\rm s}$.
Thus the orbital depairing effect could destroy the superconductivity
while the Pauli depairing effect is suppressed in the presence of the Rashba
spin-orbit interaction \cite{Frigeri, Fujimoto1}.
If the spin-triplet amplitude is dominant and the $s$-wave pair amplitude is
small enough, 
the condition (\ref{eq:condition}) can be met for a weak magnetic field
without destroying the superconducting gap \cite{SF09}. 
On the other hand, if the $s$-wave pair amplitude is dominant, 
one always needs a strong magnetic field larger than the superconducting gap.
Several schemes to realize the non-Abelian
topological phase in this case has been proposed,
(i) $s$-wave superfluids in neutral ultracold fermionic atoms with laser
generated spin-orbit interaction \cite{STF09,STF10},
(ii) heterostructure semiconductor device \cite{SLTD10,Alicea10},
and (iii) heavy fermion systems \cite{STF10}.

\begin{figure}[h]
\begin{center}
\includegraphics[width=8cm]{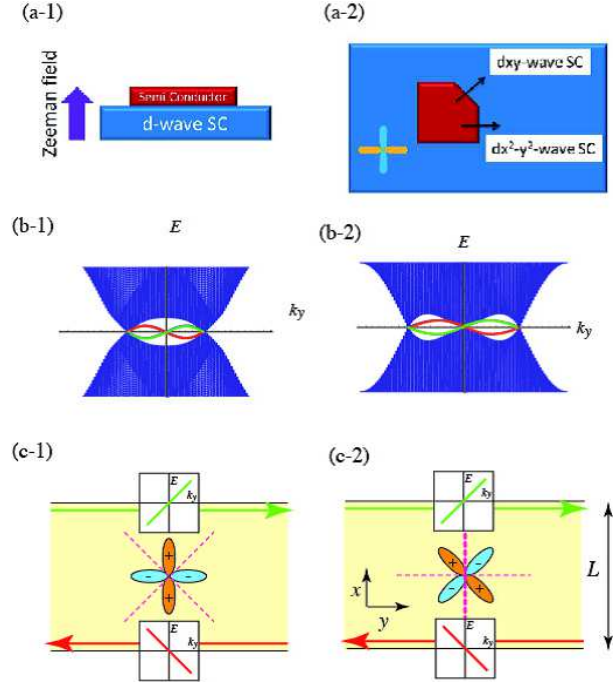}
\caption{(Color online) Majorana fermions in nodal superconductors.
(a) Possible realization scheme to realize nodal superconductor with
 Majorana fermion. Due to the proximity effect and the potential
 gradient at the interface, $d$-wave superconducting state with the
 Rashba spin-orbit coupling is realized in the interface between the
 high $T_c$ cuprate and the semiconductor.
(a-1) Side view (a-2) Top view.  
(b) Energy spectra of the systems with open boundaries at $x=0$ and $x=L$
in the $x$-direction
and the periodic boundary condition in the $y$-direction for
(b-1) $d_{x^2-y^2}$-wave pairing, and (b-2) $d_{xy}$-wave pairing. 
Majorana gapless edge modes at $x=0$ and $x=L$ are depicted, respectively, in
 green and red curves.
(c) Schematic illustration of Majorana edge modes counterpropagating on
 two opposite edges for (c-1) $d_{x^2-y^2}$-wave pairing and
(c-2) $d_{xy}$-wave pairing.
[(b) and (c): Reproduced from Fig.1 of 
 Phys. Rev. Lett. {\bf 105}, 217001 (2010) by Sato and Fujimoto]
}
\label{fig:edgemodes2}
\end{center}
\end{figure}

If we consider $d_{x^2-y^2}$-wave or $d_{xy}$-wave pairing instead of
the $s$-wave pairing, the condition (\ref{eq:condition}) changes as \cite{SF10}
\begin{eqnarray}
h^2>\varepsilon^2_{{\bm k}=0}, 
\label{eq:dxycond}
\end{eqnarray}
since these pair amplitudes vanish at ${\bm
k}={\bm 0}$.
Thus now the condition (\ref{eq:dxycond}) is independent of the
superconducting gap.
This means that the Majorana fermion and the non-Abelian topological
phase could be realized in a weak magnetic field without destroying
superconductivity for systems with 
small $\varepsilon_{\bm
k}={\bm 0}$.
In Fig.\ref{fig:edgemodes2}, we illustrate the Majorana edge modes
in this case. 
The Majorana fermion state is topologically protected in spite of the
presence of bulk gapless nodal excitations.
Due to the existence of bulk nodes, one cannot obtain a well-defined
Chern number, but the particle-hole symmetry makes the parity of
the Chern number well-defined \cite{SF10}.
The non-Abelian
nodal superconductor is realizable in an interface
between a centrosymmetric nodal superconductor such as
high-$T_c$ cuprates and a semiconductor.
In such a system, because of the considerably large
superconducting gap, the experimental detection of
Majorana modes may be easier.


\subsubsection{Dispersionless Majorana fermion}

For a certain type of noncentrosymmetric superconductors, novel Majorana
fermion is possible \cite{TMYYS10,YSTY10,STYY11}.
It preserves the time-reversal symmetry, however, unlike the helical Majorana
fermions, it consists of a single branch of flat dispersion.

\begin{figure}[h]
\begin{center}
\includegraphics[width=6cm]{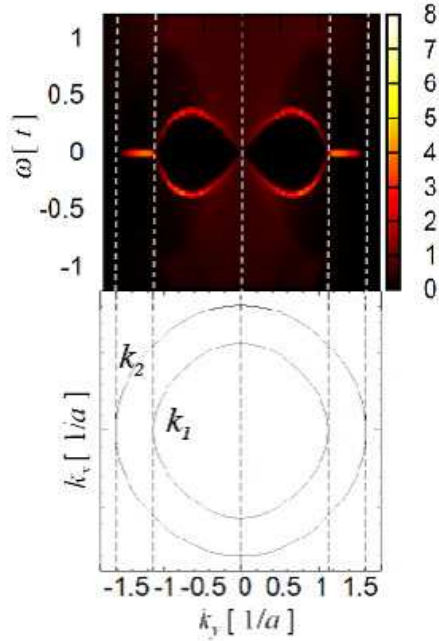}
\caption{(Color online) (Top) Angle-resolved LDOS for $d_{xy}+p$-wave
 pairing as a function of $k_y$. $\Delta_t\neq 0$ and
 $\Delta_s=0$ .
(Bottom) The corresponding Fermi surfaces of the
 system. The Fermi surfaces are split into two by the Rashba spin-orbit
 interaction. Single branch of zero energy ABS exist in the region
 of $k_1<|k_y|<k_2$.}
\label{fig:fig9}
\end{center}
\end{figure}

The time-reversal invariant Majorana fermion is realized for the
$d_{xy}+p$-wave Rashba noncentrosymmetric superconductor.
Such a pairing symmetry has been suggested for heterointerface
LaAlO$_3$/SrTiO$_{3}$ \cite{YOTI09}. 
Figure \ref{fig:fig9} illustrates the quasiparticle spectrum for the
semi-infinite $d_{xy}+p$-wave Rashba noncentrosymmetric superconductor
on $x>0$.
Due to the Rashba spin-orbit interaction, the Fermi surface is split
into two with the Fermi momenta $k_1$ and $k_2$ ($k_1<k_2$).
See Fig.\ref{fig:fig9}.
There exists a single branch of zero energy edge  state in the momentum region
$k_1<|k_y|<k_2$ between the Fermi surfaces. 
The wave function for the zero energy edge state 
$\Psi_{m}(k_{y})$ can be written as  
 $^{T}\Psi_{m}(k_{y}) =
 (u_{1}(k_{y}),u_{2}(k_{y}),v_{1}(k_{y}),v_{2}(k_{y}))$ with
$
u_{1}(k_{y})=v_{1}^{*}(-k_{y}),
$
$ 
u_{2}(k_{y})=v_{2}^{*}(-k_{y}).
$
The Bogoliubov quasiparticle creation operator for this state is constructed in the usual way as $\gamma^\dagger(k_{y}) 
= u_1(k_y) c_\uparrow^\dagger(k_y) + u_2(k_y) c_\downarrow^\dagger(k_y) 
+ v_1(k_y) c_\uparrow(-k_y) + v_2(k_y) c_\downarrow(-k_y)$, and
satisfies the Majorana condition
$\gamma^{\dagger}(k_{y})=\gamma(-k_{y})$.

Unlike Majorana fermions discussed in previous sections, the
present single Majorana bound state at $k_y$ is realized with time-reversal
symmetry. The time-reversal invariant Majorana bound state has the
following three characteristics. (a) It has a unique flat
dispersion: To be consistent with the time-reversal invariance,
the single branch of zero mode should be symmetric
under $k_y\rightarrow -k_y$. Therefore, by taking into account the
particle-hole symmetry as well, the flat dispersion is required.
On the other hand, the conventional time-reversal
breaking Majorana bound state has a linear dispersion.
(b) The spin-orbit coupling is necessary to obtain the time-reversal invariant
Majorana bound state. Without spin-orbit coupling, the
time-reversal invariant Majorana bound state vanishes. (c) The time-reversal invariant
Majorana bound state is topologically stable under small
deformations of the Hamiltonian preserving $k_y$.
The topological stability is ensured by the topological invariant $W$
\cite{YSTY10,STYY11} 
\begin{eqnarray}
W(k_y)=\frac{1}{2\pi}{\rm Im}
\left[
\int dk_x \partial_{k_x}\ln{\rm det}\hat{q}({\bm k})
\right],
\label{eq:Wdef} 
\end{eqnarray}
with $\hat{q}({\bm k})=[\varepsilon_{\bm k}-i\psi({\bm k})]
\sigma_y+[\alpha{\bm {\mathcal L}}_{\bm k}
-i{\bm d}({\bm k})]
\cdot{\bm \sigma}\sigma_y$.
The flat dispersion of the Majorana fermion is terminated in a gap node.
The last property is closely related to the topological stability of
line nodes of noncentrosymmetric superconductors \cite{Sato06,Beri10}.

It is noted that 
the appearance of surface flat bands in three-dimensional noncentrosymmetric 
superconductors or topological media have been also 
discussed recently \cite{Schnyder10,Brydon11,HV11,Heikkila}. 

\subsection{Topological phase in odd-parity superconductors}
\label{sec:tos}

In \S\ref{sec:TRBSandMF}, we have seen that the topological properties of the 
spinless superconductor changes as the topology of the Fermi surface changes.
This is a general property of spin-triplet odd-parity superconductors
\cite{Sato09,Sato10}.

To see this, consider a general Hamiltonian $H$
for odd-parity superconductors,
\begin{eqnarray}
&&H=\frac{1}{2}\sum_{{\bm k}\alpha\alpha'}
(c_{{\bm k}\alpha}^{\dagger}, c_{-{\bm k}\alpha})
H({\bm k})
\left(
\begin{array}{c}
c_{{\bm k}\alpha'} \\
c^{\dagger}_{-{\bm k}\alpha'}
\end{array}
\right),
\nonumber\\
&&H({\bm k})=\left(
\begin{array}{cc}
{\cal E}({\bm k})_{\alpha\alpha'} &\Delta({\bm k})_{\alpha\alpha'} \\
\Delta^{\dagger}({\bm k})_{\alpha\alpha'} & -{\cal E}^{T}(-{\bm
 k})_{\alpha\alpha'}
\end{array}
\right),
\label{eq:hamiltonian}
\end{eqnarray}
where $c^{\dagger}_{{\bm k}\alpha}$ ($c_{{\bm k}\alpha}$) denotes the
creation (annihilation) operator of electron with momentum ${\bm k}$.
The suffix $\alpha$ labels other degrees of freedom for electron such as
spin, orbital degrees of freedom, sub-lattice indices, and so on. 
${\cal E}({\bm k})$ is the Hamiltonian of the electron in the normal
state.  The system in the normal
state is supposed to be invariant under the inversion  
$
c_{{\bm k}\alpha}\rightarrow \sum_{\alpha'}
P_{\alpha \alpha'}c_{-{\bm k}\alpha'}$ with
$P^2=1$, 
$
P^{\dagger} {\cal E}({\bm k}) P={\cal E}(-{\bm k})
$.
We also assume that the parity of the gap function $\Delta({\bm k})$
is odd,
$
P^{\dagger}\Delta({\bm k})P^{*}=-\Delta(-{\bm k}). 
$

First, consider the case with the time-reversal symmetry breaking.
The BdG Hamiltonian (\ref{eq:hamiltonian}) has the following
particle-hole symmetry,
\begin{eqnarray}
CH({\bm k})C^{\dagger}=-H^{*}(-{\bm k}), 
\quad
C=\left(
\begin{array}{cc}
0 &1 \\
1 & 0
\end{array}
  \right). 
\label{eq:particle-hole}
\end{eqnarray}
From this, we can say that if $|u_n({\bm k})\rangle$
is a quasiparticle state with positive energy $E_n({\bm k})>0$ satisfying 
$
H({\bm k})|u_n({\bm k})\rangle=E_n({\bm k})|u_n({\bm k})\rangle, 
$
then $C|u_n^{*}(-{\bm k})\rangle$ is a quasiparticle state with
negative energy $-E_n(-{\bm k})<0$.
We use a positive (negative) $n$ for  $|u_n({\bm k})\rangle$
to represent
a positive (negative) energy quasiparticle state, and set 
\begin{eqnarray}
|u_{-n}({\bm k})\rangle =C |u_n^{*}(-{\bm k})\rangle. 
\label{eq:negative energy state}
\end{eqnarray}
Let us define the gauge fields
\begin{eqnarray}
A^{(+)}_i({\bm k})=i\sum_{n>0}
\langle u_n({\bm k})|\partial_{k_i}|u_n({\bm k})\rangle,
\quad
A^{(-)}_i({\bm k})=i\sum_{n< 0}
\langle u_n({\bm k})|\partial_{k_i}|u_n({\bm k})\rangle,
\end{eqnarray}
where the relation
\begin{eqnarray}
A^{(+)}_i({\bm k})=A^{(-)}_i(-{\bm k}) 
\label{eq:a+a-}
\end{eqnarray}
holds from (\ref{eq:negative energy state}).
In addition, we have
\begin{eqnarray}
A_i^{(-)}({\bm k})+A_i^{(+)}({\bm k})
=i{\rm tr}\left[
U^{\dagger}({\bm k})\partial_{k_i}
U({\bm k})
\right]
=i\partial_{k_i}\ln\det U({\bm k}),
\label{eq:U}
\end{eqnarray}
with the unitary matrix $U({\bm k})$ given by
\begin{eqnarray}
U_{nm}({\bm k})=u_n^{m}({\bm k}) 
\end{eqnarray}
where $u_n^m({\bm k})$ is the $m$-th component of 
$|u_n({\bm k})\rangle=(\cdots,u_n^1({\bm k}),u_n^{2}({\bm k}),
\cdots)^{T}$.

\begin{figure}[h]
\begin{center}
\includegraphics[width=7cm]{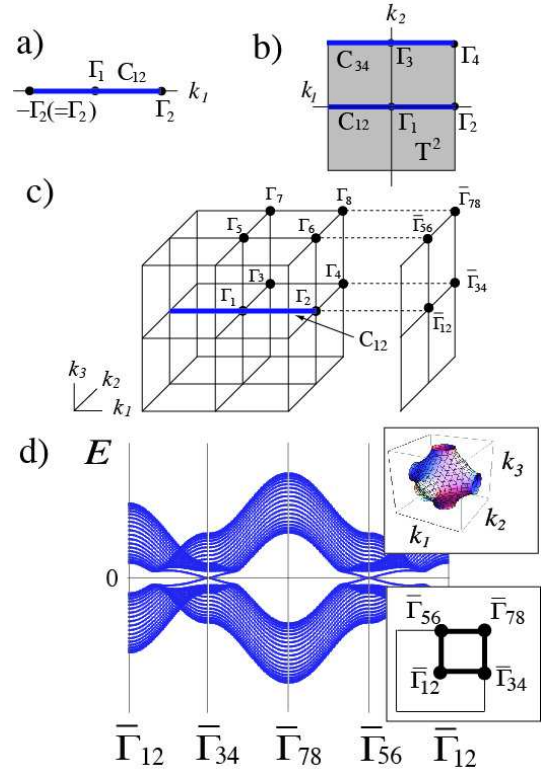}
\caption{(Color online) The time-reversal invariant momenta $\Gamma_i$, and the time-reversal invariant closed path ${\rm
 C}_{ij}$ passing through $\Gamma_i$ and $\Gamma_j$ in the BZ. a) 1D BZ.
The solid line denotes ${\rm C}_{12}$.
b) 2D BZ $T^2$. c) 3D BZ and the surface BZ of a (100) face. d) 
2D band structure for a
 slab with a (100) face for the 3D time-reversal invariant odd parity superconductor that has
 the Fermi surface with $(-1)^{\tilde{\nu}[{\rm
 C}_{34}]}=(-1)^{\tilde{\nu}[{\rm C}_{56}]}=-1$ and the gap function
 $\Delta({\bm k})=i{\bm d}({\bm k})\cdot{\bm \sigma}\sigma_y$ with
 $d_i({\bm k})=\Delta\sin k_i$. The insets show the Fermi surface (top)
 and the 2D surface BZ (bottom). [Reproduced from Fig.1 of
 Phys. Rev. B {\bf 81}, 220504(R) (2010) by Sato]}
\label{fig:bz2}
\end{center}
\end{figure}

Now consider the time-reversal invariant path ${\rm C}_{ij}$ in the
BZ, which passes
through the time-reversal invariant momenta $\Gamma_i$ and
$\Gamma_j$. See Fig.\ref{fig:bz2}. 
Note that ${\rm C}_{ij}$ is a closed path because of
the periodicity of the BZ. 
The periodicity of the BZ implies that 
the integration of $A_i^{(-)}({\bm k})$ along ${\rm
C}_{ij}$ is quantized as,
\begin{eqnarray}
\oint_{\rm C_{ij}}dk_i A_i^{(-)}({\bm k})
&=&\frac{1}{2}\oint_{\rm C_{ij}}dk_i\left[A_i^{(-)}({\bm k})
+A_i^{(+)}({\bm k})
\right]
\nonumber\\
&=&\frac{1}{2}\oint_{\rm C_{ij}}dk_i i\partial_{k_i}\ln\det U({\bm k})
\nonumber\\
&=&N\pi,
\quad
(\mbox{$N$: integer }).
\label{eq:wilsonloop}
\end{eqnarray}
Thus the Berry phases $e^{i\oint_{{\rm C}_{ij}}A_i^{(-)}({\bm
k})}$ takes only two possible values, and 
we use it as the topological index,
\begin{eqnarray}
(-1)^{\nu[{\rm C}_{ij}]}=
e^{i\oint_{{\rm C}_{ij}}A_i^{(-)}({\bm k})}
=\pm 1. 
\end{eqnarray} 
Like the one-dimensional winding number $w_{\rm 1d}$ in \S \ref{sec:3c}, 
the topological index
$(-1)^{\nu[{\rm C}_{ij}]}$ characterizes the ``one-dimensional gapped
superconductor'' along the path ${\rm C}_{ij}$. 
When $(-1)^{\nu[{\rm C}_{ij}]}=-1$, 
the bulk-edge correspondence yields that there exists a zero energy edge state
at the corresponding momentum in the surface BZ. 
On the other hand, when $(-1)^{\nu[{\rm C}_{ij}]}=1$, no ZEABS is
required at the corresponding momentum in the surface BZ. 

Up to now, we have not used the specific property of odd-parity
superconductors. 
For odd-parity superconductors,
the combination of the inversion
and the electromagnetic $U(1)$ gauge symmetry,  
$
c_{{\bm k}\alpha}\rightarrow iP_{\alpha\alpha'}c_{{\bm k}\alpha'},
$
is manifestly preserved.
Thus, $H({\bm k})$
has the following symmetry
\begin{eqnarray}
\Pi^{\dagger} H({\bm k}) \Pi=H(-{\bm k}),
\quad
\Pi=
\left(
\begin{array}{cc}
P & 0\\
0 & -P^{*}
\end{array}
\right). 
\end{eqnarray}
From this, we have $[H({\Gamma_i}), \Pi]=0$ for the time-reversal
invariant momentum $\Gamma_i$. Thus, the quasiparticle state
$|u_n(\Gamma_i)\rangle$ at $\Gamma_i$ is simultaneously an eigenstate of $\Pi$,
\begin{eqnarray}
\Pi |u_n(\Gamma_i)\rangle=\pi_n(\Gamma_i)|u_n(\Gamma_i)\rangle.  
\end{eqnarray}
Then, the product of $\pi_n(\Gamma_i)$ for occupied states, 
$\prod_{n<0}\pi_n(\Gamma_i)$, has the following interesting properties:
(a) it takes only discrete values  $\prod_{n<0}\pi_n(\Gamma_i)=\pm 1$, and 
(b) its value can change only when the gap of the system closes at $\Gamma_i$.
These properties are what we expect for the topological index, and
suggest that $\prod_{n<0}\pi_n(\Gamma_i)$ is a kind of topological
index. 
Indeed, we can relate $\prod_{n<0}\pi_n(\Gamma_i)$ to the topological
index $(-1)^{\nu[{\rm C}_{ij}]}$ as \cite{Sato10},  
\begin{eqnarray}
(-1)^{\nu[{\rm C}_{ij}]}=\prod_{n<0}\pi_n(\Gamma_i)\pi_n(\Gamma_j).
\label{eq:piformula}
\end{eqnarray}

For ordinary superconductors, the superconducting gap is much smaller than the
Fermi energy.
Therefore, we reasonably assume that  the typical energy scale of
the gap function $\Delta(\Gamma_i)$ at the time-reversal invariant
momentum is much smaller than that of ${\cal E}(\Gamma_i)$.  
Under this assumption, we can deform the gap function $\Delta({\bm k})$
adiabatically
so as $\Delta({\Gamma_i})\rightarrow 0$
without closing the bulk energy gap. 
Because of the topological nature of $(-1)^{\nu[{\rm C}_{ij}]}$, this
adiabatic process does not change the value of $(-1)^{\nu[{\rm C}_{ij}]}$.
In the process $\Delta(\Gamma_i)\rightarrow 0$, the BdG Hamiltonian at
$\Gamma_i$ reduces to
$H(\Gamma_i)\rightarrow {\rm diag}({\cal E}(\Gamma_i),-{\cal E}^{T}(\Gamma_i))$.
For this simple Hamiltonian, we can evaluate $\pi_n(\Gamma_i)$ 
rather easily, and we obtain the final expression \cite{Sato10},  
\begin{eqnarray}
(-1)^{\nu[{\rm C}_{ij}]}=\prod_{\alpha}{\rm
sgn}\varepsilon_{\alpha}(\Gamma_i)
{\rm sgn}\varepsilon_{\alpha}(\Gamma_j),
\label{eq:epsilonformula}
\end{eqnarray}
where $\varepsilon_{\alpha}(\Gamma_i)$ is an eigenvalue of ${\cal
E}({\Gamma_i})$ and  the product of $\alpha$ is taken for all eigenstates of
${\cal E}({\Gamma_i})$.
This formula implies that for odd-parity superconductors the gapless
boundary state can be predicted by the Fermi surface structure.

When the odd-parity superconductors are fully-gapped in two dimensions, 
the Chern number
$C_1$ also characterizes the topological phases.
By using eqs. (\ref{eq:C1}) and (\ref{eq:a+a-}),   
it can be linked to the topological index $\nu[{\rm C}_{ij}]$ as
\begin{eqnarray}
C_1&=&\frac{1}{2\pi}\int_{T^2}d^2k{\cal B}_z({\bm k}) 
= \frac{1}{\pi}\int_{T^2_{+}}d^2k{\cal B}_z({\bm k}) 
\nonumber\\
&=& \frac{1}{\pi}\oint_{\partial T_{+}^2}dk_i A_i^{(-)}({\bm k})
=\nu[{\rm C}_{12}]-\nu[{\rm C}_{34}],
\label{eq:chw}
\end{eqnarray}
where $T_{+}^2$ is the upper half of $T^2$, and ${\rm C}_{ij}$ is
the time-reversal invariant path illustrated in Fig.\ref{fig:bz2}(b). 
Therefore, eq.(\ref{eq:epsilonformula}) yields
\begin{eqnarray}
(-1)^{C_1}
=\prod_{\alpha, i=1,2,3,4}{\rm
sgn}\varepsilon_{\alpha}(\Gamma_i).
\label{eq:chformula}
\end{eqnarray}

One can confirm the validity of the formulas (\ref{eq:epsilonformula}) and
(\ref{eq:chformula}) by applying them to the spinless
superconductors in Fig.\ref{fig:edgestate_chiral}. 
For example, from the Fermi surface structures illustrated in the top row, 
we find that only the case of Fig.\ref{fig:edgestate_chiral} (b) has an odd
Chern number.
Correspondingly, a single edge mode appears on each edge in
Fig.\ref{fig:edgestate_chiral} (b). 
Furthermore, it is found that the difference between
Fig.\ref{fig:edgestate_chiral} (a) and
(c) can be understood as the
difference in $(-1)^{\nu[{\rm C}_{ij}]}$ for each case:
One find that $(-1)^{\nu[{\rm C}_{12}]}=(-1)^{\nu[{\rm C}_{34}]}=-1$
for Fig. \ref{fig:edgestate_chiral} (c), while $(-1)^{\nu[{\rm
C}_{12}]}=(-1)^{\nu[{\rm C}_{34}]}=1$ for (a). 
The zero energy edge states in (c) are originated from the non-trivial
values of these topological indices.

Now consider the time-reversal invariant case.
The time-reversal invariance implies that $(-1)^{\nu[{\rm C}_{ij}]}$ is
always trivial, {\it i.e.} $(-1)^{\nu[{\rm C}_{ij}]}=1$. 
However, use of the time-reversal
invariance as well as the particle-hole symmetry 
makes it possible to define another topological index.

Because of the time-reversal invariance, 
the quasiparticle states form Kramers pairs,
$|u_n^{s}({\bm k})\rangle$ $(s={\rm I},{\rm II})$,
$
|u_n^{\rm I}({\bm k})\rangle =T |u_n^{{\rm II}}(-{\bm k})\rangle, 
$
with the time-reversal operator $T$.
The new topological index is defined by
the gauge field for the ``half'' of the Kramers doublets, say,
$A_i^{{\rm I}(-)}({\bm k})=i\sum_{n<0}\langle u_n^{{\rm I}}({\bm k})|
\partial_{k_i}|u_n^{\rm I}({\bm k})\rangle$.
Then the topological index $(-1)^{\tilde{\nu}[{\rm C}_{ij}]}$ 
is defined as 
\begin{eqnarray}
(-1)^{\tilde{\nu}[{\rm C}_{ij}]}=
e^{i\oint_{{\rm C}_{ij}}A_i^{I(-)}({\bm k})}.
\end{eqnarray}
In a manner similar to the above, for odd-parity
superconductors, $(-1)^{\tilde{\nu}[{\rm
C}_{ij}]}$ are determined 
by the Fermi surface structure \cite{Sato10},
\begin{eqnarray}
(-1)^{\tilde{\nu}[{\rm C}_{ij}]}
=
\prod_{\alpha}{\rm sgn}\varepsilon_{2\alpha}(\Gamma_i)
{\rm sgn}\varepsilon_{2\alpha}(\Gamma_j),
\label{eq:epsilonformula2}
\end{eqnarray}
where $\varepsilon_{\alpha}(\Gamma_i)$ is an eigenvalue of ${\cal
E}(\Gamma_i)$, and we have set
$\varepsilon_{2\alpha}(\Gamma_i)=\varepsilon_{2\alpha+1}(\Gamma_i)$ by
using the Kramers degeneracy. 

Furthermore, 
other topological indices characterizing time-reversal invariant
superconductors can be linked to the Fermi surface topology. 
For the full-gapped
two-dimensional case, their topological nature is
characterized by the ${\bm Z}_2$ number $(-1)^{\nu_{\rm 2dTI}}$
introduced in \S \ref{sec:3e} and \S \ref{sec:3f}. 
For odd-parity superconductors, it satisfies \cite{Sato09, Sato10}
\begin{eqnarray}
(-1)^{\nu_{\rm 2dTI}}=\prod_{\alpha,i=1,2,3,4}{\rm sgn}
\varepsilon_{2\alpha}(\Gamma_i),
\label{eq:2dTI}
\end{eqnarray}
where $\Gamma_i$ is the time-reversal invariant momenta in
Fig.\ref{fig:bz2}(b).
Thus, we can predict the helical Majorana edge mode from
the knowledge of the Fermi surface.
In addition, for the full-gapped three-dimensional case, we have another
topological index known as  the three-dimensional winding number $w_{\rm 3d}$
\cite{GV88,Volovikbook,SRFL08,RSFL10}. 
(When $w_{\rm 3d}$ takes a nonzero integer $n$, there exist $n$ Majorana
cones on the surface.)
We can also connect this to the Fermi surface
structure as \cite{Sato09,Sato10,FB10}
\begin{eqnarray}
(-1)^{w_{\rm 3d}}= \prod_{\alpha, i=1,\cdots,8}{\rm
 sgn}\varepsilon_{2\alpha}(\Gamma_i),
\label{eq:winding}
\end{eqnarray}
with $\Gamma_i$ in Fig.\ref{fig:bz2} (c).
The last formula determines the parity of the three-dimensional winding
number, thus it gives a
sufficient condition for nonzero $w_{\rm 3d}$.
In Fig.\ref{fig:bz2} (d), we illustrate surface states of a
three-dimensional time-reversal invariant topological superconductor. 
The Fermi surface of this model leads to $(-1)^{w_{\rm 3d}}=1$, but
$(-1)^{\tilde{\nu}[{\rm C}_{34}]=-1}$ and $(-1)^{\tilde{\nu}[{\rm C}_{56}]=-1}$.
Consistently, we have two Majorana cones on the surface at
the time-reversal invariant points $\bar{\Gamma}_{34}$ and $\bar{\Gamma}_{56}$ in the surface Brillouin
zone. (For the definition of $\bar{\Gamma}_{ij}$, see Fig \ref{fig:bz2}(c).)

We would like to emphasis here that the
correspondence between the Fermi surface and the
gapless surface state discussed in the above are inherent to
spin-triplet/odd-parity superconductors.
Thus 
we may identify the spin-triplet/odd-parity superconductivity through the
direct measurement of the surface state,
irrespective of the details of the gap function \cite{Sato09,Sato10}.
For instance, the formula (\ref{eq:winding}) has been applied to
Cu$_x$Bi$_2$Se$_3$ in order to identify its possible
odd-parity superconductivity \cite{FB10}. 
Detailed information on ABS for such a material can be
predicted by using the formulas (\ref{eq:epsilonformula2}) and
(\ref{eq:2dTI}) as well.

\subsection{Majorana fermion on the surface of topological insulator}
\label{sec:3-8}

As discussed in \S \ref{sec:3e}, the topological insulator is a 
new state of matter with time-reversal ($T-$symmetry)
\cite{KM05a,KM05b,FK06,BZ06,BHZ06}.
Especially, on the surface of the 3D strong topological 
insulator (STI), there appears odd number of 2D Dirac fermions.
Even with a disorder scattering etc., at least one Dirac fermion
is protected by the gap and topology of the bulk states and 
is stable as long as the perturbation is $T-$symmetric and
does not destroy the bulk gap.
In the case of Bi$_2$Se$_3$ and Bi$_2$Te$_3$ \cite{TI}, 
there appears only one surface Dirac fermion, which can be
regarded as the "half" of the 2D electrons because the 
spin direction is determined by the momentum.
This "fractionalization" of the electrons is one of
the key concepts in the topologically ordered states.
This situation is similar to the case of spinless fermion, and
when the Cooper paring occurs, there is a possibility that
Majorana fermions will appear \cite{Kitaev}.
Fu and Kane were the first to study the superconductivity
of the surface Dirac fermions of the STI 
induced by the proximity effect by the superconductors 
attached to it~\cite{Fu1}. 
Also they considered the ferromagnetic insulator (FI) put on the
STI, and its interface with the conventional superconductor (S).
They predicted the emergence of the one-dimensional chiral 
Majorana  mode as an Andreev bound state \cite{Fu1}. 
This situation is similar to that discussed in 
\S \ref{sec:TRBSandMF}. 
Also by applying the magnetic field, the
vortices penetrate the sample. At the core of each
vortex, the Majorana bound state is realized at zero energy \cite{Fu1,
Majoranas}.

\begin{figure}[h]
\begin{center}
\includegraphics[width=5cm]{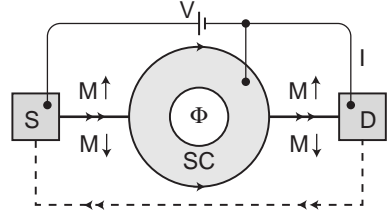}
\caption{Majorana interferometer on top of a strong topological
 insulator. 
[Reproduced from Fig. 1 of Phys. Rev. Lett. {\bf 102} 216403 (2009) by 
Fu and Kane. ] }
\label{fig:FuKane}
\end{center}
\end{figure}

Majorana interferometry has been proposed to detect the 
Majorana fermions
\cite{FK09,ANB09,LLN09,BP10}.
The schematic configuration is shown in 
Fig.\ref{fig:FuKane}  . Everything is on top of the STI, and 
there appears the chiral fermion channel at the
interface of two magnetic domains with $\uparrow$ and 
$\downarrow$ magnetization, while the chiral Majorana
fermions appear at the interface between FI and S.
Let $c^\dagger$ and $c$ 
being the creation and annihilation operators of this 
chiral fermion. At the branching points, this chiral fermion
is connected to the two Majorana edge channels for
upper and lower semicircles.  
Let $\gamma_1$ ($\gamma_2$) be the operator of
the Majorana fermion in the upper (lower) semicircle. 
Let us consider the process that an electron is entering 
from the left to the branching point described by
\begin{equation}
c \to \gamma_1 + i \gamma_2.
\end{equation}
At the right branching point, these two Majorana 
fermions merge into an electron as
\begin{equation}
\gamma_1, \gamma_2 \to c = \gamma_1 + i \gamma_2
\end{equation}
or 
\begin{equation}
\gamma_1, \gamma_2 \to c^\dagger = \gamma_1 - i \gamma_2
\end{equation}
depending on the even or odd number of vortices $\Phi/\phi_0$
($\phi_0 = hc/2e$) trapped by the superconducting donut
according to the Aharanov-Bohm effect. 
This leads to strong $\Phi$-dependence of the conductance $G$ 
between the left and right leads.
Namely, $G= 2e^2/h$ when $\Phi= 2n \times \phi_0$ while
$G=0$ when  $\Phi= (2n+1) \times \phi_0$ with 
$n$ being an integer.  This prediction is based on the 
fact that the Majorana fermion contains the particle and
hole components with equal weight, or an electron is composed of
two Majorana fermions. 

\begin{figure}[htb]
\begin{center}
\scalebox{0.8}{
\includegraphics[width=11.0cm,clip]{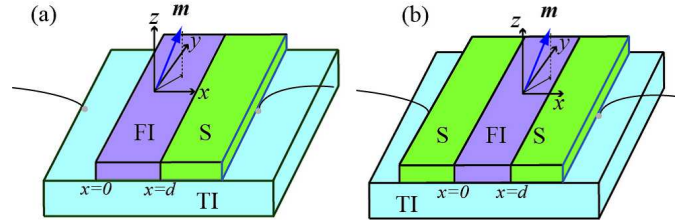}
}
\end{center}
\caption{(Color online)
Schematic illustration of the junction. 
(a) Normal metal (N) / Ferromagnetic insulator (FI) / 
Superconductor (S) junction and 
(b) S/FI/S junction formed  on the surface of 
3D topological insulator (STI). The current is flowing on the surface of STI. 
[Reproduced from Fig. 1 of Phys. Rev. Lett. {\bf 103} 107002 (2009) by Tanaka $et.$ $al$.] }
\label{fig:Majorana}
\end{figure}

 The other direction of the research is to study 
the possible manipulation of the Majorana fermion and
its relevance to the transport properties in the 
transverse direction to the interface \cite{TYN09,Linder10a,Linder10b2}, 
For that purpose we consider the Hamiltonian 
for the surface state on STI  
influenced by the superconductor and ferromagnet ( Fig. \ref{fig:Majorana} )
as given by 
\begin{eqnarray}
\check{H}_{S} = \left( {\begin{array}{*{20}c}
{\hat H( {\bm k}) + \hat{M} } & {\hat \Delta } \\
{ - \hat \Delta ^*  } & { - \hat H^ * \left( 
{ - {\bm k}} \right) - \hat{M}^{*} } \\
\end{array}} \right)
\label{Hamiltonian}
\end{eqnarray}
where
$\hat{H}(\bm{ k})=v_F(\hat{\sigma}_{x}k_{x}+\hat{\sigma}_{y}k_{y}) 
-\mu[\Theta(-x)+\Theta(x-d)]$   
describes the 2D Dirac fermion, 
and 
$\hat{M}= {\bm m} \cdot \hat{\bm \sigma}\Theta(d-x) \Theta(x)$
is the magnetic exchange interaction 
with 
${\bm m} \cdot \hat{\bm \sigma}= m_{x}\hat{\sigma}_{x} +
m_{y}\hat{\sigma}_{y} + m_{z}\hat{\sigma}_{z}$. 
Here, $\mu$, 
$\hat{\bm{\sigma}}$, $v_F$, ${\bm m}$ denote
chemical potential, Pauli matrices, velocity, and 
magnetization (times the exchange coupling 
constant which we assume to be 1), respectively \cite{Fu1}. 
Interestingly, ${\bm m}$ couples to the Dirac fermion as an effective 
vector potential ${\bm A}$ of the electromagnetic field. 
For the moment, we assume that the pairing symmetry of superconductor is  
spin-singlet $s$-wave and $\hat{\Delta}$ is given as  
$\hat{\Delta} = i \hat{\sigma}_{y} 
\Delta \Theta(x-d)$ and  
$\hat{\Delta} = i \hat{\sigma}_{y} 
[\Delta \Theta(x-d) + \Delta \Theta(-x)\exp(i\varphi)]$  
for N/FI/S (Fig. \ref{fig:Majorana} (a)) and S/FI/S  (Fig. \ref{fig:Majorana} (b)) 
junctions formed on the surface of STI, respectively, 
where $\varphi$ denotes the macroscopic phase difference between 
left and right superconductor. 
(N represents the normal metal
which is the 2D Dirac fermion in the present case.)
We also assume that the magnitude of $\Delta$ is smaller than 
that of the bulk energy gap of superconductor deposited on STI 
because the S/TI interface is not ideal \cite{Fagas}. 

%

\begin{figure}[htb]
\begin{center}
\scalebox{0.8}{
\includegraphics[width=8cm,clip]{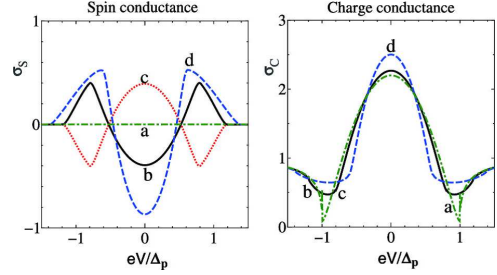}
}
\end{center}
\caption{(Color online) 
Left Panel: Chiral Majorana mode energy dispersion $E_{b}$ 
as a function of the incident angle $\theta$. 
Right Panel: 
Normalized tunneling conductance $\sigma$ in N/FI/S junctions. 
$m_{z}d/v_F=1$ and $m_{y}/m_{z}=0$. 
a: $\mu/m_{z}=1$, b: $\mu/m_{z}=2$ and c: $\mu/m_{z}=0.5$. 
[Reproduced from Fig. 2 of Phys. Rev. Lett. {\bf 103} 107002 (2009) by Tanaka $et.$ $al$.] }
\label{fig:Andreev}
\end{figure}

First, we solved the Andreev bound state (ABS) at 
the interface of FI/S. 
As an example, we show in Fig.  \ref{fig:Andreev} 
the energy dispersion as a function of the angle 
$\theta$ (corresponding
to the momentum along the interface) in 
the right panel, and the corresponding 
conductance $\sigma$ of the  N/FI/S junction 
in the left panel for several values of $\mu/m_z$. 
Note that the slope of the
energy dispersion affects the conductance $\sigma$
sensitively. Therefore, it is concluded that the 1D Majorana channel
can be controlled by the magnetization of the FI and the chemical 
potential $\mu$.
 
\begin{figure}[htb]
\begin{center}
\scalebox{0.8}{
\includegraphics[width=5cm,clip]{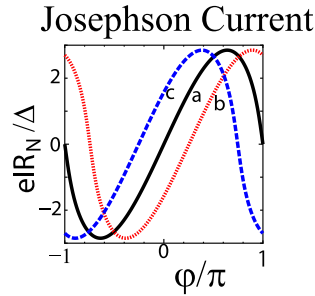}
}
\end{center}
\caption{(Color online) 
Josephson current in S/FI/S junctions 
is plotted  with a: $m_{x}/m_{z}=0$, b: $m_{x}/m_{z}=0.4$ 
and c: $m_{x}/m_{z}=-0.4$. 
We choose 
$m_{z}d/v_F=1$, $\mu/m_{z}=1$ and $m_{y}/m_{z}=0$. $T=0.05T_{C}$ with 
transition temperature $T_{C}$. 
[Reproduced from Fig. 4 from Phys. Rev. Lett. {\bf 103} 107002 (2009) by Tanaka $et.$ $al$.] }
\label{fig:Josephson}
\end{figure}

Next, we consider the Josephson effect in S/FI/S junction 
(Fig. \ref{fig:Majorana}(b)).  This is mainly due to the 
tunneling process between the
two Majorana edge channels at the two interfaces between S and FI.
One interesting prediction is that the current-phase relation can be shifted
by the $m_x$. Figure \ref{fig:Josephson} shows the Josephson current
as a function of the  phase different $\varphi$ between the two superconductors
for different values of $m_x$, where the parallel shift of the curves is seen.

The generalization to the anisotropic pairing superconductors on STI has been 
discussed in ref.\cite{Linder10a,Linder10b2} .
It is found that the proximity effect to the superconductivity is strongly dependent 
on the pairing symmetry. 
For spin-triplet $p$-wave superconductor, there occurs no proximity effect because the spin direction of
the 2D Dirac electrons is not consistent with that of the $p$-wave pairing. 
In the cases of spin-singlet $s$-wave and $d$-wave pairings, there occurs the proximity effect, 
and the Majorana fermions appear at the interface of FI and S. 
For spin-singlet $d_{xy}$-pairing, the zero energy states appear along the edge of the sample even without
the magnetization of FI, {\it i.e.}, without $T-$symmetry breaking, which has been well-known.
However, the new aspect here is that the zero energy dispersionless bound states
are Majorana fermions due to the proximity to STI.
With the $z$-component of the magnetization, the Majorana fermion obtains the 
dispersion, which results in the very different behavior of the conductance 
compared with the case of spin-singlet $s$-wave superconductors. \par

\section{Discussion}
\label{sec:4}
\subsection{Interplay between symmetry and topology}

\begin{table}
\begin{center} 
\begin{tabular}[h]{|c|c|c|c|c|c|c|}
\hline
 & ZEABS & dispersion & top. index &
 odd-$\omega$ pairing & Majorana &section\\
\hline
$s$ & No & - & - & No &-&2.2,\,3.3\\
\hline
$d_{xy}$ & Yes & No &$w_{\rm 1d}$ & OSO& double&2.2,\,3.3\\
\hline
$p_x$ & Yes & No & $w_{\rm 1d}$ &OTE &double&2.2,\,3.3 \\
\hline
chiral $p$ (spinless) & Yes& Yes &$w_{\rm 2d}$ & OTE & chiral&3.4\\
\hline
chiral $p$ (spinful) & Yes& Yes &$w_{\rm 2d}$ &OTE & double&3.4\\
\hline
$s+p$ ($p>s$)&Yes &Yes &${\bm Z}_2$ &OTE+OSO & helical&3.6\\
\hline
$s+p$ ($p<s$)&No& - &- &No &-&3.6\\
\hline
$d_{xy}+p$ & Yes &No &$W$ &OSO+OTE &flat&3.6\\
\hline
 \end{tabular} 
\end{center}
\caption{Edge states of various superconductors. Each column describes
 the type of bulk superconductor, zero energy
 Andreev bound state (ZEABS), the presence or absence of the energy
 dispersion, topological index characterizing the ZEABS, symmetry of the
 induced odd-frequency pairing, the type of Majorana fermion, and the
 corresponding sections in the text, respectively.}
\label{table:summary}
\end{table}

Up to now, we have studied induced odd-frequency pairings in \S \ref{sec:2}
and topological bulk-edge correspondence in \S \ref{sec:3}. 
In this section, we discuss the relation between these two.

First, all the known zero energy edge states at the surface of superconductors
can be characterized by the topological indices, and are topologically
protected. 
These zero energy states
are always associated with the pure odd-frequency pairing at the
surface. (When the ABS has the energy
dispersion, this statement applies only to the sector of $k_y=0$.) 
In Table \ref{table:summary}, we summarize the results for various
situations studied thus far.

For $s$-wave superconductor, which belongs to ESE, there is no ZEABS
because it is topologically trivial. At the same time, odd-frequency
pairing is absent at the surface.
However, $d_{xy}$-wave superconductor, which also belongs to ESE, has ZEABS
characterized by the topological index $w_{\rm 1d}$. 
$w_{\rm 1d}$ is the winding number defined for the one-dimensional
systems when $k_y$ is fixed as discussed in \S \ref{sec:3c}.
The formation of the ZEABS stems from the sign change of the pair
potential felt by injected and reflected quasiparticles.
The similar situation occurs for $p_x$-wave superconductor which belongs
to ETO pairing. 
In these cases, pure odd-frequency pair amplitudes (OSO for $d_{xy}$, OTE
for $p_x$) exist at the surface. 
The ZEABSs can be regarded as double Majorana modes satisfying
\begin{eqnarray}
\gamma_{\sigma}(k_y)
=\gamma_{-\sigma}(-k_y),
\quad (\sigma=\uparrow, \downarrow).
\end{eqnarray}
Next, we discuss the spinless chiral $p$-wave superconductor where the
time-reversal symmetry is broken in the bulk.
There exists ABS, however, in contrast to the cases above,
it has a dispersion linear in $k_y$.  
Because of the broken time-reversal symmetry, it is characterized by
the Chern number $w_{\rm 2d}$ in a manner similar to the quantum
Hall states.
The induced odd-frequency pair amplitude at the surface belongs to OTE
symmetry.
The present ABS is a chiral Majorana mode, which is a genuine Majorana
fermion. 
The case of the spinful chiral $p$-wave superconductor can be regarded
as the two copies of the spinless case discussed above.
Lastly, the noncentrosymmetric superconductors (NCSs) are discussed.
In these superconductors, the parity of the pair amplitudes are mixed.
For the case of $s$-wave and $p$-wave mixture, there are two
topologically distinct states depending on the relative magnitudes of
the corresponding pair potentials.
When $p$-wave component is dominant, the ABS is generated as a helical
Majorana edge mode. 
The pair amplitude induced at the surface has OTE+OSO symmetry. 
The topological index characterizing the ABS is the ${\bm Z}_2$ invariant. 
On the other hand, when $s$-wave component is dominant, no ABS is generated
because it is smoothly connected to the topologically trivial $s$-wave
superconductor case without closing the bulk gap.
For ($d_{xy}+p$)-wave  pair potential case, a single branch of
dispersionless ABS appears if $k_y$ satisfies some condition. 
This ZEABS is a single Majorana fermion.
The topological index $W$ defined by eq.(\ref{eq:Wdef}) ensures its
topological stability \cite{STYY11}. The symmetry of induced odd-frequency pair
amplitude is OSO+OTE.

\subsection{Related topics}
There are several important issues which are not addressed in this
review article due to the limited space.
We will briefly mention each of them by indicating relevant references.

In this review article, we mainly focus on  the interface and surface of
superconductors. 
There are several works on the bulk superconductors from the viewpoint of 
topology and/or odd-frequency pairings. 
Non-interacting topological insulators and topological superconductors
have been classified based on the universal class of the Anderson localization.
\cite{SRFL08,RSFL10}
The similar results have been reached independently by Kitaev \cite{Kitaev09}
in the framework of K-theory.
The classification scheme also has been applied for solitons in
topological insulators and topological superconductors.\cite{Teo} 
Possible bulk odd-frequency superconductors have been an important issue
in many-body physics. 
Recently, there are several proposals of odd-frequency gap functions in
quasi one-dimensional systems \cite{Shigeta,Shigeta2}, 
magnetic superconductors\cite{Shigeta3}, heavy fermion 
systems \cite{Fuseya,Hotta}, and strongly coupled electron-phonon
systems \cite{Kusunose2}.
However, it is noted that the thermodynamic stability 
and physical properties of the bulk 
odd-frequency pairing superconductors are still  controversial issues \cite{Solenov,Fuseya2011,Kusunose}.

Since superfluid $^{3}$He is a well established spin-triplet $p$-wave 
superfluid, it is natural to expect ABS \cite{ABS,ABSb}. 
ABSs in superfluid $^{3}$He have been predicted 
in the 1980's \cite{ABS,ABSb}. 
However, the detection of them is not easy since 
Cooper pairs of superfluid $^{3}$He do not have electric charge. 
Surface ABS of $^{3}$He B phase
has been observed experimentally 
by the transverse acoustic impedance \cite{Aoki2005} in 2005 
and specific heat measurements \cite{Choi06} in 2006.  
In the actual surface of $^{3}$He B phase, 
diffusive scattering by surface roughness plays an important 
role \cite{Nagato98,Nagai08}. 
When a surface is specular, SABS has a Dirac like linear dispersion 
which have been recently interpreted  as 
Majorana fermion \cite{QHRZ09,Chung}. 
There are several theoretical discussions about SABS 
from the viewpoint of topological 
superconductivity and Majorana fermions 
\cite{SRFL08,RSFL10,QHRZ09,Nagato09,Volovik091,Volovik092,Tsutsumi10,TsutsumiPRB10}.
It is noted that surface Majorana cone has been experimentally 
observed for highly specular wall of  $^{3}$He B phase 
\cite{Murakawa09,Murakawa11}. 
Surface Majorana ABS (SMABS) has been also discussed in slab geometry of 
$^{3}$He A and B phases \cite{Tsutsumi10,TsutsumiPRB10}. 
An analogous system of diffusive normal metal/spin-triplet
superconductor can be
realized in superfluid aerogel/$^3$He hybrid system. 
Anomalous proximity effects discussed in \S \ref{sec:2d} are also expected
in the aerogel, since it can be regarded as a diffusive 
Fermi liquid \cite{Higashitani09}. \par

Recently, promising candidate of topological superconductor
has been discovered in doped topological insulator
Cu$_{x}$Bi$_{2}$Se$_{3}$ \cite{Hor10}.
Several experimental \cite{Wray10,Kriener11}
and theoretical \cite{FB10,Hao11} researches have started.
MSABS are also expected in the present new superconductors.

Majorana fermion can be a platform of topological
quantum computation. 
Various systems have been proposed to produce Majorana fermion; 
spin-triplet $p$-wave superconductors / superfluids \cite{Ivanov01,TDNZZ07,Mizushima08,ZTLD08,CS09,Mizushima10,Tsutsumi10,Kawakami09,Kawakami11}, 
$\nu=5/2$ fractional quantum Hall system \cite{MR91}, 
spin singlet $s$-wave superconductor on topological
insulator surface with vortex or the interface with ferromagnet 
\cite{FK08,ANB09,NOF10}, noncentrosymmetric superconductors with Zeeman
field \cite{Fujimoto08, STF10, SF10}, $s$-wave superfluid of cold atoms
with laser generated spin-orbit interaction \cite{STF09, STF10},  
semiconductor/SC systems including Rashba spin-orbit coupling with
FM \cite{SLTD10} or Zeeman field \cite{Alicea10},
one-dimensional system made of 2D QSHS\cite{FK09} or
nanowire\cite{Kitaev01,LSD11,ORO10,AOROF10, BDRO11, BDRO11b, SLHS11}, 
quantum anomalous Hall/
SC hybrid system\cite{QHZ10,IYST11}, and Half metal /SC system
\cite{CZQZ10,Lee09}.

The detection of Majorana fermion is not a trivial issue since it is a
neutral particle. 
Majorana fermions in condensed matter physics are emergent particles
resulting from the many-body interactions of electrons which are charged.
Thus, they can be detected through the conversion processes to electron
influenced by the phases of the superconducting order parameters.  
In addition to the Majorana interferometry on topological insulator
discussed in \S \ref{sec:3-8} \cite{FK09,ANB09,LLN09,BP10}, there are many
other proposals for this purpose: (i) Josephson effect
\cite{Kitaev01,FK09,LSD10,TYN09,ATN10,PN11}, (ii) charge transport
thorough Majorana fermion \cite{LLN09,Flensberg10,SRM10,Fu10,Beri10L,Beri11}, 
(iii)
transport noise \cite{CQMZ11,GH11}, and (iv) spin anisotropy
\cite{SF09,Chung, Nagato09,SFN10,YSTY10}. \par

\subsection{Perspectives}

The two major fields opened up in 80's are the 
quantum Hall effect and the high-$T_c$ superconductors.
The former introduced the concept of topology and the latter 
the electron correlation in condensed matter physics.
The idea of the topology extended to the time-reversal 
symmetric systems and the topological insulators have been
discovered. On the other hand, the electron correlation 
naturally leads to the anisotropic Cooper pairings with 
various symmetries. These two streams are now merging 
into one producing many interesting new directions in the 
current condensed matter physics. 

It is now recognized that there are 3 distinct classes
of currents in solids; (i) One is the conventional Ohmic 
currents with dissipation. (ii) The second is the 
superconducting currents and/or the superfluid flow 
with broken gauge symmetry and macroscopic quantum 
coherence. (iii) The third one is the ``topological 
currents'' induced by the nontrivial topology of
wavefunctions, and does not require the broken 
gauge symmetry. The last two have the 
common feature of dissipationless currents.
The last one has not been paid attention 
thus far, but will be very unique because it can 
operate at room temperature when the gap protecting
the topology is larger than its energy scale.

These three classes have been considered to be rather
clearly separated. However, as we have described in this
review article, class (ii) and class (iii) are 
overlapping each other. Namely, there is a class of
topological supercurrent at the edge/surface of the 
superconductors/superfluid. Therefore, the interplay
of the broken gauge symmetry and topology
offers an intriguing new arena for the 
future condensed matter physics. 

These fundamental understandings of the currents in 
solids will be essential to develop the  electronics 
in the next generation. The dissipationless operations
of devices are the key ingredient for the 
future applications to electronics, and hopefully
the topological supercurrent will serve this goal.
The other important issue is the quantum computing.
Majorana fermion is regarded as the promising 
candidate for fault-tolerant qubit.
In general, the topological orders bring about 
the electron fractionalization, which gives  
the robustness of the quantum states. 
The three degrees of freedom of electrons, {\it i.e.},
spin, chirality, particle-antiparticle,  can be fractionalized
at the edge/surface of topologically ordered bulk states, and
chiral Majorana fermions can be regarded as the 
ultimate fractionalized quantum state. 
On the other hand, the other direction is to recombine
these fractionalized particles into new types of 
particles \cite{TN09}. The more the degrees of freedom,
the more the system is susceptible to the external 
stimuli, and hence we can adjust the 
robustness and sensitivity of the system at will by 
manipulating the fractionalization/recombination.

To verify these theoretical proposals, experimental progresses are
indispensable.
There are several relevant experimental techniques such as $\mu$SR, NMR,
neutron scattering, X-ray diffraction, and magneto transport. 
In particular, the direct observation of edge/surface states by the tunneling
spectroscopy and ARPES is highly desired. 
For more details, the readers are referred to other contributions in
this special volume.

Finally, we would like to stress that physics of 
interface/surface superconductivity includes unexplored quantum
phenomena and will be a central topic in
condensed matter physics in the decades to come.

\begin{acknowledgements}
The authors thank T. Akazaki, Y. Ando, Y. Asano, A.V. Balatsky, Y. Fominov,
S. Fujimoto, A. Furusaki, Y. Fuseya, A.A. Golubov, N. Hayashi,
 S. Higashitani, M. Ichioka, J. Inoue, S. Kashiwaya, M. Kohmoto, H. Kusunose, 
J. Linder, Y. Maeno, K. Miyake, T. Mizushima, K. Nagai, S. Onari, M. Oshikawa, 
X.L. Qi, A. P. Schnyder, R. Shindou, A. Sudbo, Y. Takahashi, Y.Tanuma, M. Ueda, K.Yada, T. Yokoyama, S.C. Zhang for valuable discussions. 
This work is supported by Grant-in-Aid for Scientifc Research 
(Grants No. 17071007, No. 17071005, No.19048008, No. 19048015,
No. 20654030, 
No. 22103005(Innovative Areas
"Topological Quantum Phenomena"),
No. 22340096, No. 21244053 and No.22540383) from the Ministry of Education, Culture,
Sports, Science and Technology of Japan, Strategic International Cooperative Program
(Joint Research Type) from Japan Science and Technology Agency, and
Funding Program for World-Leading Innovative RD on Science and Technology (FIRST Program).
\end{acknowledgements}

\bibliographystyle{jpsj}
\bibliography{topologicalSC}

\begin{thebibliography}{100}
\bibitem{exchange}
It is noted that the exchange of the two time variables does not mean the time
  reversal. Therefore the odd-frequency pairing does not mean necessarily the
  time reversal symmetry breaking.
\bibitem{QHE}
Note that the plateau observed experimentally is explained by the disorder and
  consequent Anderson localization effect. See for example Chap. 3 of ref.1.
\bibitem{winding1d}
Note that $w_{\rm 1d}$ is different from $w$ in ref.\cite{STYY11} by a factor
  $-1$ while they are essentially the same topological index.
\bibitem{winding2d}
Here we implicitly assume that ${\rm sgn}[{\rm det}\partial_{k_j}\Delta_j({\bm
  k})]\neq 0$ at ${\bm k}$ with $\Delta({\bm k})=0$. If this is not the case,
  one needs to deform the gap function slightly to satisfy the assumption.
  After this regulaization, one can apply the formula. This prescription does
  not change the value of the topological index.
\bibitem{Majoranas}
A similar Majorana bound state in a vortex and its non-Abelian statistics for
  an $s$-wave superconducting state have been discussed for the first time in
  ref.\cite{Sato03} in a different context. For the relation to Fu-Kane model,
  see also appendix E in ref.\cite{STF10}.
\end{thebibliography}

\end{document}